\documentclass[a4paper,10pt]{article}

\pdfoutput=1
\usepackage{jheppub}
\usepackage{array}
\usepackage{multirow}
\usepackage{amsmath}
\usepackage{makecell}

\definecolor{green}{rgb}{0.1,0.8,0.2}

\usepackage{cancel}
\usepackage{amsmath,amssymb} 
\usepackage{graphicx} 
\usepackage[dvipsnames,x11names]{xcolor} 

\usepackage{hyperref}
\usepackage{cleveref}
\usepackage{float}
\usepackage{subcaption}

\makeatletter
\newcommand{\footnoteref}[1]{\protected@xdef\@thefnmark{\ref{#1}}\@footnotemark}
\makeatother

\begin{document}

	\preprint{}
	\title{Non-minimal coupling of scalar and gauge fields with gravity: an entropy current and linearized second law}
	\author[a]{Parthajit Biswas}
    \affiliation[a]{Department of Physics, Indian Institute of Technology Kanpur, Kalyanpur, Kanpur 208016, India}

    \author[a]{, Prateksh Dhivakar}
    \author[a]{, and Nilay Kundu}
    
	\emailAdd{parthajitbiswas8@gmail.com, prateksh@iitk.ac.in, nilayhep@iitk.ac.in}

	\abstract{This work extends the proof of a local version of the linearized second law involving an entropy current with non-negative divergence by including the arbitrary non-minimal coupling of scalar and $U(1)$ gauge fields with gravity. In recent works, the construction of entropy current to prove the linearized second law rested on an important assumption about the possible matter couplings to gravity: the corresponding matter stress tensor was assumed to satisfy the null energy conditions. However, the null energy condition can be violated, even classically, when the non-minimal coupling of matter fields to gravity is considered. Considering small dynamical perturbations around stationary black holes in diffeomorphism invariant theories of gravity with non-minimal coupling to scalar or gauge fields, we prove that an entropy current with non-negative divergence can still be constructed. The additional non-minimal couplings that we have incorporated contribute to the entropy current, which may even survive in the equilibrium limit. We also obtain a spatial current on the horizon apart from the entropy density in out-of-equilibrium situations. We achieve this by using a boost symmetry of the near horizon geometry, which constraints the off-shell structure of a specific component of the equations of motion with newer terms due to the non-minimal couplings. The final expression for the entropy current is $U(1)$ gauge-invariant for gauge fields coupled to gravity. We explicitly check that the entropy current obtained from our abstract arguments is consistent with the expressions already available in the literature for specific model theories involving non-minimal coupling of matter with higher derivative theories of gravity. Finally, we also argue that the physical process version of the first law holds for these theories with arbitrary non-minimal matter couplings.
}


\maketitle

\section{Introduction} \label{sec:intro}

It is well known that black hole solutions in Einstein's General Relativity (GR), a two-derivative classical theory of gravity, obey the laws of thermodynamics \cite{Hawking:1971tu, Bardeen:1973gs, Bekenstein:1973ur, Hawking:1974sw}. However, GR cannot provide a complete description. Upon taking the low energy limit of any UV complete theory of quantum gravity, one generically expects to obtain higher derivative corrections to the leading two derivative theory. Motivated by this, it is, therefore, quite natural to consider modifications of Einstein's GR by allowing arbitrary higher derivative terms in the Lagrangian of the theory \footnote{When we say modifications of GR, we have in mind a situation allowed by the relevant length scales, where gravity is still weakly coupled so that we can treat it classically ignoring quantum effects. However, the higher derivative corrections to the Einstein-Hilbert piece in the Lagrangian cannot be ignored. We assume such a window will be available, at least in principle, in a UV complete theory of gravity.}, and further wonder whether black holes in these diffeomorphism invariant higher curvature theories of gravity would obey the laws of thermodynamics. Surprisingly enough, we still do not entirely understand black hole thermodynamics in the arbitrary diffeomorphism invariant theory of gravity. 

For arbitrary diffeomorphism invariant theories of gravity, the first law was shown to be obeyed in \cite{Wald:1993nt, Iyer:1994ys} \footnote{A different version of the first law, known as the physical process version of it, has also been discussed in the literature, see \cite{Jacobson:1995uq, Gao:2001ut, Amsel:2007mh, Bhattacharjee:2014eea, Chakraborty:2017kob, Chatterjee:2011wj, Kolekar:2012tq, Bhattacharyya:2021jhr}.}. This construction also defines an equilibrium entropy for stationary black holes \footnote{Stationary black holes possess Killing horizons, a null hypersurface where the norm of a Killing vector vanishes.}, known as the Wald entropy. An understanding of the zeroth law for arbitrary higher derivative theories of gravity is also available in the literature. For arbitrary higher curvature theory of gravity, using specific space-time symmetries, this was shown explicitly in \cite{Zerothwald}. Recently in \cite{Bhattacharyya:2022nqa}, the zeroth law has been abstractly argued for diffeomorphism invariant theories of gravity where higher derivative terms could be considered as corrections to a leading two derivative theory of Einstein's GR \footnote{Previously, for specific theories of higher derivative gravity this has been worked out in \cite{Ghosh:2020dkk, Reall:2021voz, Xie:2021bur, Sang:2021rla}. }. 

Our focus will be on the second law for dynamical black holes in this work. It is well known that when out of equilibrium dynamical configurations are taken into account, the Wald entropy suffers from possible ambiguities, known as the JKM ambiguities \cite{Jacobson:1993xs, Jacobson:1993vj, Jacobson:1995uq}, and we still lack a complete proof of the second law for arbitrary higher curvature theory of gravity. Various attempts have been made to extend the equilibrium definition of Wald entropy to configurations involving non-stationary dynamics such that the second law is satisfied in such higher derivative theories of gravity, see \cite{Iyer:1994ys, Jacobson:1993xs, Jacobson:1993vj,  Jacobson:1995uq,  Wall:2011hj, Sarkar:2013swa, Bhattacharjee:2015yaa, Wall:2015raa, Bhattacharjee:2015qaa, Bhattacharyya:2016xfs}. In a given theory, one generally considers the following setup: an initial stationary black hole configuration is perturbed slightly with an external source, and then the black hole finally settles down to another stationary configuration. The amplitude of the external time-dependent source is assumed to be very small so that one can perform an approximate linearized expansion in that small parameter, known as the linearized amplitude expansion. The main idea of all the works mentioned above was to argue that a particular way of non-stationary extension of the Wald entropy, i.e., fixing the JKM ambiguities, establishes a linearized version of the second law. We refer the readers to the recent reviews \cite{Wall:2018ydq, Sarkar:2019xfd} which discuss the progress in black hole thermodynamics in theories beyond GR. 

Recently in \cite{Bhattacharya:2019qal, Bhattacharyya:2021jhr}, following the work of \cite{Wall:2015raa}, a linearized version of the second law was argued for arbitrary diffeomorphism invariant theory of gravity by constructing an entropy current with non-negative divergence on the horizon of a dynamically perturbed black hole \footnote{See the interesting recent work \cite{Hollands:2022fck} for exploration of the second law beyond linearized approximation.}. As explained above, these analyses were carried out working within the approximation of linearized amplitude expansion around stationary black hole space-time. Since we will be extensively using the construction of the entropy current in our present calculations, let us briefly point out some highlighting features of its basic working principle; for details, see \cite{Bhattacharya:2019qal, Bhattacharyya:2021jhr}. 

The small amplitude of the dynamical perturbation is denoted by $\epsilon$. The discussion will focus on terms up to linear order in $\epsilon$, i.e., $\mathcal{O}(\epsilon)$, whereas all terms of the quadratic order $\mathcal{O}(\epsilon^2)$ and higher will be ignored. Without any loss of generality, a particular choice of gauge for the stationary black hole metric is fixed. Once this gauge is fixed, there are certain residual coordinate transformations under which the near horizon space-time of the stationary black hole is invariant: we call this the boost symmetry. Of course, dynamical fluctuations slightly break this boost symmetry of the stationary black hole. However, within the linearized amplitude expansion, one can systematically organize the breaking in a perturbative expansion in the amplitude of the perturbation, i.e., $\epsilon$, up to linearized order. Furthermore, in order to obtain the entropy current, the main object one needs to study is the off-shell structure of the equations of motion (EoM) that follows from the diffeomorphism invariant Lagrangian of the theory following the Noether charge construction of \cite{Iyer:1994ys}. The boost symmetry, along with its small breaking due to fluctuations arranged in the $\epsilon$-expansion, is powerful enough to predict this off-shell structure of the EoM up to $\mathcal{O}(\epsilon)$, when evaluated on the null horizon. Consequently, the off-shell structure of specific components of the EoM, involving the metric coefficient functions and various derivatives acting on them, gets manifestly expressed in terms of the divergence of an entropy current defined on the horizon. At this point, it is essential to emphasize that these statements about the structural form of the EoM hold true for any diffeomorphism invariant theory of gravity, without any reference to the second law of black hole thermodynamics. 

In the next step, to construct the proof for the linearized second law using the result mentioned above about the off-shell structure of the EoM, one needs to consider the energy-momentum tensor corresponding to the matter sector present in the Lagrangian for the theory. One may restrict oneself to situations where only pure gravity theories are studied and a geometric version of the second law can still be devised. However, that is not very physically appealing since the energy-momentum tensor is required as a source in the gravity equations to initiate the dynamical fluctuations. An important assumption is made at this point: the energy-momentum tensor satisfies the classical null energy condition (NEC). Within the setup of linearized amplitude ($\epsilon$) expansion, this is equivalent to demanding that up to $\mathcal{O}(\epsilon)$ the null projected components of the energy-momentum tensor would vanish \footnote{In other words, this requires that the components of the energy-momentum tensor would have to be $\mathcal{O}(\epsilon^2)$. This is because depending on the sign of the amplitude $\epsilon$, which is arbitrary, any $\mathcal{O}(\epsilon)$ piece can never guarantee a definite sign for the energy-momentum tensor components relevant for the NEC.}. Implementing the NEC for the energy-momentum tensor, in turn, fixes the sign of the relevant component of the EoM. This is where one needs to use the fact that the dynamical black hole configuration satisfies the EoM in the $\epsilon$-expansion; however, it is used in a formal sense without ever explicitly solving it. From this, one readily obtains that the divergence of the entropy current is non-negative up to $\mathcal{O}(\epsilon)$, and hence the linearized second law is proved. 

The main goal of this paper is to critically examine this assumption of the NEC, which was a crucial input in the construction of the entropy current in \cite{Bhattacharya:2019qal, Bhattacharyya:2021jhr}. We should recollect that the classical energy-momentum tensor is obtained by varying the matter part of the Lagrangian with respect to the metric. Therefore, while varying the Lagrangian of a given diffeomorphism invariant theory of gravity involving additional matter fields, we need to understand how to separate the terms corresponding to the EoM of the metric from the energy-momentum tensor for the matter fields. Once we have decided what the energy-momentum tensor is, we should be able to verify that it satisfies the classical NEC, or the relevant components of it vanish up to $\mathcal{O}(\epsilon)$. In this context, the most important and subtle issue that needs special attention is the following: how the matter fields are coupled to gravity. The simplest possible scenario is when the matter fields are minimally coupled to gravity. We will be working with the following prescription for minimal coupling given in \cite{carroll_2019,blau}: given a Lorentz invariant expression of matter fields coupled to the flat Minkowski metric in the Lagrangian, the minimal coupling is defined by simply lifting the Minkowski metric $\eta_{\mu\nu}$ to the metric of the gravitational field $g_{\mu\nu}$ and the partial derivative $\partial_{\mu}$ to covariant derivative $D_{\mu}$ compatible with the metric $g_{\mu\nu}$. Any other couplings between the matter fields and the metric tensor, which does not originate following the mentioned prescription, will be termed non-minimal coupling. 

However, it is known in the literature that if one considers the diffeomorphism invariant theory of gravity with non-minimal couplings with matter fields, the classical NEC can get violated, \cite{Barcelo:2000zf, Flanagan_1996, Wall:2018ydq, Chatterjee:2015uya}. In other words, for such non-minimally coupled Lagrangian, we may get a non-vanishing contribution to the EoM or energy-momentum tensor even at $\mathcal{O}(\epsilon)$, for which the NEC may easily get violated for some choice of the sign of the amplitude $\epsilon$ of the dynamical perturbation. In that case, the analysis of \cite{Bhattacharyya:2021jhr} will not be applicable, and the validity of the linearized second law will break down. However, the status of the linearized second law has already been examined in the literature focussing on particular models of non-minimal coupling. In the recent works \cite{Wang:2020svl, Wang:2021zyt}, the authors have looked at models of theories where a scalar or a $U(1)$ gauge field has non-minimal couplings with the space-time metric, but the couplings are restricted to only four derivative terms in the Lagrangian. In these references, a brute force calculation of such problematic terms mentioned above in those specific models argued that the second law holds true. However, just like \cite{Wall:2015raa}, their calculations missed the spatial components of the entropy current as they were arguing for the increase of the entropy integrated over a compact spatial slice of the horizon. 

In our present work, motivated by the results mentioned above, we focus on theories with non-minimal coupling between scalar or $U(1)$ gauge fields with gravity. Very recently, in \cite{Hollands:2022fck}, the authors have considered scalar fields coupled to gravity as an effective field theory (EFT), and they have argued for the second law up to quadratic order in perturbations in an EFT approximation. However, we consider the validity of the second law only up to the linear order in perturbations. Thus we do not explicitly mention any restrictions on the size of the coupling of higher derivative terms \footnote{Our final result for the scalar field coupled to gravity is more or less evident given what has been recently reported in \cite{Hollands:2022fck}. Nevertheless, our analysis is done in a slightly different language which is more aligned with the methodology followed in \cite{Bhattacharyya:2021jhr}. Additionally, we will explicitly construct the algorithm to obtain the entropy current once a Lagrangian with arbitrary scalar field coupling is provided. Furthermore, it will set up the context for discussing the gauge field coupling, which has not been considered in \cite{Hollands:2022fck}.}. For gauge fields coupled to gravity, we focus on Lagrangians which are manifestly gauge-invariant involving only the field strength tensor but arbitrary otherwise. Firstly, once we vary the Lagrangian with respect to the metric $g_{\mu\nu}$, there will be contributions that will follow from the purely gravitational part of the Lagrangian, which we will consider as the EoM for the space-time metric. These contributions will, in principle, be relevant at $\mathcal{O}(\epsilon)$. However, they are trivial because they have already been taken care of in \cite{Bhattacharyya:2021jhr} leading to an entropy current resulting from the purely gravitational terms in the Lagrangian. Then, for the minimally coupled matter fields in the Lagrangian, we will receive contributions classified as the energy-momentum tensor. However, they will always be of $\mathcal{O}(\epsilon^2)$ and thus will be taken care of by the NEC \cite{Chatterjee:2015uya}. Finally, we will have to determine the contributions that the energy-momentum tensor receives from the non-minimal couplings. As we argued before, they can, in principle, be relevant at $\mathcal{O}(\epsilon)$, and thus cannot be thrown away with the assumption of the NEC. 

Following the same setup prepared in detail in \cite{Bhattacharyya:2021jhr}, we will analyze these non-minimal energy-momentum tensor components in this paper. More precisely, using the Noether charge construction of \cite{Iyer:1994ys}, and the boost-symmetry of the near horizon space-time of the black hole \cite{Bhattacharyya:2021jhr}, we will abstractly argue that all such contributions to the energy-momentum tensor from the non-minimal couplings will also have the exact off-shell structure, up to $\mathcal{O}(\epsilon)$, required to construct an entropy current when evaluated on the horizon. Thus we will be able to show that these non-minimal terms would simply lead to contributions to the entropy current in addition to what comes from the purely gravitational piece in the Lagrangian. Although these additional non-minimal terms can violate the NEC, they have the same off-shell structure as the purely gravitational terms already considered in \cite{Bhattacharyya:2021jhr}. In other words, the conclusions regarding constructing an entropy current and the linearized second law obtained in \cite{Bhattacharyya:2021jhr} can be generalized to theories involving arbitrary non-minimal coupling of scalar and gauge fields with gravity. This is the main result of this paper. 

As compared to coupling the scalar field, for gauge fields non-minimally coupled to gravity, one needs to be more careful in defining a stationary configuration for the gauge field. Since every diffeomorphism invariant (or covariant) statement can possibly suffer from redundancies involving the $U(1)$ gauge invariance. Once this issue is taken care of in developing the formalism, it is not guaranteed that the components of the entropy current that one constructs will be invariant under the $U(1)$ gauge transformations. One approach would be to fix the $U(1)$ gauge, much like choosing the horizon adopted coordinates for the diffeomorphism. However, it will then become even trickier to examine the gauge invariance of the components of entropy current. Instead, we will work without fixing the $U(1)$ gauge and, interestingly enough, will find that the resulting entropy current is gauge-invariant. 

Next, we consider the issue of the ``physical process first law" (PPFL) for theories with gauge fields non-minimally coupled to gravity. Following the analysis of section 5 of \cite{Bhattacharyya:2021jhr}, one sees that if we can construct an entropy current for these theories, the physical process version of the first law follows immediately. This is a consequence of the fact that the setup of PPFL is the same as that of the linearized second law. While it is clear that the proof should follow through for the scalar fields non-minimally coupled to gravity once we construct the entropy current, minor subtleties remain when we consider the gauge field case. This is because, in the gauge field case, there is an additional work term in the first law relation due to the change in the charge of the black hole. The proof of PPFL for the simplest minimally coupled theory, namely the Einstein-Maxwell theory, was given in \cite{Gao:2001ut}. It is interesting to examine if their results can be generalized for arbitrary non-minimally coupled gauge theories following the strategy outlined in \cite{Gao:2001ut}. Our results will iron out these issues, and we will explicitly construct a proof of the PPFL along the lines of \cite{Gao:2001ut}.

Before we move on, let us mention how the rest of this paper is structured. We will start with a brief description of the setup and a precise statement of the problem in \S\ref{setup}. Here we will also present a particular example showing that the non-minimal couplings can contribute to the entropy current at linear order in $\epsilon$-expansion. In the next section \S\ref{sec:stationarity}, we will define the meaning of stationarity in the presence of scalar and gauge fields. In the following section \S\ref{sec:strategy}, we will briefly outline our strategy of constructing entropy current for arbitrary higher curvature theory of gravity non-minimally coupled to scalar and gauge field. In the technical heart of our paper, we will present the construction of the entropy current for the non-minimally coupled scalar field in section \S\ref{sec:scalar} and for the non-minimally coupled gauge field in section \S\ref{sec:NMeinmax}. We will discuss the proof of the PPFL for the non-minimally coupled gauge field in section \S\ref{sec:ppflproof}. We will end with a summary and outlook in \S\ref{sec:summary}. Important technical details as supplementary materials will be presented in the appendices \S\ref{app:boost}-\S\ref{app:detailTH}.

\section{A review of the set-up and statement of the current problem} \label{setup}

In this section, we would like to review the basic setup for constructing an entropy current and the conventions and notations used in the following sections. We will be very brief here, and the interested reader is referred to \cite{Bhattacharyya:2021jhr} for the details. Once the setup is ready, we will provide an explicit example of non-minimal matter coupling where one gets contributions to the entropy current at the linear order in amplitude expansion. Finally, we will lay out a precise mathematical statement regarding the main results derived in this current paper before getting involved in the technical details.

\subsection{Brief review of the set-up to construct an entropy current} \label{ssec-setup}

We are primarily interested in dynamical black hole solutions of diffeomorphism invariant theories of gravity coupled to matter fields. These black hole solutions necessarily have a horizon, a co-dimension one null hyper-surface. Such black hole horizons, dynamically perturbed by external sources, can be suitably described by a coordinate system adapted to the horizon. We will be working with a $d$-dimensional space-time spanned by the coordinates $x^{\mu} = \{v,r,x^i\}$. The coordinates $\{v,x^i\}$ ($i=\{1,\dots,d-2\}$) span the horizon, a co-dimension one null hyper-surface. Here, $v$ is an affine parameter along the null generators $\partial_v$ of the horizon, and $x^i$'s ($i=\{1, \dots, d-2\}$) are the coordinates that run along the spatial tangents, i.e., $\partial_i$'s, of the horizon \footnote{The Greek indices in $x^{\mu}$ will be used to denote the full space-time coordinates and the lower case Latin indices in $x^{i}$ will denote the spatial directions tangent to the null horizon.}. Also, $r$ is an affine parameter that runs along the null geodesics $\partial_r$ that pierce the horizon at specific angles with the null generators and the spatial tangents. The horizon is chosen to be positioned at $r=0$. A gauge choice for the space-time metric is made as follows
\begin{equation} \label{metgag1}
ds^2 = g_{\mu\nu} dx^\mu dx^\nu, \quad \text{where} \quad g_{rv} =1\, , ~g_{rr} = \, g_{ri} =\, g_{vv} \big\vert_{r=0}  =\, \partial_r g_{vv} \big\vert_{r=0}  =\, g_{vi} \big\vert_{r=0} = 0 \, ,
\end{equation}
and thus the near horizon space-time for the most general dynamical black hole can be always expressed by the following metric \footnote{This fact has been explicitly argued for in the appendix A of \cite{Bhattacharyya:2016xfs}.}
\begin{equation}\label{eq:metric}
ds^2=2\, dv\, dr-r^2 \, X(r,v,x^i) \, dv^2+2\, r \, \omega_i(r,v,x^i) \, dv \, dx^i+h_{ij}(r,v,x^i) \, dx^i \, dx^j \, .
\end{equation}
A stationary black hole, with a space-time metric $g^{(eq)}_{\mu\nu}$ in our gauge eq.\eqref{metgag1} is given by
\begin{equation}\label{eq:metequl} 
	ds^2 = g^{(eq)}_{\mu\nu} \, dx^\mu \, dx^\nu = 2 \, dv \, dr - r^2 \, X(rv, \, x^i)\, dv^2 + 2 \, r \,\omega_i(rv, \, x^i)\, dv \, dx^i + h_{ij}(rv, \, x^i) \, dx^i \, dx^j \, ,
\end{equation}
where the metric coefficient functions ($X, \, \omega_i$ and $h_{ij}$) are functions of the product of the coordinates $r v$. 
It is easy to check that the Lie derivative of $g^{(eq)}_{\mu\nu}$ with respect to the vector field $\xi^{\mu}$ vanishes, i.e. $\mathcal{L}_{\xi}g^{(eq)}_{\mu\nu} =0 $, for 
\begin{equation} \label{BT_generator}
	\xi = \xi^\mu\partial_\mu = v\partial_v - r\partial_r \, .
\end{equation}
This $\xi^{\mu}$ is the generator of the transformation (the infinitesimal version of it) given by 
\begin{equation} \label{boosttransf}
	r\rightarrow\tilde r = \lambda \, r, \quad \text{and} \quad v\rightarrow\tilde v = v / \, \lambda \, ,
\end{equation}
which will be called the boost transformation, which, as one can check easily, corresponds to a residual symmetry after one fixes the gauge as in eq.\eqref{metgag1}, eq.\eqref{eq:metric} \footnote{The nomenclature follows that of \cite{Wall:2015raa, Bhattacharya:2019qal, Bhattacharyya:2021jhr}. It should be noted that in eq.\eqref{boosttransf} we are focussing on a sub-class of a bigger residual symmetry.} . The vector field $\xi^{\mu}$ is also the Killing vector \footnote{The norm of the Killing vector $\xi^{\mu}$ also vanishes on $r=0$, thus making it a Killing horizon.} which generates the Killing horizon of this stationary space-time eq.\eqref{eq:metequl}. 

To describe the dynamics, we will employ a perturbative expansion in the amplitude of the dynamics and we will be working up to linearized order in the dynamics. Mathematically, this means that we can decompose the metric $g_{\mu\nu}$ given in eq.\eqref{eq:metric} as follows
\begin{equation}\label{eq:metdecomp}
    g_{\mu\nu} = g^{(eq)}_{\mu\nu}(rv, \, x^i) + \epsilon \, \delta g_{\mu\nu} (r, \, v, \, x^i)\, ,
\end{equation}
where $g^{(eq)}_{\mu\nu}$ is the equilibrium metric given by eq.\eqref{eq:metequl}, and $\epsilon$ is the small parameter denoting the amplitude of the fluctuations $\delta g_{\mu\nu}$. 

The boost symmetry for the near horizon metric gets slightly broken by the dynamical fluctuation $\delta g_{\mu\nu}$, which is evident from the fact that $g^{(eq)}_{\mu\nu}$ depends on the product of coordinates $rv$, but $\delta g_{\mu\nu}$ can have arbitrary dependence on both $r$ and $v$ individually. 
We will need to decide how to fix whether any covariant tensor quantity will contribute to $\mathcal{O}(\epsilon)$ or to $\mathcal{O}(\epsilon^2)$, when evaluated on the perturbed metric in eq.\eqref{eq:metdecomp}. For this, we need to know how any covariant tensor would transform under the boost transformation given in eq.\eqref{boosttransf}. The quantity that signifies this is called the boost weight, defined as the difference in the number of lower $v$-indices and the lower $r$-indices of any particular tensor. 
We can also learn that any quantity with positive boost weight will always vanish when dynamics is switched off. This is explained in detail in \cite{Bhattacharyya:2021jhr}; also see Appendix-\ref{app:boost} for a brief discussion on this and related issues. 

The gravity theories which we will consider will have an action of the following form,
\begin{equation} \label{action}
\mathcal{I} =  \dfrac{1}{4\pi} \int d^d x \, \sqrt{-g} \, \left( \mathcal{L}_{grav} + \mathcal{L}_{mat}\right) \, ,
\end{equation}
where $\mathcal{L}_{grav}$ is the Lagrangian for the degrees of freedom corresponding purely to the gravity part without involving any matter fields \footnote{One can argue that, (see section $2$ of \cite{Iyer:1994ys}), any diffeomorphism invariant Lagrangian will have the specific functional dependence as mentioned here.},
\begin{equation} \label{diff_inv_L}
\mathcal{L}_{grav}  = \mathcal{L}_{grav}  \left(g_{\alpha\beta}, \,  R_{\alpha\beta\mu\nu},  \, D_\lambda R_{\alpha\beta\mu\nu},  \cdots \right)  \, ,
\end{equation}
where $D_\lambda$ is the covariant derivative compatible with $g_{\alpha\beta}$, and the dots signify arbitrary number of derivatives on $R_{\alpha\beta\mu\nu}$. Also, $\mathcal{L}_{mat}$ is the Lagrangian corresponding to the matter part, which will obviously include $g_{\alpha\beta}$ coupled to matter fields. 

The total equation of motion, denoted  as ${\cal E}_{\mu\nu}$, follows from the variation of the full Lagrangian in eq.\eqref{action}, which contains gravity as well as matter part. We decompose ${\cal E}_{\mu\nu}$ as follows
\begin{equation} \label{EgravTvv}
{\mathcal E}_{\mu\nu}=E_{\mu\nu}+T_{\mu\nu} \, ,
\end{equation}
where, $E_{\mu\nu}$ is the part of the equation of motion coming from the variation of $L_{grav}$ with respect to the space time metric $g_{\mu\nu}$ and $T_{\mu\nu}$ is the part of the equation of motion coming from the variation of $L_{mat}$ with respect to $g_{\mu\nu}$ \footnote{$T_{\mu\nu}$ is defined as $T_{\mu\nu} = \dfrac{1}{\sqrt{-g}}\dfrac{\delta L_{mat}}{\delta g^{\mu\nu}}$.}. 
It is important to note that this split of $E_{\mu\nu}$ and $T_{\mu\nu}$ crucially depends on the definition of the energy-momentum tensor. In \cite{Bhattacharya:2019qal}, and \cite{Bhattacharyya:2021jhr} the same action as in eq.\eqref{action} was considered along with the following convention: All the terms which involve only the metric components were grouped in $E_{\mu\nu}$ and the terms which involved both the metric and matter field components were grouped together in $T_{\mu\nu}$. Thereafter, following \cite{Wall:2015raa}, one knows that a certain component of the equations of motion, namely $\mathcal{E}_{vv}$ (in our gauge eq.\eqref{eq:metric}) will be crucial in proving the second law. It was shown in \cite{Bhattacharyya:2021jhr} that $E_{vv}$ in eq.\eqref{EgravTvv} has the following structure when evaluated on the horizon $r=0$
\begin{equation}\label{eq:main1a}
   \begin{split}
    E_{vv} \big\vert_{r=0} = 
    % &\, \partial_v \bigg[{1\over\sqrt h}\partial_v \left({\sqrt h} \, {\cal J}^v\right) +{1\over\sqrt h}\partial_i \left({\sqrt h} \, {\cal J}^i\right)\bigg] + \mathcal{O} (\epsilon^2)
    \, \partial_v \bigg[\dfrac{1}{\sqrt{h}}\partial_v \left({\sqrt h} \, {\cal J}^v\right) +\nabla_i{\cal J}^i\bigg] + \mathcal{O} (\epsilon^2) \, ,
   \end{split}
\end{equation}
where $h$ is the determinant of the induced metric $h_{ij}$ on the horizon, and $\nabla_i$ is the covariant derivative compatible with $h_{ij}$. This eq.\eqref{eq:main1a} serves as the definition for the objects ${\cal J}^v$ and ${\cal J}^i$, as they are obtained algorithmically expressed in terms of the basic building blocks: the metric coefficient functions $X, \, \omega^i, \, h_{ij}$ and derivatives $\partial_v, \, \partial_r, \, \nabla_i$ acting on them. 

In the next step, the analysis in \cite{Bhattacharyya:2021jhr} assumes that $T_{\mu\nu}$ satisfies the NEC, namely $T_{vv} \geq 0$ \footnote{In our gauge eq.\eqref{eq:metric}, the generator of the null horizon is given by $\chi = \partial_v$ and therefore $T_{\mu\nu} \chi^\mu \chi^\nu \sim T_{vv}$.}. Thus, using the NEC together with fact that for on shell configurations $\mathcal{E}_{vv} = 0$ in eq.\eqref{EgravTvv}, one readily obtains $E_{vv} \leq 0$. Further using this in eq.\eqref{eq:main1a} one can show that \footnote{We have used the physical boundary condition that the dynamics settles down to some equilibrium configuration at the end, i.e. we assumed $\dfrac{1}{\sqrt{h}}\partial_v \left({\sqrt h}~{\cal J}^v\right) +\nabla_i{\cal J}^i \rightarrow 0$ as $v \rightarrow \infty$.} 
\begin{equation} \label{cond3}
\dfrac{1}{\sqrt{h}}\partial_v \left({\sqrt h}~{\cal J}^v\right) +\nabla_i{\cal J}^i \geq 0~~\text{for all finite $v$}\, .
\end{equation}
Thus, we can interpret $\mathcal{J}^v$ and $\mathcal{J}^i$ as components of an entropy current, and there is local entropy production at each point on the black hole horizon. We should also note that both $\mathcal{J}^v$ and $\mathcal{J}^i$ as defined in eq.\eqref{eq:main1a} get contribution only from $E_{vv}$, without involving the matter fields.

\subsection{An example of non-minimal couplings contributing to entropy current at $\mathcal{O}(\epsilon)$} \label{ssec-exNM}

As discussed in the previous sub-section \S\ref{ssec-setup}, a couple of key ingredients that go into the proof devised in \cite{Bhattacharyya:2021jhr} are essentially the definition of the energy-momentum tensor and the assumption that it satisfies the NEC. Now, since $T_{vv} \geq 0$ and it vanishes in the equilibrium limit \footnote{This tensor component has positive boost weight.}, in our amplitude approximation, it essentially vanishes up to $\mathcal{O}(\epsilon)$, i.e. $T_{vv} \sim \mathcal{O}(\epsilon^2)$. So within the assumptions and the setup of \cite{Bhattacharyya:2021jhr}, the matter couplings do not contribute to the entropy current. This assumption is based on the premise that such couplings satisfy the NEC. However, there might be matter couplings that do not satisfy the NEC. In particular, it is known that non-minimally coupled matter fields do violate the NEC, see \cite{Flanagan_1996, Barcelo:2000zf, Wall:2018ydq, Chatterjee:2015uya}. Then within our approximation, these terms are potentially contributing to $\mathcal{O}(\epsilon)$ and apriori, there is no reason to justify that they will have the same entropy current structure of the purely gravitational terms given in eq.\eqref{eq:main1a}. Therefore, based on these arguments, the proof presented in \cite{Bhattacharyya:2021jhr} seemingly does not work for such non-minimally coupled theories. 

In other words, we are trying to argue that non-minimal couplings present in the Lagrangian can give rise to additional terms, say we denote them by $\widetilde{T}_{vv}$, in a form as follows
\begin{equation}\label{eq:nonmin1}
   \begin{split}
   \mathcal{E}_{vv}\big\vert_{r=0} =& \left(E_{vv} + T_{vv} + \widetilde{T}_{vv} \right)\big\vert_{r=0} \\
   =& \left. \left( \partial_v \bigg[\dfrac{1}{\sqrt{h}}\partial_v \left({\sqrt h} \, {\cal J}^v\right) +\nabla_i{\cal J}^i\bigg] + T_{vv} + \widetilde{T}_{vv} \right) \right\vert_{r=0}+\, \mathcal{O} (\epsilon^2) \, .
   \end{split}
\end{equation}
One can now argue that if the contribution from the non-minimal couplings contribute to linear order in the amplitude expansion
\begin{equation} \label{eq:nonmin2}
\widetilde{T}_{vv} \sim \mathcal{O} (\epsilon) \, ,
\end{equation}
we do not get eq.\eqref{cond3} even if we get $T_{vv} \geq 0$
\footnote{It is worth highlighting that our subsequent analysis concerning $\widetilde{T}_{vv}$ will be important even if we consider non-minimal coupling terms as small corrections (within an EFT perspective) to the leading NEC satisfying matter field terms. One might then wonder that the $\mathcal{O}(\epsilon)$ contributions from non-minimal coupling (denoted by $\widetilde{T}_{vv}$ in eq.\eqref{eq:nonmin1}) can never be as important as the NEC satisfying terms (denoted by $T_{vv}$ in eq.\eqref{eq:nonmin1}). However, there might be non-trivial scenarios where a particular field configuration results in the leading order minimal coupling contribution to $T_{vv}$ being zero. In such extreme cases, the $\mathcal{O}(\epsilon)$ contribution of $\widetilde{T}_{vv}$ dominates and can potentially violate NEC, and thus also violate the linearized second law.}.

Let us first show by example that non-minimal couplings can indeed give rise to contributions such that eq.\eqref{eq:nonmin2} is true. Consider the following action of a scalar field non-minimally coupled to gravity
\begin{equation} \label{exmpl_nonmin}
S_{non-min}
%=\int d^d x \, {\cal L}_{scalar}
=\int d^d x \sqrt{-g}~ R_{\mu\nu}(D^\mu\phi)(D^\nu \phi) \, .
\end{equation}
It is easy to derive the EoM from this Lagrangian giving us the $\widetilde{T}_{\mu\nu}$ for this particular example as written below
\begin{equation}\label{eq:Emunuphi}
\begin{split}
\widetilde{T}_{\mu\nu} =\frac{1}{2} & g_{\mu\nu}R_{\alpha\beta}(D^\alpha\phi)(D^\beta\phi)-R_{\mu\alpha}(D_\nu\phi)(D^\alpha\phi)-R_{\beta\nu}(D^\beta\phi)(D_\mu\phi)\\
+\frac{1}{2} & \Big[D^\alpha D_\nu\{(D_\alpha\phi)(D_\mu\phi)\}+D^\alpha D_\mu\{(D_\alpha\phi)(D_\nu\phi)\}-D^\gamma D_\gamma\{(D_\nu\phi)(D_\mu\phi)\}\\
&~~-g_{\mu\nu}D^\alpha D^\beta\{(D_\alpha\phi)(D_\beta\phi)\}\Big] \, .
\end{split}
\end{equation}
We can now evaluate $\widetilde{T}_{vv}$ at the horizon $r=0$ in our gauge eq.\eqref{eq:metric} in a brute-force fashion and after a tedious but straightforward manipulations we get 
\begin{equation} \label{entcur_NM}
\widetilde{T}_{vv}|_{r=0}=\partial_v\left[\frac{1}{\sqrt{h}}\partial_v\left(\sqrt{h} \, \left(\partial_r\phi \, \partial_v\phi\right) \right)+\nabla_i \left(\nabla^i\phi \, \partial_v\phi\right)\right]+{\cal O}(\epsilon^2) \, .
\end{equation}
It is obvious from eq.\eqref{entcur_NM} that we do indeed get a non trivial $\widetilde{T}_{vv} \sim \mathcal{O} (\epsilon)$, for the example given in eq.\eqref{exmpl_nonmin}. 

Very interestingly, we must note that the off-shell structure of this term eq.\eqref{entcur_NM} is precisely of the form eq.\eqref{eq:main1a}, which is required to construct an entropy current. Therefore, the proof of \cite{Bhattacharyya:2021jhr} goes smoothly, but the entropy current receives an additional contribution from the non-minimal coupling.

\subsection{The statement of the current problem} \label{ssec-SOP}

In the previous sub-section, for an example, we have seen that the additional contributions coming to the equations of motion from non-minimal couplings have the desired structure that felicitates the proof of a linearized second law constructed in \cite{Bhattacharyya:2021jhr}.  However, there is just the following modification: we get an entropy current with non-negative divergence as in eq.\eqref{cond3}, but it gets contribution both from the purely gravitational term as well as the non-minimal coupling. Additionally, in some of the recent works \cite{Wang:2020svl, Wang:2021zyt}, it has also been argued that for some particular models of non-minimally coupled matter interactions, the linearized second law holds \footnote{The example in \cite{Wang:2020svl} considered a specific non-minimally coupled scalar field Lagrangian and \cite{Wang:2021zyt} looked at a non-minimally coupled $U(1)$ gauge field Lagrangian. But both of them considered only specific models of four derivative theories of gravity.}. However, in these references, the version of the second law was local in time but not in the spatial directions of the constant $v$-slices of the horizon as the spatial components of the entropy current were missed \footnote{Here, the spatial constant $v$-slices were considered to be compact, similar to the construction in \cite{Wall:2015raa}. Consequently, one compares an expression of Wald entropy integrated over the spatial slices between the initial and final equilibrium. Because of this integration over the spatial slices, the total derivative terms as $\nabla_i \widetilde{\cal J}^i$ vanish, and the locality in spatial directions is lost.}.

Motivated by these examples, our primary goal is to analyze whether the most general Lagrangian, which contains scalars and gauge fields non-minimally coupled to gravity, has this desired structure in the equations of motion. Automatically, this would give us an entropy current such that a proof of the linearized second law, along the lines of \cite{Bhattacharyya:2021jhr}, is available to us.

We will be working with systems where scalar fields and $U(1)$ gauge fields can have all possible coupling with diffeomorphism invariant theory of gravity such that the Lagrangian would have the following form
\begin{equation} \label{action_gen}
\mathcal{L}=\mathcal{L}\big(g_{\alpha\beta}, \, R_{\alpha\beta\rho\sigma}, \, D_{\alpha_1}R_{\alpha\beta\rho\sigma},\cdots, \, \phi, \, D_{\alpha_1}\phi,\cdots, \, F_{\mu\nu}, \, D_{\alpha_1} F_{\mu\nu}, \, D_{(\alpha_1}D_{\alpha_2)}F_{\mu\nu}, \cdots\big) \, ,
\end{equation}
where $F_{\mu\nu} (= \partial_\mu A_\nu - \partial_\nu A_\mu)$ is the field strength tensor for the gauge fields $A_\mu$, and the dots represent arbitrary number of derivatives acting on $R_{\alpha\beta\rho\sigma}, \, \phi$ or $F_{\mu\nu}$. 

In this paper we will explicitly prove the following statement: \\
\emph{If one considers all possible couplings (including non-minimal interactions) of scalar fields and $U(1)$ gauge fields with arbitrary diffeomorphism invariant theory of gravity (with all possible higher derivative terms), it can be abstractly argued that the full equations of motion will have the following off-shell structure
\begin{equation}\label{eq:nonmin3}
   \begin{split}
   \mathcal{E}_{vv}\big\vert_{r=0} = \left. \left( \partial_v \bigg[{1\over\sqrt h}\partial_v \left({\sqrt h} \, ( {\cal J}^v + \widetilde{\cal J}^v )\right) +\nabla_i({\cal J}^i + \widetilde{\cal J}^i )\bigg] + T_{vv} \right) \right\vert_{r=0}+\, \mathcal{O} (\epsilon^2) \, ,
   \end{split}
\end{equation}
where possible generic non-minimal couplings may in principle have $\mathcal{O} (\epsilon)$ contributions (denoted by $\widetilde{T}_{\mu\nu}$) to the equations of motion but they can always be written as 
\begin{equation} \label{entcur_nonmin}
\widetilde{T}_{vv}|_{r=0}=\partial_v\left[\frac{1}{\sqrt{h}}\partial_v\left(\sqrt{h} \, \widetilde{\cal J}^v\right)+\nabla_i \widetilde{\cal J}^i\right]+{\cal O}(\epsilon^2) \, ,
\end{equation}
such that it adds to the entropy current exactly in the same way (denoted by $\widetilde{\cal J}^v$ and $\widetilde{\cal J}^i$) as the purely gravitational terms (denoted by ${\cal J}^v$ and ${\cal J}^i$). As a consequence, we will always get a linearized second law given by 
\begin{equation} \label{cond3modf}
\dfrac{1}{\sqrt{h}}\partial_v \left({\sqrt h} \, ( {\cal J}^v + \widetilde{\cal J}^v )\right) +\nabla_i({\cal J}^i + \widetilde{\cal J}^i )\geq 0~~\text{for all finite $v$}\, ,
\end{equation}
once we assume the other contributions from matter sector satisfies the NEC, $T_{vv} \geq 0$. Additionally, we will also find that the components of the entropy current are $U(1)$ gauge-invariant for theories involving the gauge field coupled to gravity.} This is our main result \footnote{To be very precise, in our framework we are only analysing quantities upto linear order in $\epsilon$-expansion. Therefore, we will effectively argue that the RHS of eq.\eqref{cond3modf} will vanish at ${\cal O}(\epsilon)$.}.

In mathematical terms, in the rest of this paper, our aim would thus be to focus on just the generic non-minimal coupling of scalar and gauge fields with gravity and to show that the EoM for such terms, $\widetilde{T}_{\mu\nu}$, would have an off-shell structure given by eq.\eqref{entcur_nonmin}, 
and to figure out an abstract algorithm to compute $\widetilde{\cal J}^v$ and $\widetilde{\cal J}^i$ given any Lagrangian.

\section{Stationarity and its breaking in the presence of scalar and gauge fields} \label{sec:stationarity}

As we have discussed in \S\ref{ssec-setup}, for theories with the space-time metric $g_{\mu\nu}$ as the only degrees of freedom, the stationarity of black hole solutions is defined by the vanishing of the Lie derivative of $g^{(eq)}_{\mu\nu}$ with respect to the generator of boost transformations, eq.\eqref{boosttransf}, i.e., $\xi$ given in eq.\eqref{BT_generator}. In this paper, our primary focus is on scalar and gauge fields coupled to gravity, and, therefore, it is essential to understand the statement of stationarity involving them. As a consequence of stationarity, we must also understand the implications of the boost symmetry and its breaking for scalar and gauge fields present in theory. This will be useful in discussing the dynamics of matter fields like scalar or gauge fields. We will address these issues in this present section.

Let us start by discussing a scalar field $\phi$ on the background space-time given by eq.\eqref{eq:metequl}. The scalar field inherits the symmetry of the background space-time and thus the stationarity is implemented as
\begin{equation}
    \mathcal{L}_{\xi} \phi = 0 \, ,
\end{equation}
where $\xi$ is the Killing vector given by eq.\eqref{BT_generator}. In dynamical situations, the scalar field thus has a similar decomposition as the metric field (see eq.\eqref{eq:metdecompAP} in Appendix-\ref{app:boost})
\begin{equation}
    \phi(r,v,x^i) = \phi^{eq}(rv,x^i) + \epsilon \, \delta \phi(r,v,x^i) \, .
\end{equation}
Here, the form of the equilibrium configuration $\phi^{eq}$ is fixed by
\begin{equation}
    \mathcal{L}_{\xi} \phi^{eq} = ( v \partial_v - r \partial_r ) \phi^{eq} = 0 \, ,
\end{equation}
which can be solved to obtain
\begin{equation} \label{eqphisol}
\phi^{eq} = \phi^{eq}(rv,x^i) \, .
\end{equation}
Let us now comment on the non-equilibrium structures for quantities involving the scalar field. Most importantly, we must be able to decide which quantities involving the scalar field (possibly with various derivatives acting on it) would contribute because of only dynamics or would it be non-zero when evaluated on equilibrium configurations. For example, based on our analysis of boost weights of various quantities, we would learn that the quantity $\partial^{m_r}_r \partial^{m_v}_v \phi(r,v,x^i)$ will be non-zero solely due to dynamics and that too it will be of $\mathcal{O}(\epsilon)$ when evaluated on the horizon for $m_v > m_r$ (see eq.\eqref{defEQvsNEQAP} in Appendix-\ref{app:boost}), 
\begin{equation}
 \partial^{m_r}_r \partial^{m_v}_v \phi(r,v,x^i)|_{r=0} \sim \mathcal{O}(\epsilon) ~~ \text{whenever} ~~ m_v > m_r \, .
\end{equation}

The statement of stationarity for gauge fields is more involved because of the gauge redundancy of the theory. For simplicity, we consider a $U(1)$ gauge field $A_{\mu}$ \footnote{We believe that the analysis is easily generalizable to higher gauge groups.}. The naive expectation of stationarity given by demanding
\begin{equation} \label{eq:gaugestat0}
    \mathcal{L}_{\xi} A_{\mu} = 0 \, ,
\end{equation}
is not completely consistent because the above equation is not invariant under a gauge transformation $A_{\mu} \rightarrow A_{\mu} + D_{\mu} \Lambda$ \footnote{Here $D_{\mu}$ denotes the covariant derivative compatible with the full dynamical metric eq.\eqref{eq:metric}.}. This issue has been addressed in its various guises in the literature \cite{Gao:2001ut, Gao:2003ys, Prabhu:2015vua, Elgood:2020svt}. Basically, the main issue is that when one has fields that transform non-trivially under some internal gauge transformation, one cannot distinguish the action of a diffeomorphism on the space-time from the action of a diffeomorphism followed by an internal gauge transformation. In the following, we will adopt the following convention for defining the stationary configurations of a gauge field: $A^{eq}_\alpha$ is said to be stationary if there exist a $\Lambda$ such that
\begin{equation}\label{eq:gaugestat}
    \mathcal{L}_{\xi} A^{eq}_{\alpha} + D_{\alpha} \Lambda = 0 \, .
\end{equation}
From this, one can derive a statement equivalent to the zeroth law for the electrostatic potential, as follows
\begin{equation}\label{eq:genzerothlaw}
    (A^{eq}_{\alpha}\xi^{\alpha} + \Lambda) \big\vert_{r=0} = 0 \, , 
\end{equation}
where as usual $\xi$ is the Killing vector generating the boost transformation, given in eq.\eqref{BT_generator}.  A justification for the condition in eq.\eqref{eq:genzerothlaw} is provided in Appendix-\ref{app:genzerothlaw}. 

It is important to note that this is an extended notion of symmetry in the presence of fields that transform non-trivially under some internal gauge transformation. This definition is gauge invariant in the following sense: If one does the gauge transformation $A^{eq}_{\mu} \rightarrow A^{\prime eq}_{\mu}= A^{eq}_{\mu} + D_{\mu} \lambda$, then $A^{\prime eq}_{\mu}$ satisfies the equation
\begin{equation}
 \mathcal{L}_{\xi} A^{\prime eq}_{\alpha} + D_{\alpha} \Lambda^{\prime} = 0 \, ,
\end{equation}
where, $\Lambda^\prime$ is given by $\Lambda^\prime=\Lambda-(\xi\cdot\partial)\lambda$, which implies the invariance of the stationarity condition \eqref{eq:gaugestat} under gauge transformation. 

From eq.\eqref{eq:genzerothlaw} we note that $A^{eq}_{\alpha}\xi^{\alpha} + \Lambda$ evaluates to zero on the horizon. Any quantity that vanishes when evaluated on the horizon $r=0$ for equilibrium configurations is necessarily $\mathcal{O}(\epsilon)$ in dynamical situations. This follows directly from the boost weight analysis mentioned above, see Appendix-\ref{app:boost} for an explanation. Hence, we have
\begin{equation}
    \begin{split}
  (A_{\alpha}\xi^{\alpha} + \Lambda)|_{r=0} = \mathcal{O}(\epsilon) \, , 
    \end{split}
\end{equation}
which, by using eq.\eqref{BT_generator}, can be equivalently written as the following 
\begin{equation} \label{eq:STgauge1}
    \begin{split}
(v A_v + \Lambda)|_{r=0} = \mathcal{O}(\epsilon) \, .
    \end{split}
\end{equation}
We will work with this definition of stationarity for the $U(1)$ gauge fields.

\section{Outlining the strategy to construct entropy current with non-minimal coupling} \label{sec:strategy}

To construct an entropy current at linearized order in amplitude expansion for the non-minimal coupling of scalar and gauge fields with gravity we will essentially follow the formalism developed in \cite{Bhattacharyya:2021jhr} based on Noether charge analysis of \cite{Iyer:1994ys}. In the following sections we will execute this in details by considering scalar and gauge fields one at a time. Before we get into the technical details, in this section our aim is to provide a schematic sketch of the main steps involved in that technical derivation. This will be brief as the purpose of this section is to present a guideline highlighting only the important features of the construction by referring to the corresponding portions in \cite{Bhattacharyya:2021jhr} for the details. We summarize them as follows
\begin{itemize}
	\item The starting point will be to relate an arbitrary variation in the Lagrangian $\mathcal{L} (g_{\mu\nu}, \, \psi)$ to the equations of motion (EoM) and thus defining the total derivative term $\Theta ^\mu [\delta g, \, \delta \psi]$
	\begin{equation}\label{eq:varl}
		\begin{split}
			\delta\left[\sqrt{-g} \, \mathcal{L} \right]&= \sqrt{-g} \, \mathcal{E}^{\mu\nu} \, \delta g_{\mu\nu}  +\sqrt{-g} \, G_{\psi} \, \delta \psi  + \sqrt{-g} \, D_\mu \Theta ^\mu [\delta g, \, \delta \psi] \, ,
		\end{split}
	\end{equation}
	where $\psi$ denote the matter fields collectively, $\mathcal{E}^{\mu\nu}$ is the EoM for $g_{\mu\nu}$, and $G_{\psi}$ is the EoM for the matter fields $\psi$. 
	\item Next, we would show that  the EoM can be related to $\Theta^{\mu}$ and the Noether charge $Q^{\mu\nu}$. Starting from eq.\eqref{eq:varl}, and considering variations that are symmetries of the theory, we can arrive at the following equation
	\begin{equation}\label{hijibiji-4}
		\begin{split}
			& \zeta^2 \mathcal{L} -\Theta^\mu \zeta_\mu+ \zeta_\mu \, D_\nu  Q^{\mu\nu} =2 \,  \zeta^\mu\zeta^\nu \mathcal{E}_{\mu\nu} \, + \mathcal{T}[G_{\psi}],
		\end{split}
	\end{equation}
	where $\zeta^{\mu}$ is an arbitrary diffeomorphism, $x^\mu \rightarrow x^\mu + \zeta^\mu $. The derivation of eq.\eqref{hijibiji-4} can be understood following the steps outlined in section 2.5 of \cite{Bhattacharyya:2021jhr}. However, we must be careful that the EoM for $\psi$, i.e. the second term involving $G_{\psi}$, was not considered in \cite{Bhattacharyya:2021jhr}. But this term can now, in principle, contribute and that is exactly what we have indicated as $\mathcal{T}[G_{\psi}]$ in RHS of eq.\eqref{hijibiji-4}. As we will explicitly show later, for scalar field coupled to gravity, there will be no extra contribution to eq.\eqref{hijibiji-4} in the form of $\mathcal{T}[G_{\psi}]$. But, for gauge fields coupled to gravity, we will get a non-vanishing contribution
	\begin{equation}\label{hijibiji-5}
		\mathcal{T}[G_{\psi = A_{\mu}}] \sim G^{\mu} \zeta_{\mu} ( A^{\nu} \zeta_{\nu} + \Lambda ),
	\end{equation}
	where $\Lambda$ is a parameter for the $U(1)$ gauge transformation. 
	\item Choosing $\zeta^{\mu}$ to be the Killing vector $\xi^{\mu}$ given in eq.\eqref{BT_generator} and evaluating the expression of eq.\eqref{hijibiji-4} on the horizon $r=0$ in our metric gauge eq.\eqref{eq:metric}, we get to our main equation that relates $vv$-component of EoM to $\Theta^r$ and $Q^{r \mu}$
	\begin{equation}\label{keyeqn}
		\begin{split}
			2\, v \, (E_{vv}+\widetilde{T}_{vv})\big\vert_{r=0}=~&\big(-\Theta^r+ D_\mu Q^{r \mu}\big)\big\vert_{r=0}\, + \mathcal{O}(\epsilon^2) .
		\end{split}
	\end{equation} 
	This is our key equation and the intermediary steps to derive eq.\eqref{keyeqn} are detailed in section 2.5 of \cite{Bhattacharyya:2021jhr}. 
	\begin{itemize}
		\item The last term on the RHS of eq.\eqref{hijibiji-4}, which was a new contribution due to explicitly keeping track of the matter fields in our present analysis as compared to \cite{Bhattacharyya:2021jhr}, does not contribute to linear order in $\epsilon$-expansion.  As we will show (see \S\ref{ssec:scalEvv} for scalar coupling and \S\ref{ssec:gaugeEvv} for gauge field coupling), using eq.\eqref{eq:STgauge1}, the extra term $\mathcal{T}[G_{\psi = A_{\mu}}]$ written in eq.\eqref{hijibiji-5} will be negligible since it will contribute to $\mathcal{O}(\epsilon^2)$. In this sense, the RHS in eq.\eqref{keyeqn} retains the same structure as was obtained in \cite{Bhattacharyya:2021jhr}.
		\item As is known from \cite{Bhattacharyya:2021jhr}, all contributions to $E_{vv}$ from the purely gravitational terms will have the desired structure shown in eq.\eqref{eq:main1a} to construct the entropy current. 
		\item However, it is important to note that in the LHS of eq.\eqref{keyeqn}, the contributions from the non-minimal couplings, which we have denoted as $\widetilde{T}_{\mu\nu}$ in \S\ref{ssec-exNM} (see eq.\eqref{eq:nonmin1}), appears explicitly. But, we are not including the part $T_{vv}$ on the LHS of the same equation, as it will satisfy the NEC and hence will be of $\mathcal{O}(\epsilon^2)$. 
		\item Interestingly, through the eq.\eqref{keyeqn}, the $\widetilde{T}_{\mu\nu}$ term also gets related to $\Theta^r$ and $Q^{r \mu}$. These quantities now get contributions from $\psi$ and $\delta \psi$. Thus, eq.\eqref{keyeqn} would be crucial to establish that $\widetilde{T}_{\mu\nu}$ also has the required structure as in eq.\eqref{entcur_nonmin}. 
	\end{itemize}
	\item Following the same strategy that was adopted in \cite{Bhattacharyya:2021jhr}, our next step would be to analyze both the terms $\Theta^r$ and $Q^{r \mu}$ on the RHS of eq.\eqref{keyeqn} and extract out their exact dependence on the coordinate $v$. We need to remember that our main interest in this paper is to capture the matter field contributions, and hence we will primarily focus on those contributions in $\Theta^r$ and $Q^{r \mu}$. After some manipulations, we will be able to argue eq.\eqref{entcur_nonmin}.  
	\item The definitions of $\Theta^\mu$ and $Q^{\nu \mu}$ in terms of $g_{\mu\nu}, \, \delta g_{\mu\nu}, \, \psi,$ and $\delta \psi$ can be obtained from \cite{Iyer:1994ys}. For example, $\Theta^\mu$ is defined in Lemma-(3.1) of \cite{Iyer:1994ys} and can also be obtained from eq.(3.16), (3.17), and (3.17) in \cite{Bhattacharyya:2021jhr}. The $Q^{\nu \mu}$ is defined in Proposition-(4.1) of \cite{Iyer:1994ys}, and is also written in eq.(3.58) in \cite{Bhattacharyya:2021jhr}. 
	\item Starting with these definitions, we will also use the boost symmetry of the near horizon metric in our chosen gauge while manipulating the terms $\Theta^r$ and $Q^{r \mu}$. More specifically, we must remember that these quantities will have a definite boost weight determined by how they would transform under boost transformation eq.\eqref{boosttransf}. For concreteness, from the discussion in \S\ref{ssec-setup} and the discussion in Appendix-\ref{app:boost}, it would be clear that both $\Theta^r$ and $Q^{r i}$ will have boost weight equal to $+1$, and $Q^{r v}$ will be boost invariant.
	\item Another important technical ingredient that we will need to borrow from the construction in \cite{Bhattacharyya:2021jhr} is the following: one can determine the general off-shell structure of any covariant tensor quantity, having a definite boost weight in terms of the basic building blocks. This was called ``Result: 1", and was given in eq.(3.14) in \cite{Bhattacharyya:2021jhr}. 
	\begin{itemize}
		\item This was, for example, very crucial in determining the structure of the $vv$-component of the EoM eq.\eqref{eq:main1a}, which we write here schematically
		\begin{equation} \label{Evvschm}
			E_{vv} \big\vert_{r=0} \sim \partial_v \left[\partial_v \left({\cal J}^v\right) +\nabla_i{\cal J}^i\right] + \mathcal{O} (\epsilon^2).
		\end{equation}
		On the LHS of the equation above, $E_{vv}$ has boost-weight $=2$, on the RHS each of the $\partial_v$ has boost weight $=1$, so ${\cal J}^v$ is boost invariant, and ${\cal J}^i$ has boost-weight $=1$. 
		\item In the context of \cite{Bhattacharyya:2021jhr}, where the degrees of freedom are only the metric of pure gravity theory, the basic building blocks are the metric coefficient functions ($X, \, \omega_i, \, h_{ij}$) and various derivatives ($\partial_v, \, \partial_r, \, \nabla_i$) acting on them (see Table-1 in \cite{Bhattacharyya:2021jhr}). It is also important to recollect that each one of $X, \, \omega_i, \, h_{ij}$ are boost invariant. One can only generate positive boost weight by acting with $\partial_v$ on these boost invariant objects. This fact was important in deriving the ``Result:1", and also in deciding the structure on the RHS of eq.\eqref{Evvschm}: all the boost weight of $E_{vv}$ (which is $=2$) is generated by the action of two $\partial_v$s on the RHS. If we had another element with positive boost weight as the basic building block, that might have also appeared, and we could not have concluded the RHS of eq.\eqref{Evvschm}. 
		\item However, in our present context, where we have scalar or gauge fields as additional degrees of freedom in addition to the metric, the set of ``basic building blocks" must be expanded. In addition to the metric coefficient functions and their derivatives, we must consider the matter fields along with the derivatives ($\partial_v, \, \partial_r, \, \nabla_i$) acting on them as eligible candidates for the basic building blocks. The main point is to examine if we are getting any new elements in the basic building blocks with positive boost weight other than $\partial_v$. Suppose we have one such new object due to the matter fields with a positive boost weight. In that case, we must also appropriately modify the ``Result: 1" since the generic structure of any tensor quantity now will certainly involve these newer elements.
		\item For scalar fields coupled to gravity, we indeed have a new basic element, i.e., the scalar field $\phi$ itself and other possible objects that we can form by acting with $\partial_v, \, \partial_r, \, \nabla_i$ on $\phi$. From our discussions in \S\ref{ssec-setup} and \S\ref{sec:stationarity} we learn that $\phi$ is boost invariant. Therefore, all new terms with positive boost weight can only be built by acting with $\partial_v$'s on $\phi$. So, we do not need to modify ``Result: 1", and the construction of \cite{Bhattacharyya:2021jhr} can trivially be extended to incorporate contributions from scalar fields into the construction of entropy current.
		\item For coupling gauge fields ($A_\mu$) with gravity, the situation is not so trivial as the scalar fields. We certainly get newer basic elements such as the components of the gauge field and various derivatives of them. However, most importantly, there is now one component of the gauge field, i.e. $A_v$, that has a boost weight $=+1$ ($A_r$ has a boost weight $=-1$, and $A_i$ is boost invariant), see \S\ref{ssec-setup} and \S\ref{sec:stationarity}. Therefore, we now encounter a situation where apart from operating with $\partial_v$ on a boost invariant object (say $\partial_v X$, which also has boost weight $=+1$), we have $A_v$ that has a positive boost weight ($=1$). So we need to modify ``Result: 1" keeping in mind these possible extra contributions, which we will do in \S\ref{ssec:modresult1}. Therefore, it is indeed a non-trivial task to argue abstractly that the RHS of eq.\eqref{Evvschm} will still hold. This paper aims to extend the formalism developed in \cite{Bhattacharyya:2021jhr} to take care of these extra contributions involving the gauge fields such that an entropy current can still be constructed.
	\end{itemize}
	\item We have now successfully developed the technical toolbox, mostly duplicating from \cite{Bhattacharyya:2021jhr}, but also mentioning the additional modifications needed to address the extended context of matter coupling being studied in the present paper. It is now straightforward, but may be tedious, exercise to manipulate the abstract definitions of $\Theta^r$ and $Q^{r \mu}$. As we have seen before, with the knowledge of their boost weights and the modified ``Result: 1" (the modified one is needed for gauge field coupling only, as we discussed before), we can abstractly argue that the $\mathcal{O}(\epsilon)$ contributions from the non-minimal couplings, denoted by $\widetilde{T}_{\mu\nu}$, indeed has the off-shell structure given in eq.\eqref{entcur_nonmin}. In that process, we will also be able to give an algorithm to construct the components of entropy current ($\widetilde{\cal J}^v$ and $\widetilde{\cal J}^i$). This will be done in \S\ref{ssec:scalJvJi} for scalar fields and in \S\ref{ssec:gaugeJvJi} for gauge fields coupled to gravity. 
\end{itemize}

\section{Scalar fields non-minimally coupled to gravity}
\label{sec:scalar}

This section aims to consider a diffeomorphism invariant theory of gravity coupled to a scalar field $\phi$, with arbitrary non-minimal coupling between the scalar field and gravity being allowed. In such theories with non-minimal coupling, one may get contributions to the energy-momentum tensor that can potentially violate the NEC. However, as we would argue in this section, the off-shell structure of those contributions can always be arranged in a form that will contribute to the entropy current. That way, we can also argue for a linearized version of the second law for such theories
\footnote{As we discussed before in \S\ref{sec:intro}, the entropy current for scalar fields coupled to gravity has recently been studied in \cite{Hollands:2022fck}. In \S 4.2 of \cite{Hollands:2022fck}, an effective Lagrangian has been considered where the higher derivative terms are small in the sense of an effective field theory. However, following the setup of \cite{Bhattacharyya:2021jhr}, we will not explicitly use any restrictions on the size of the couplings of the higher derivative terms involving scalar fields. We will also obtain an explicit algorithm to construct the components of the entropy current, consistent with the generic arguments in \cite{Hollands:2022fck}.
}. 
We will divide the discussion into various sub-sections highlighting essential steps in the derivation. 

To start with, let us mention that we are working with theories with the action given as
\begin{equation}\label{eq:action}
I=\frac{1}{4\pi}\int d^dx \sqrt{-g} \, \mathcal{L}_{(\phi \, g_{\mu\nu})} \, .
\end{equation}
In \eqref{eq:action}, $\mathcal{L}_{(\phi \, g_{\mu\nu})}$, containing pure gravity as well as the interactions terms, will have the following general structure, see \cite{Iyer:1994ys}, 
\begin{equation} \label{eq:phiLag}
\mathcal{L}_{(\phi \, g_{\mu\nu})}=\mathcal{L}_{(\phi \, g_{\mu\nu})}(g_{\mu\nu},R_{\mu\nu\rho\sigma},D_{\alpha_1}R_{\mu\nu\rho\sigma},D_{(\alpha_1}D_{\alpha_2)}R_{\mu\nu\rho\sigma}, \cdots, \phi,D_{\alpha_1}\phi,D_{(\alpha_1}D_{\alpha_2)}\phi, \cdots) \, .
\end{equation}

\subsection{Relating the equations of motion ($\mathcal{E}_{vv}$) with the Noether charges ($\Theta^{\mu}$ and $Q^{\mu\nu}$)} \label{ssec:scalEvv}

Here, following the discussion in \S\ref{sec:strategy}, we will first express the components of equations of motion (EoM) in terms of $\Theta^\mu$ and $Q^{\mu\nu}$, by appropriately defining them. The variation of the Lagrangian due to arbitrary variations of the metric and the scalar field, denoted by $\delta g_{\mu\nu}$ and $\delta \phi$ respectively, is given by
\begin{equation}\label{eq:varL}
\delta[\sqrt{-g} \, \mathcal{L}_{(\phi \, g_{\mu\nu})}]=\sqrt{-g} \, \mathcal{E}^{\mu\nu}\delta g_{\mu\nu}+\sqrt{-g} \, G \, \delta \phi+\sqrt{-g}\, D_\mu \Theta^\mu \, ,
\end{equation}
where, ${\cal E}^{\mu\nu}$ and $G$ are the EoMs for $g_{\mu\nu}$ and $\phi$ respectively, and $\Theta^\mu$ is locally constructed out of $g_{\mu\nu}$, $\phi$, $\delta g_{\mu\nu}$, $\delta \phi$ and their derivatives. We can consider eq.\eqref{eq:varL} as definitions for $\mathcal{E}^{\mu\nu}, \, G, \, \Theta^\mu$. Also, $\Theta^\mu$ is linear in the field variations $\delta g_{\mu\nu}$ and $\delta \phi$. Next, we consider that $\delta g_{\mu\nu}$ and $\delta \phi$ are produced by an infinitesimal diffeomorphism, $x^\mu\rightarrow x^\mu +\zeta^\mu$, such that
\begin{equation}\label{eq:delta}
\begin{split}
\delta g_{\mu\nu}={\cal L}_\zeta g_{\mu\nu}=D_\mu \zeta_\nu+D_\nu \zeta_\mu \, , \quad \text{and} \quad 
\delta \phi={\cal L}_\zeta \phi=\zeta^\alpha D_\alpha \phi \, ,
\end{split}
\end{equation}
where, ${\cal L}_\zeta$ is Lie derivative along $\zeta$.
Independently, one can argue that the variation of the Lagrangian due to the above coordinate transformation is given by
\begin{equation}\label{eq:deltaL1}
\delta[\sqrt{-g} \mathcal{L}_{(\phi \, g_{\mu\nu})}]=\sqrt{-g}D_\mu(\zeta^\mu \mathcal{L}_{(\phi \, g_{\mu\nu})}) \, .
\end{equation}
Substituting eq.\eqref{eq:deltaL1} in \eqref{eq:varL} we can obtain
\begin{equation}\label{eq:totald}
\begin{split} 
D_\mu (\zeta^\mu \mathcal{L}_{(\phi \, g_{\mu\nu})}-2 \mathcal{E}^{\mu\nu}\zeta_\nu-\Theta^\mu)=-2\zeta_\nu D_\mu \mathcal{E}^{\mu\nu}+G \, \zeta^\nu D_\nu \phi \, .
\end{split}
\end{equation}
Now, without any loss of generality, let us assume that $\zeta^\mu$ is non-zero over a small region ${\cal R}$. Then, if we integrate eq.\eqref{eq:totald} over the full space-time, left hand side would vanish since it would integrate to a boundary term at infinity where $\zeta^\mu$ vanishes. So, we get
\begin{equation}
\int_{\text{full-space}} \zeta^\nu\Big[2 D^\mu \mathcal{E}_{\mu\nu}-G\, D_\nu\phi\Big]=0 \, .
\end{equation} 
The above relation holds true for arbitrary $\zeta$ as long as it has non-zero support over a small region ${\cal R}$. Therefore, we see the following relation holds identically
\begin{equation}\label{eq:identity}
2 D^\mu \mathcal{E}_{\mu\nu}-G\, D_\nu\phi=0 \, .
\end{equation}
Substituting \eqref{eq:identity} in \eqref{eq:totald} we see that the RHS of eq.\eqref{eq:totald} vanishes, for any $\zeta$,
\begin{equation}
D_\mu (\zeta^\mu \mathcal{L}_{(\phi \, g_{\mu\nu})}-2 \mathcal{E}^{\mu\nu}\zeta_\nu-\Theta^\mu)=0 \, .
\end{equation}
An identically conserved vector can always be written as the divergence of a rank-two antisymmetric tensor, and hence we get 
\begin{equation}\label{eq:muL}
\zeta^\mu \mathcal{L}_{(\phi \, g_{\mu\nu})}-2 \mathcal{E}^{\mu\nu}\zeta_\nu-\Theta^\mu=-D_\nu Q^{\mu\nu} \, .
\end{equation}
This can be considered as the definition of $Q^{\mu\nu}$. 
Contracting the free index in eq.\eqref{eq:muL} with another $\zeta_\mu$ we get
\begin{equation}\label{eq:zeta2L}
\zeta^2 \mathcal{L}_{(\phi \, g_{\mu\nu})}-\zeta_\mu\Theta^\mu+\zeta_\mu D_\nu Q^{\mu\nu}=2\zeta_\mu \mathcal{E}^{\mu\nu}\zeta_\nu \, .
\end{equation}
The above equation is true for any $\zeta^\mu$, but, we will choose $\zeta^\mu$ to be the Killing vector $\xi$ as given in eq.\eqref{BT_generator} and evaluate eq.\eqref{eq:zeta2L} on the horizon $r=0$, such that we finally obtain
\begin{equation}\label{eq:import}
2 \, v \, \mathcal{E}_{vv}|_{r=0}=(-\Theta^r+D_\mu Q^{r\mu})|_{r=0} \, .
\end{equation}
Note that this is the same relation obtained in \cite{Bhattacharyya:2021jhr}, but now we are re-deriving this result with scalar fields coupled to gravity. 

In deriving eq.\eqref{eq:import}, we have basically gone through the same steps which were worked out in \S2.5 of \cite{Bhattacharyya:2021jhr}, but it was essential to go through it again here to confirm that the EoM for the scalar field, i.e., $G$ as shown in eq.\eqref{eq:varL}, is not going to contribute at all to our analysis.

In the LHS of eq.\eqref{eq:import}, the quantity $\mathcal{E}_{vv}$ contains the contributions from the pure gravity terms (denoted by $E_{vv}$), contributions from minimally coupled terms (denoted by ${T}_{vv}$), as well as the contributions from the possible non-minimal couplings (denoted by $\widetilde{T}_{vv}$), see the discussions in \S\ref{ssec-exNM}. We know that $T_{vv}$ satisfies null energy condition implying $T_{vv}\sim{\cal O}(\epsilon)^2$. From the analysis in \cite{Bhattacharyya:2021jhr}, we already know that $E_{vv}$ will have the desired structure to give us an entropy current in the form of eq.\eqref{eq:main1a}. To conclude the same for $\widetilde{T}_{vv}$, we need to focus on the contributions to  
$\Theta^{\mu}$ and $Q^{\mu\nu}$ from the possible non-minimal coupling of the scalar fields, which we will do in the following sub-section.

\subsection{Constructing the entropy current for non-minimally coupled scalar fields} \label{ssec:scalJvJi}

In order to justify that $\mathcal{E}_{vv}$ has an off-shell structure given by eq.\eqref{eq:nonmin3}, our primary task in this sub-section would be to determine the explicit $v$-dependence of the two terms in the RHS of \eqref{eq:import} and then compare different terms with same explicit $v$-dependence from both sides. Hence, we will also argue that the non-minimal coupling contribution $\widetilde{T}_{vv}$ can be written as eq.\eqref{entcur_nonmin}.

In figuring out the structure of $\mathcal{E}_{vv}$ we must remember that we are working in a particular coordinate system by fixing the gauge as eq.\eqref{eq:metric}. Following the same philosophy advocated in \cite{Bhattacharyya:2021jhr} (see \S2.1 there) we should view $\mathcal{E}_{vv}$ as a component of a covariant tensor of rank two and boost weight $=+2$, built out of some elementary variables: which are known as ``the basic building blocks" and for the case of scalar field coupled to gravity they are the following 
\begin{equation} \label{eq:phibasicvar}
\begin{split}
1.& \,  \text{the metric variables:} \quad X, \, \omega_i, \, h_{ij} \, ,\\
2.& \,  \text{the scalar field:} \quad \phi \, , \\
3. & \, \text{various derivatives acting on them:} \quad \partial_v, \, \partial_r, \, \nabla_i \, .
\end{split}
\end{equation}
As we have discussed in \S\ref{sec:strategy}, we should note that $\phi$ is a new entry in eq.\eqref{eq:phibasicvar},  compared to what we listed in Table-1 in \cite{Bhattacharyya:2021jhr}. 

The next question that one should ask here is: What would be the generic structure of a boost weight $=+2$ tensor in terms of these basic elements? The answer to this is given by ``Result:1" in eq.(3.14) in \cite{Bhattacharyya:2021jhr}. According to this, we must carefully analyze the distribution of $\partial_v$ derivatives because this is needed to get the structure of a covariant tensor with a positive boost weight. These will be important in constraining the structure of $\mathcal{E}_{vv}$ and this is sketched in section 3.1 of \cite{Bhattacharyya:2021jhr}.
The main point in establishing this ``Result:1" was to identify that (at least when we did not consider matter couplings in \cite{Bhattacharyya:2021jhr}) the only way to generate positive boost weight was to apply $\partial_v$ on boost invariant metric functions $X, \, \omega_i, \, h_{ij}$. From our discussion in \S\ref{sec:stationarity} of equilibrium scalar field configurations in eq.\eqref{eqphisol}, we note that the new entry in the list of basic building blocks written above, $\phi$, is boost invariant \footnote{This is obvious since from eq.\eqref{eqphisol} we see that $\phi$ can be a function of the product of $vr$ only, which is invariant under the boost transformation eq.\eqref{boosttransf}.}. Therefore, for scalar couplings with gravity, we do not need to change the statement of ``Result:1" \footnote{It does not, however, mean that we would not get new contributions when we consider non-minimal couplings of the scalar field with gravity. }. This confirms that we do not need to modify ``Result:1" (eq.(3.14) in \cite{Bhattacharyya:2021jhr}) and the same analysis of \S3 in \cite{Bhattacharyya:2021jhr} can be followed to analyze the possible non-minimal coupling terms of scalar fields which are of interest in this present work. 

Next, in order to extract their explicit $v$-dependence, we will use the definitions of $\Theta^\mu$ and $Q^{\mu\nu}$ as given in \cite{Iyer:1994ys} and follow the exactly similar manipulations as in \S3 in \cite{Bhattacharyya:2021jhr} \footnote{For $\Theta^\mu$ see Lemma-(3.1) of \cite{Iyer:1994ys}, and eq.(3.16), (3.17), and (3.18) in \cite{Bhattacharyya:2021jhr}. Whereas, for $Q^{\nu \mu}$ see Proposition-(4.1) of \cite{Iyer:1994ys}, and eq.(3.58) in \cite{Bhattacharyya:2021jhr}}. Following the arguments of \S 3.2 of \cite{Bhattacharyya:2021jhr}, we can write down the final results for the expressions of the entropy current for scalar fields non-minimally coupled to gravity.
This should also be viewed as an algorithm to derive the components of the entropy current once we are given the particular Lagrangian of the theory with all possible non-minimal coupling of scalar fields with gravity. Using the expressions in \S 3.2.4 of \cite{Bhattacharyya:2021jhr}, one can derive the following form of $\mathcal{E}_{vv}$ 
\footnote{Let us remind ourselves that, as argued in section 3.1 of \cite{Bhattacharyya:2021jhr}, we first extract the explicit dependence of $v$ of the terms in $\Theta^r$ and $Q^{r\mu}$ on the RHS of eq.\eqref{eq:import}, and then compare the powers of $v$ on both sides of it. In that process, we schematically obtain (see eq.(3.9) to eq.(3.11) in \cite{Bhattacharyya:2021jhr})
\begin{equation*}
 2\, v \, \mathcal{E}_{vv} |_{r=0}
        = \mathcal{U}_{(1)} + v \, \mathcal{V}_{(2)} + \mathcal{O}(\epsilon^2) \quad \Rightarrow \quad \mathcal{U}_{(1)} = \mathcal{O}(\epsilon^2), \quad \text{and} \quad 2 \mathcal{E}_{vv}|_{r=0} = \mathcal{V}_{(2)} + \mathcal{O}(\epsilon^2)  \, .
\end{equation*}
Further manipulations of the second set of equations lead us to eq.\eqref{eq:Evvfin}. 
It is important to note that this analysis is just a \emph{structural rearrangement of the off-shell form} of $\mathcal{E}_{vv}$ involving various metric components $X,\omega_i,h_{ij}$ and their derivatives. For example, the relation $\mathcal{U}_{(1)} = \mathcal{O}(\epsilon^2)$ is identically true, independent of considering  what could possibly be the explicit forms of $X,\omega_i,h_{ij}$ as functions of $v,r,x^i$. In other words, $\mathcal{U}_{(1)} = \mathcal{O}(\epsilon^2)$ should not be thought of as differential equations constraining $X,\omega_i,h_{ij}$ as functions of $v,r,x^i$. One can check that, for any given theory, $\mathcal{U}_{(1)}$ will identically vanish.
}, 
\begin{equation}\label{eq:Evvfin}
2 ~{\cal E}_{vv}|_{r=0}=-\partial_v\left(\frac{1}{\sqrt{h}}\partial_v\left[\sqrt{h}\left(\tilde{Q}^{rv}+{\cal B}_{(0)}\right)\right]+\nabla_i\left[\tilde{Q}^{ri}-J_{(1)}^i\right]\right)+{\cal O}(\epsilon^2) \, . 
\end{equation}
We will now mention how one can obtain ${\cal B}_{(0)}, \, \tilde{Q}^{rv}, \, \tilde{Q}^{ri}$ and $J_{(1)}^i$ once the Lagrangian eq.\eqref{eq:phiLag} is known. For that we first need to obtain $\Theta^r$ and $Q^{r\mu}$ for the given Lagrangian. Firstly, ${\cal B}_{(0)}$ is defined through the following equation
\begin{equation} \label{defB}
\Theta^r=(1+v\partial_v){\cal A}_{(1)}+v\partial_v^2 {\cal B}_{(0)}+{\cal O}(\epsilon^2) \, .
\end{equation}
Similarly, $Q^{r\mu}$ can be expressed by explicitly working out the $v$ dependence as follows
\begin{equation} \label{defQtW}
Q^{r\mu}=\tilde{Q}^{r\mu}+v W_v^{r\mu} \, ,
\end{equation}
where $\tilde{Q}^{r\mu}$ and $W_v^{r\mu}$ have no explicit $v$ dependence. Thus, eq.\eqref{defQtW} defines $\tilde{Q}^{rv}$ and $\tilde{Q}^{ri}$ for us. Finally, $J^i_{(1)}$ is defined through the equation
\begin{equation} \label{defJi}
W^{ri}_v=\partial_v J^i_{(1)}+{\cal O}(\epsilon^2) \, .
\end{equation}
From \eqref{eq:Evvfin}, using the following relation $\mathcal{E}_{vv} =  E_{vv} +\widetilde{T}_{vv} +T_{vv} $ and also the NEC, $T_{vv} \sim \mathcal{O}(\epsilon^2)$, we can write
\begin{equation}
(E_{vv}+\widetilde{T}_{vv})|_{r=0}=\partial_v\left[\frac{1}{\sqrt{h}}\partial_v\left(\sqrt{h}\, ( {\cal J}^v + \widetilde{\cal J}^v )\right)+\nabla_i ( {\cal J}^i + \widetilde{\cal J}^i)\right]+{\cal O}(\epsilon^2) \, ,
\end{equation}
where $( {\cal J}^v + \widetilde{\cal J}^v )$ and $( {\cal J}^i + \widetilde{\cal J}^i )$ are given by
\begin{equation} \label{finJvJiscal}
\begin{split}
( {\cal J}^v + \widetilde{\cal J}^v ) =-\frac{1}{2}\left(\widetilde{Q}^{rv}+{\cal B}_{(0)}\right)\, ,\quad \text{and} \quad ( {\cal J}^i + \widetilde{\cal J}^i)=-\frac{1}{2}\left(\widetilde{Q}^{ri}-J^i_{(1)}\right) \, ,
\end{split}
\end{equation}
where it is obvious that in the RHS of eq.\eqref{finJvJiscal}, we have contributions from both purely gravitational terms and the possible non-minimal couplings as well. 
This completes the proof that $\widetilde{T}_{vv}$ has the required structure even if we have a non-minimally coupled scalar field.

\subsection{Verification of the abstract proof: An example} \label{ssec:checkscal}

In this sub-section, we would like to consider a particular example of a theory with an explicit non-minimal coupling of scalar fields with gravity. Following the abstract algorithm to construct the components of entropy current $\widetilde{\cal J}^v$ and $\widetilde{\cal J}^i$, we compute them explicitly for this particular model. 

We considering a Lagrangian of the form
\begin{equation}
\mathcal{L}_{(\phi \, g_{\mu\nu})}= R_{\mu\nu}(D^\mu\phi)(D^\nu \phi) \, .
\end{equation}
The standard variation of the Lagrangian gives
\begin{equation}
    \begin{split}
        \delta (\sqrt{-g} \, \mathcal{L}_{(\phi \, g_{\mu\nu})}) &= \sqrt{-g}\Bigg[ \frac{1}{2}g^{\mu\nu}R_{\alpha\beta}(D^\alpha\phi)(D^\beta\phi)-R^{\mu\alpha}(D^\nu\phi)(D_\alpha\phi)-R^{\beta\nu}(D_\beta\phi)(D^\mu\phi)\\
        &+\frac{1}{2}\Big[D_\alpha D^\nu\{(D^\alpha\phi)(D^\mu\phi)\}+D^\alpha D^\mu\{(D_\alpha\phi)(D^\nu\phi)\}-D^\gamma D_\gamma\{(D^\nu\phi)(D^\mu\phi)\}\\
        &-g^{\mu\nu}D^\alpha D^\beta\{(D_\alpha\phi)(D_\beta\phi)\}\Big] \Bigg] \delta g_{\mu\nu} -2 \sqrt{-g} D_\mu\left(R^{\mu\nu}D_\nu\phi\right) \delta \phi + \sqrt{-g} D_{\mu} \Theta^{\mu} \, ,
    \end{split}
\end{equation}
where $\Theta^{\lambda}$ is given by
\begin{equation}
    \begin{split}
        \Theta^\lambda&=2E_R^{\lambda\nu\alpha\beta}(D_\beta\delta g_{\nu\alpha})+2R^{\mu\lambda}(D_\mu \phi) \delta\phi+\frac{1}{2}\delta g^{\beta\alpha}D_\alpha(D^\lambda\phi D_\beta\phi)\\
        &+\frac{1}{2}\delta g^{\beta\alpha}D_\beta(D^\lambda\phi D_\alpha \phi)-\frac{1}{2}\delta g^{\beta\alpha}D^\lambda(D_\alpha\phi D_\beta \phi)-\frac{1}{2}\delta g^{\rho\nu}g_{\rho\nu}D^\beta(D^\lambda\phi D_\beta\phi) \, ,
    \end{split}
\end{equation}
where 
\begin{equation}
    E_R^{\lambda\nu\alpha\beta} = \frac{1}{4}g^{\lambda\alpha}D^\nu\phi D^\beta\phi-\frac	{1}{4}g^{\nu\alpha}D^\lambda\phi D^\beta\phi-\frac	{1}{4}g^{\lambda\beta}D^\nu\phi D^\alpha\phi+\frac	{1}{4}g^{\nu\beta}D^\lambda\phi D^\alpha\phi \, .
\end{equation}
Using the equation $\Theta^\mu-\xi^\mu L=-2 E^{\mu\nu}\zeta_\nu+D_\nu Q^{\mu\nu}$, we can calculate $Q^{\mu\nu}$
\begin{equation}
\begin{split}
Q^{\rho\nu}&=2E_R^{\rho\nu\alpha\mu}D_\mu \xi_\alpha+D^\rho(D^\nu\phi D^\alpha\phi)\xi_\alpha-D^\nu(D^\rho\phi D^\alpha\phi)\xi_\alpha\\
&+g^{\nu\alpha}D_\sigma(D^\rho\phi D^\sigma\phi)\xi_\alpha-g^{\rho\alpha}D_\sigma(D^\nu\phi D^\sigma\phi)\xi_\alpha \, .\\
\end{split}
\end{equation}
From these expressions, and using eq.\eqref{defB}, eq.\eqref{defQtW}, and eq.\eqref{defJi}, we can readily obtain the following quantities 
\begin{equation}
\begin{split}
{\cal B}_{(0)} &= 0 \, ,\quad \widetilde{Q}^{rv} = -2 (\partial_v \phi)(\partial_r \phi) \, ,\quad \widetilde{Q}^{ri} = -(\nabla^i \phi)(\partial_v \phi) \, ,\quad J^i_{(1)} = (\nabla^i \phi)(\partial_v \phi) \, .
\end{split}
\end{equation}
Using them in eq.\eqref{finJvJiscal} we can find 
\begin{equation} \label{eq:scalJvJiEx}
\begin{split}
      {\cal J}^v=(\partial_r\phi)(\partial_v\phi) \, , \quad \text{and} \quad 
      {\cal J}^i=(\nabla^i\phi)(\partial_v\phi) \, .
\end{split}
\end{equation}
We are now in a position to verify that the expressions of the entropy current obtained in eq.\eqref{eq:scalJvJiEx} exactly matches with the brute force calculation of ${\cal J}^v$ and ${\cal J}^i$ from the EoM, which we have worked out in \S\ref{ssec-exNM}, see eq.\eqref{entcur_NM}. 

Interestingly, $\mathcal{J}^v$ is purely $\mathcal{O}(\epsilon)$ with no equilibrium contribution. Thus, there is no equilibrium Wald Entropy for this Lagrangian, and $\mathcal{J}^v$ is purely a JKM type ambiguity.

\section{Gauge fields non-minimally coupled to gravity} \label{sec:NMeinmax}

In this section, our aim is to construct an entropy current for diffeomorphism invariant theories of gravity coupled to a $U(1)$ gauge field $A_\mu$. Consider an action of the form, \begin{equation}\label{eq:gaugeaction}
I=\frac{1}{4\pi}\int d^dx \, \sqrt{-g}\, \mathcal{L}_{(A_\mu g_{\mu\nu})} \, ,
\end{equation}
where the Lagrangian $\mathcal{L}_{(A_\mu g_{\mu\nu})}$ contains pure gravity terms as well as gauge fields non-minimally coupled to gravity and we consider the most general gauge invariant Lagrangian of the following form
\begin{equation}\label{eq:gaugelagrangian}
\mathcal{L}_{(A_\mu g_{\mu\nu})}=\mathcal{L}_{(A_\mu g_{\mu\nu})} (g_{\alpha\beta},R_{\alpha\beta\rho\sigma},D_{\alpha_1}R_{\alpha\beta\rho\sigma},\cdots,F_{\mu\nu},D_{\alpha_1} F_{\mu\nu},D_{(\alpha_1}D_{\alpha_2)}F_{\mu\nu},\cdots) \, ,
\end{equation}
where $R_{\alpha\beta\gamma\delta}$ is the Riemann tensor and $F_{\mu\nu}$ is the electromagnetic field strength tensor given by $F_{\mu\nu} = D_{\mu} A_{\nu} - D_{\nu} A_{\mu}$ \footnote{By considering this form of the Lagrangian, we inherently make a choice to only work with gauge-invariant Lagrangians.}. 

The Lagrangian $\mathcal{L}_{(A_\mu g_{\mu\nu})}$ may, in principle, include generic terms with non-minimal couplings between the field strength tensor and the Riemann tensor (also, each of them can have arbitrary derivatives acting on them). In this paper, our main focus is on such non-minimal interaction terms which can potentially contribute at linear order in amplitude expansion at $\mathcal{O}(\epsilon)$ to the $T_{\mu\nu}$ so that the NEC may get violated, see \S\ref{ssec-exNM}. In the rest of this section, we will show that these linear order terms with a potential threat of violating NEC can always be written in the form required for constructing an entropy current as in eq.\eqref{entcur_nonmin}, and thus a linearized second law eq.\eqref{cond3modf} would automatically follow. Additionally we will also see that in particular, the components of the entropy current derived from eq.\eqref{eq:gaugelagrangian} will always turn out to be gauge invariant. We will present our derivation by dividing it into a few sub-sections for the sake of clarity. Also, in our discussions below, we will focus on contributions that will come from the coupling terms between gauge fields and metric in eq.\eqref{eq:gaugelagrangian}. Since the pure gravity terms have already been taken care of in \cite{Bhattacharyya:2021jhr}.

\subsection{Relating the equations of motion ($\mathcal{E}_{vv}$) with the Noether charges ($\Theta^{\mu}$ and $Q^{\mu\nu}$)} \label{ssec:gaugeEvv}

Following the Noether charge analysis of \cite{Iyer:1994ys}, the variation of the Non-minimal Lagrangian, eq.\eqref{eq:gaugelagrangian}, is given by
\begin{equation}\label{eq:Lvariation}
    \delta [ \sqrt{-g} \, \mathcal{L}_{(A_\mu g_{\mu\nu})} ] = \sqrt{-g} \, \mathcal{E}^{\mu\nu} \delta g_{\mu\nu} + \sqrt{-g} \, G^{\mu} \delta A_{\mu} + \sqrt{-g} \, D_{\mu} \Theta^{\mu}  \, .
\end{equation}
where $G^{\mu}$ is the equations of motion (EoM) for the gauge field, and $\Theta^\mu$ should be viewed as locally constructed out of $g_{\mu\nu}$, $A_\mu$, $\delta g_{\mu\nu}$, $\delta A_\mu$ and their derivatives.
Let us now consider the variation induced by a diffeomorphism $x^{\mu} \rightarrow x^{\mu} + \zeta^{\mu}$ and a gauge transformation $A_{\mu} \rightarrow A_{\mu} + D_{\mu} \Lambda$, then we have
\begin{equation}\label{eq:Fvariation}
    \begin{split}
        \delta g_{\mu\nu} &= \mathcal{L}_{\zeta} g_{\mu\nu} = D_{\mu} \zeta_{\nu} + D_{\nu} \zeta_{\mu} \, , \\
        \delta A_{\mu} &= \mathcal{L}_{\zeta} A_{\mu} + D_{\mu}\Lambda = \zeta^{\alpha} D_{\alpha} A_{\mu} + A_{\alpha} D_{\mu} \zeta^{\alpha} + D_{\mu} \Lambda
        = \zeta^{\alpha} F_{\alpha\mu} + D_{\mu} ( A_{\alpha}\zeta^{\alpha} + \Lambda ) \, .
    \end{split}
\end{equation}
Since diffeomorphisms and gauge transformations are symmetries of the theory, the variation of the Lagrangian should be written as a total derivative
\begin{equation}\label{eq:diffsym}
    \delta [ \sqrt{-g} \,\mathcal{L}_{(A_\mu g_{\mu\nu})} ] = \sqrt{-g} \, D_{\mu} [ \zeta^{\mu} \mathcal{L}_{(A_\mu g_{\mu\nu})}] \, .
\end{equation}
Substituting eq.\eqref{eq:Fvariation} and eq.\eqref{eq:diffsym} in eq.\eqref{eq:Lvariation}, and after some manipulation we get,
\begin{equation}\label{eq:preidentity}
    \begin{split}
        D_{\mu} \left[ \zeta^{\mu} \mathcal{L}_{(A_\mu g_{\mu\nu})}- 2 \mathcal{E}^{\mu\nu}\zeta_{\nu} \right. & \left. - \, G^{\mu} ( A^{\nu}\zeta_{\nu} + \Lambda ) - \Theta^{\mu}\right] \\
        & = - 2 \, \zeta_{\nu} D_{\mu} \mathcal{E}^{\mu\nu} - (\zeta^{\nu} A_{\nu} + \Lambda ) D_{\mu} G^{\mu} + G^{\mu}\zeta^{\nu} F_{\nu\mu} \, .
    \end{split}
\end{equation}
The LHS of eq.\eqref{eq:preidentity} is a total derivative. If we make a choice for $\zeta^{\mu}$ and $\Lambda$ such that it is non-zero only in a small region $\mathcal{R}$, and if we integrate on both sides over the full space-time, LHS would vanish since it would integrate to a pure boundary term at infinity where $\zeta^\mu$ and $\Lambda$ vanishes. Hence, we obtain 
\begin{equation}
    \int_{\text{full spacetime}} \zeta_{\nu} \left[ -2 \, D_{\mu}\mathcal{E}^{\mu\nu} + G_{\mu} F^{\nu\mu} \right] - \int_{\text{full spacetime}} (\zeta_{\nu} A^{\nu} + \Lambda ) D_{\mu} G^{\mu}  = 0 \, .
\end{equation}
The above relation is true as long as $\zeta^{\nu}$ and $\Lambda$ has non-zero support over a finite region $\mathcal{R}$, which can only happen if the integrand in the above equation itself vanishes. Since $A^{\nu}$ is an independent field, we see that the following relations, known as the Bianchi identities, hold identically
\begin{equation}\label{eq:bianchiidentity}
     \begin{split}
         -2 D_{\mu} \mathcal{E}^{\mu\nu} + G_{\mu} F^{\nu\mu} = 0 \, , \quad \text{and} \quad 
         D_{\mu} G^{\mu} = 0 \, .
     \end{split}
\end{equation}
Substituting eq.\eqref{eq:bianchiidentity} in eq.\eqref{eq:preidentity}, and realizing that we get an identically conserved vector, which can always be written as the divergence of a rank-two antisymmetric tensor (say, $Q^{\mu\nu}$), we have
\begin{equation} \label{eq:zetareln1}
      \Theta^{\mu} - \zeta^{\mu} \mathcal{L}_{(A_\mu g_{\mu\nu})} = - 2 \mathcal{E}^{\mu\nu}\zeta_{\nu} - \, G^{\mu} (A^{\nu}\zeta_{\nu} + \Lambda ) + D_{\nu} Q^{\mu\nu} \, .
\end{equation}
Contracting both sides of the eq.\eqref{eq:zetareln1} by $\zeta_{\mu}$ we get
\begin{equation} \label{eq:zetareln2}
    -2 \mathcal{E}^{\mu\nu} \zeta_{\mu} \zeta_{\nu} - G^{\mu} \zeta_{\mu} ( A^{\nu} \zeta_{\nu} + \Lambda ) + \zeta_{\mu} D_{\nu} Q^{\mu\nu} = \zeta_{\mu} \Theta^{\mu} - \zeta^2 \mathcal{L}_{(A_\mu g_{\mu\nu})} \, .
\end{equation}
Here we notice that the gauge field EoM, i.e. $G^{\mu} $, has appeared in eq.\eqref{eq:zetareln2}, which we have previously alluded to in eq.\eqref{hijibiji-4} and eq.\eqref{hijibiji-5} in \S\ref{sec:strategy} \footnote{We should remember that this was not the case when scalar field couplings were considered, see eq.\eqref{eq:zeta2L} in \S\ref{ssec:scalEvv}.}.

Next, we choose $\zeta$ to be $\xi = v \partial_v - r \partial_r$ (the Killing vector given by eq.\eqref{BT_generator}), and evaluating both sides using the gauge eq.\eqref{eq:metric} on the horizon $r=0$, we get \footnote{We have used $\mathcal{E}_{vv} =  E_{vv} +\widetilde{T}_{vv} +T_{vv} $ and also the NEC, $T_{vv} \sim \mathcal{O}(\epsilon^2)$.}
\begin{equation} \label{eq:EvvAmu1}
2 v \, (E_{vv} +\widetilde{T}_{vv})+ G_v ( v A_v + \Lambda ) =\left( -\Theta^r + D_{\nu} Q^{r\nu}\right)  |_{r=0} +\mathcal{O}(\epsilon^2)\, .
\end{equation}
Now, from the discussion in \S\ref{ssec-setup}, and Appendix-\ref{app:boost}, we know that the $v$-component of the EoM for the gauge field, $G_v$, has a boost weight of $+1$ and thus it will vanish in equilibrium configurations,  or, equivalently, we will get
\begin{equation}
G_v \sim \mathcal{O}(\epsilon) \, .
\end{equation} 
Following the arguments of \S\ref{sec:stationarity}, we have seen, in eq.\eqref{eq:STgauge1}, that the quantity $ ( A^{\mu}\xi_{\mu} + \Lambda )$ is also of $\mathcal{O}(\epsilon)$ when evaluated on the horizon. Therefore, we immediately obtain that 
\begin{equation} \label{eq:Gvrel}
    G_v ( v A_v + \Lambda ) |_{r=0} \sim \mathcal{O}(\epsilon^2) \, .
\end{equation} 
This important relation in eq.\eqref{eq:Gvrel} establishes that the contribution from the EoM of $A_\mu$, although it appears in eq.\eqref{eq:EvvAmu1}, is going to disappear from our final key equation that relates $\mathcal{E}_{vv}$ with $\Theta^r$ and $Q^{r\nu}$ up to linearized order in $\epsilon$. So, using eq.\eqref{eq:Gvrel} in eq.\eqref{eq:EvvAmu1}, we are left with the usual relation given by eq.\eqref{keyeqn}, which we write here again for convenience
\begin{equation}\label{maineq}
   2 v \, (E_{vv} +\widetilde{T}_{vv})=\left( -\Theta^r + D_{\nu} Q^{r\nu}\right)  |_{r=0} +\mathcal{O}(\epsilon^2) \, . 
\end{equation} 
Next, our aim will be to show that the RHS in eq.\eqref{maineq} has the desired structure mentioned in eq.\eqref{eq:nonmin3}. Alternatively, since we are focussing here on the possible non-minimal contributions, our aim is to show that $\widetilde{T}_{vv}$ has the structure predicted in eq.\eqref{entcur_nonmin}.

\subsection{New basic element with positive boost weight and modification of ``Result: 1"}
\label{ssec:modresult1}

As we are trying to establish that the $vv$-component of the EoM (more specifically, of $\widetilde{T}_{vv})$ in our present context) has the desired structure, we need to work out the explicit $v$-dependence of $\Theta^r$ and $Q^{r\nu}$. Before we start on that, in this sub-section, we will justify in what aspect the construction of entropy current for gauge fields coupled to gravity is technically different compared to the analysis with pure gravity modes, studied in \cite{Bhattacharyya:2021jhr}, and also to the analysis with scalar fields coupled to gravity, studied previously in \S\ref{sec:scalar}. 

According to the principle we are following in our construction all covariant tensor quantities (e.g., $E_{vv}, \, \widetilde{T}_{vv}, \, \Theta^r,$ or $Q^{r\nu}$) are built out of certain basic building blocks, in our chosen metric gauge. The relevant basic building blocks for gauge fields coupled to gravity are the following:
\begin{equation} \label{eq:Amubasicvar}
\begin{split}
1.& \,  \text{the metric variables:} \quad X, \, \omega_i, \, h_{ij} \, ,\\
2.& \,  \text{the gauge field components:} \quad A_v, \, A_r, \, A_i\, , \\
3. & \, \text{various derivatives acting on them:} \quad \partial_v, \, \partial_r, \, \nabla_i \, .
\end{split}
\end{equation}
Firstly, we see that the components of the gauge field are now new addition to the list of basic elements (compare with Table-1 in \cite{Bhattacharyya:2021jhr}). Most importantly, from the discussions in \S\ref{ssec-setup}, \S\ref{sec:stationarity}, and Appendix-\ref{app:boost}, one must be convinced that we now have $A_v$ as a new element  with boost weight $=+1$ \footnote{Similarly, $A_r$ and $A_i$ have boost weights equal to $-1$ and $0$ respectively.}. For future reference, let us also note down the boost weights of the components of the field strength tensors: $F_{vi}$ has boost weight $ =+1$, and both $F_{vr}$, and $F_{ij}$ are boost invariant.

This new entry in the list of basic elements presented in eq.\eqref{eq:Amubasicvar} will play an important role as we are trying to incorporate gauge field couplings with gravity in the formalism developed in \cite{Bhattacharyya:2021jhr} based on boost weights of objects. Following what we have already mentioned in \S\ref{sec:strategy}, this means that along with $\partial_v$ acting on boost invariant objects (such as $\partial_v X$, $\partial_v \omega_i$, etc. with boost-weight $=1$) we can also use $A_v$ and its various derivatives, in fixing the generic off-shell structure of any tensor quantity with positive boost weight (e.g., $\widetilde{T}_{vv}$). The important relation that provides us this most general form of any tensor quantity with a given positive boost weight in terms of the basic building blocks was called the ``Result: 1" in \cite{Bhattacharyya:2021jhr} (see eq.(3.14) there). Let us mention the statement of this result for clarity: 
any covariant tensor, denoted by $ {\mathbb T}_{(a+1)}$, of boost weight $a+1>0$ can always be expressed (when evaluated at $r=0$) as follows
\begin{equation}\label{result1Sch}
    {\mathbb T}_{(a+1)}= \partial_v^{(a+1)} C_{(0)} + A_{(0)} \, \partial_v^{(a+1)}B_{(0)} + \mathcal{O}(\epsilon^2) \, .
\end{equation}
Note that here we are presenting a schematic form of the result, and the detailed expression can be found in eq.\eqref{eq:res1detAP} in Appendix-\ref{app:ModRes1}. Also, we are not being cautious with the indices, and the numbers in the $()$ of subscripts denote the boost weights. 

In \cite{Bhattacharyya:2021jhr}, no matter fields were considered, except they contribute to an energy-momentum tensor that satisfies the NEC. In that context, each of the $A_{(0)}$, $B_{(0)}$, and $C_{(0)}$ appearing in eq.\eqref{result1Sch}, were viewed as quantities built out of basic building blocks involving only the metric functions and their derivatives. This relation in eq.\eqref{result1Sch} was derived in Appendix-E of \cite{Bhattacharyya:2021jhr}, and it was significant for this derivation that from within the basic elements available, $\partial_v$ is the only operation that can generate a positive boost weight. However, now we have seen that if one considers gauge fields coupled to gravity, one has $A_v$, which can contribute to the RHS of eq.\eqref{result1Sch} written above. So we need to consider modification of this result, keeping this possible new element in mind \footnote{Remember this was not necessary in the scalar case because although we had a new basic element in the form of $\phi$, that was boost invariant.}. 

The detailed steps to incorporate the modifications to the ``Result: 1" is presented in Appendix-\ref{app:ModRes1}, and here we will just present the final result, and that too in a schematic form, as given below
\begin{equation}\label{modresult1f}
    {\mathbb T}_{(a+1)}= \partial_v^{(a+1)} C_{(0)} + A_{(0)} \, \partial_v^{(a+1)}B_{(0)} + \partial^{(a)}_v ( F_{vi} \, D^i_{(0)} ) + E_{(0)} \, \partial^{(a)}_v ( F_{vi} \, G^i_{(0)} ) + \mathcal{O}(\epsilon^2) \, ,
\end{equation}
where the last two terms on the RHS of eq.\eqref{modresult1f} are the new modifications and note that both $F_{vi} \, D^i_{(0)} $ and $F_{vi} \, G^i_{(0)} $ are quantities with boost weight $=+1$. 

Before we end this sub-section, let us mention that in eq.\eqref{modresult1f}, we are only considering those modifications of ``Result: 1" that are relevant for our proof. For instance, we are not allowing any terms that explicitly depend on $A_v$, but it shows up implicitly via $F_{vi}$. For example, we could have had a contribution to the RHS of eq.\eqref{modresult1f} as $\partial_v^{(a)} ( A_{v} \, H_{(0)} )$, which should have been allowed given what we have said so far. However, as we will see later in the next section since our Lagrangian in eq.\eqref{eq:gaugelagrangian} depends only on gauge-invariant objects like $F_{\mu\nu}$ and no explicit dependence on $A_{\mu}$ is considered, we will always get gauge-invariant contributions in the RHS of eq.\eqref{maineq}. Hence we have ignored such contributions to the RHS of eq.\eqref{modresult1f}. We must also note that the first two terms on the RHS of eq.\eqref{modresult1f} are of the same structure as eq.(3.14) in \cite{Bhattacharyya:2021jhr}. However, the boost invariant quantities $A_{(0)}, \, B_{(0)}$ and $C_{(0)}$ may generically receive contributions involving the field strength tensor (for example, a boost invariant term like $F_{vi} {F_{r}}^i$ can arise in these quantities). But, the last two terms on the RHS of eq.\eqref{modresult1f} are genuinely new structures appearing only because we have a new basic element with boost weight $=+1$.

\subsection{Structures of $\Theta^r$ and $Q^{r\mu}$ with contributions from gauge field coupling} \label{ssec:ThQcontr}

We will now look into the structures of both $\Theta^r$ and $Q^{r\mu}$. In the presence of a gauge field $A_{\mu}$ coupled to gravity, according to Lemma 3.1 of \cite{Iyer:1994ys}, $\Theta^{\mu}$ takes the following form
\begin{equation} \label{defThAmu}
    \begin{split}
        \Theta^{\mu} = 2 E^{\mu\nu\alpha\beta}_R D_{\beta} \delta g_{\nu\alpha} + S^{\mu\alpha\beta} \delta g_{\alpha\beta} + \sum^{m-1}_{i=0} T^{\mu\alpha\beta\rho\sigma\alpha_1 \dots \alpha_i}_i \delta D_{(\alpha_1} \dots D_{\alpha_i)} R_{\alpha\beta\rho\sigma} \\
        + \sum^{l-1}_{i=0} U^{\mu\alpha_1 \dots \alpha_i \sigma}_i \delta D_{(\alpha_1} \dots D_{\alpha_i)} A_{\sigma} \, .
    \end{split}
\end{equation}
But, we would like to point out that all the three terms on the RHS of eq.\eqref{defThAmu}, will not contain any terms involving the gauge field components, i.e., $A_v$, $A_r$ or $A_i$. In other words, $E^{\mu\nu\alpha\beta}_R$, $S^{\mu\alpha\beta}$, $T^{\mu\alpha\beta\rho\sigma\alpha_1 \dots \alpha_i}_i$, $U^{\mu\alpha_1 \dots \alpha_i \sigma}_i$ are all gauge invariant structures. This is because the Lagrangian in eq.\eqref{eq:gaugelagrangian} contains only $F_{\mu\nu}$ and not explicit $A_\mu$. 
However, notice that the last term involving $\delta D_{(\alpha_1} \dots D_{\alpha_i)} A_{\sigma}$ will explicitly have $A_\mu$, but that too only linear terms. Thus, gauge invariance is apparently lost. Interestingly, as we will prove in Appendix-\ref{Ap:structureu}, there is an anti-symmetry in $\alpha_i$ and $\sigma$ indices for the following term
\begin{equation}\label{eq:structureu}
    U^{\mu\alpha_1 \dots \alpha_i \sigma}_i \delta D_{(\alpha_1} \dots D_{\alpha_i)} A_{\sigma} = -U^{\mu\alpha_1 \dots \alpha_i \sigma}_i \delta D_{(\alpha_1} \dots D_{\sigma)} A_{\alpha_i} \, .
\end{equation}
Using this for the last term on the RHS of eq.\eqref{defThAmu} we can isolate the term proportional to $\delta A_{\nu}$ and thus splitting the summation into terms which are gauge invariant and one non-gauge invariant term
\begin{equation}
    \sum^{l-1}_{i=0} U^{\mu\alpha_1 \dots \alpha_i \sigma}_i \delta D_{(\alpha_1} \dots D_{\alpha_i)} A_{\sigma} = U^{\mu\nu}_0 \delta A_{\nu} +  \sum^{l-2}_{i=0} \widetilde{U}^{\mu\alpha_1 \dots \alpha_i \rho \sigma}_i \delta D_{(\alpha_1} \dots D_{\alpha_i)} F_{\rho\sigma} \, .
\end{equation}
Furthermore, using the arguments considered in Appendix \ref{Ap:structureu} and \ref{app:ThQNoGinv}, it is also clear that $U^{\mu \nu}_0$ is anti-symmetric,
\begin{equation} \label{U0antisym}
    U^{\mu \nu}_0 = - U^{\nu \mu}_0 \, .
\end{equation}

Finally, we have the following form of $\Theta^{\mu}$ for gauge fields coupled to gravity with a gauge-invariant Lagrangian given by eq.\eqref{eq:gaugelagrangian}:
\begin{equation}\label{finalformtheta}
    \begin{split}
        \Theta^{\mu} = 2 E^{\mu\nu\alpha\beta}_R D_{\beta} \delta g_{\nu\alpha} + U^{\mu\nu}_0 \delta A_{\nu} + \Theta'^{\mu} \, ,
    \end{split}
\end{equation}
where
\begin{equation}
    \begin{split}
        \Theta'^{\mu} &= S^{\mu\alpha\beta} \delta g_{\alpha\beta} + \sum^{m-1}_{i=0} T^{\mu\alpha\beta\rho\sigma\alpha_1 \dots \alpha_i}_i \delta D_{(\alpha_1} \dots D_{\alpha_i)} R_{\alpha\beta\rho\sigma} 
        \\&~~~ 
        + \sum^{l-2}_{i=0} \widetilde{U}^{\mu\alpha_1 \dots \alpha_i \rho \sigma}_i \delta D_{(\alpha_1} \dots D_{\alpha_i)} F_{\rho\sigma} \, .
    \end{split}
\end{equation}
It is also clear that in $\Theta^{\mu}$, the only gauge non-invariant contribution comes through $U^{\mu\nu}_0 \delta A_{\nu} $, and rest are all gauge invariant. 

With the form of $\Theta^{\mu}$ given in eq.\eqref{finalformtheta}, our next job is to obtain the Noether charge, $Q^{\mu\nu}$. For that we need to use the prescription laid out in Proposition-4.1 of \cite{Iyer:1994ys}, see also \S3.2.3 in \cite{Bhattacharyya:2021jhr}. Leaving out the gauge non-invariant term (i.e. $U^{\mu\nu}_0 \delta A_{\nu}$), the other two terms on the RHS of eq.\eqref{finalformtheta} result in a Noether charge of the form \footnote{We have used eq.\eqref{eq:Fvariation} for $\zeta^\mu = \xi^\mu$.}
\begin{equation} \label{exprQmunu1}
    Q^{\mu\nu} = W^{\mu\nu\rho}\xi_{\rho} - 2 E^{\mu\nu\alpha\beta}_R D_{[\alpha}\xi_{\beta]} \, .
\end{equation}
These terms in $Q^{\mu\nu}$, obtained in eq.\eqref{exprQmunu1}, include contributions from both the metric and the gauge field, and are also gauge invariant (i.e., only involves $F_{\mu\nu})$.

\subsubsection*{No contribution from $U^{\mu\nu}_0 \delta A_{\nu} $ in eq.\eqref{finalformtheta} to $\mathcal{E}_{vv}$ at $\mathcal{O}(\epsilon)$:}
\label{sssec:amutheta}

We have already taken care of the contributions from the gauge-invariant terms in $\Theta^{\mu}$ in eq.\eqref{finalformtheta} to $Q^{\mu\nu}$ in eq.\eqref{exprQmunu1}. 
The only gauge non-invariant contribution that we need to separately analyze is $U^{\mu\nu}_0 \delta A_{\nu}$. 
One can rearrange it in the following way, using eq.\eqref{eq:Fvariation}, 
\begin{equation} \label{ThNoGinv}
    \begin{split}
    \Theta^{\mu} \sim U^{\mu\nu}_0 \delta A_{\nu} &= U^{\mu\nu}_0 ( \xi^{\alpha} D_{\alpha} A_{\nu} + A^{\alpha} D_{\nu} \xi_{\alpha} + D_{\nu} \Lambda ) \\
        % &= U^{\mu\nu}_0 \xi^{\alpha} D_{\alpha} A_{\nu} + D_{\nu} [ U^{\mu\nu}_0 ( A^{\alpha}\xi_{\alpha} + \Lambda )] - D_{\nu} ( U^{\mu\nu}_0 A^{\alpha} ) \xi_{\alpha} - D_{\nu} U^{\mu\nu}_0 \Lambda \\
        &= U^{\mu\nu}_0 \xi^{\alpha} F_{\alpha\nu} - D_{\nu} U^{\mu\nu}_0 ( A^{\alpha}\xi_{\alpha} + \Lambda ) + D_{\nu} [ U^{\mu\nu}_0 ( A^{\alpha}\xi_{\alpha} + \Lambda )] \, , 
    \end{split}
\end{equation}
such that, the contribution of this term to $Q^{\mu\nu}$ is readily obtained as 
 \footnote{Notice how the antisymmetry of $U^{\mu\nu}_0$ mentioned in eq.\eqref{U0antisym} is consistent with the antisymmetry of $Q^{\mu\nu}$.}
\begin{equation} \label{QNoGinv}
 Q^{\mu\nu} \sim   U^{\mu\nu}_0 (A^{\alpha}\xi_{\alpha} + \Lambda) \, .
\end{equation}
Next we will evaluate both $\Theta^r$ and $Q^{r\nu}$ given in eq.\eqref{ThNoGinv} and eq.\eqref{QNoGinv}, on the horizon and in our chosen metric gauge eq.\eqref{eq:metric} with $\xi$ being the Killing generator for the boost transformation, eq.\eqref{BT_generator}. The final result for various components of these two objects (only considering the $U^{\mu\nu}_0 \delta A_{\nu}$ term in eq.\eqref{finalformtheta}) are the following (see Appendix-\ref{app:ThQNoGinv})
\begin{equation} \label{ThQNoGinv1}
    \begin{split}
     \Theta^r |_{r=0} \sim & \, \partial_v \left[ U^{rv}_0 (v A_v + \Lambda) \right] + \mathcal{O}(\epsilon^2) \, , \\
     Q^{rv}|_{r=0} \sim & \, U^{rv}_0 ( v \, A_v + \Lambda ) \, , \quad \text{and} \quad Q^{ri} |_{r=0} \sim \, U^{ri}_0 ( v \, A_v + \Lambda ) \sim \mathcal{O}(\epsilon^2) \, .
    \end{split}
\end{equation}
Using them one can easily check that 
\begin{equation} \label{eq:EvvNoGinv}
   \left( -\Theta^r + D_{\mu} Q^{r\mu} \right) \vert_{r=0} \sim \, \mathcal{O}(\epsilon^2) \, .
\end{equation}
Thus we have argued that the only gauge non-invariant term (i.e., the second term on the RHS of eq.\eqref{finalformtheta}) drops out from eq.\eqref{maineq}. This result is going to be very crucial in justifying the $U(1)$ gauge invariance of the entropy current.

\subsection{Construction of the entropy current and establishing its gauge invariance} \label{ssec:gaugeJvJi}

Our next job is to argue that the structure of $\Theta^r$ and $Q^{r\nu}$ that we discussed in the previous sub-section are indeed of the form expected so that we can construct an entropy current as mentioned in eq.\eqref{eq:nonmin3} and eq.\eqref{entcur_nonmin}. Besides this, we will also write down the algorithm to derive the components, $\mathcal{J}^v$ and $\mathcal{J}^i$. 
In the following, we will study $\Theta^r$ and $Q^{r\nu}$ individually, one at a time.

\subsubsection*{Contribution of the gauge field terms in $\Theta^r$ to the entropy current:}

We have already seen that the only gauge non-invariant term present in $\Theta^\mu$, i.e., the $U^{\mu\nu}_0 \delta A_{\nu}$ term in eq.\eqref{finalformtheta}), contribute at $\mathcal{O}(\epsilon^2)$ to eq.\eqref{maineq}. Ignoring that term,  the rest of the terms in eq.\eqref{finalformtheta}) can be expressed as 
\begin{equation} \label{Thgenform1}
\begin{split}
        \Theta^{\mu} = 2E_R^{\mu\nu\alpha\beta}D_\beta\left({\cal L}_\xi g_{\nu\alpha}\right)+\sum_k {\cal T}^{\mu\alpha_1\alpha_2...\alpha_k}{\cal L}_\xi{\cal S}_{\alpha_1\alpha_2...\alpha_k} + \mathcal{O}(\epsilon^2)  \, ,
\end{split}
\end{equation}
where,
\begin{equation}
\begin{split} \label{Thgenform2}
{\cal S}_{\alpha_1\alpha_2...\alpha_k}:\Big\{g_{\alpha\beta}, \, & R_{\alpha\beta\rho\sigma},D_{\alpha_1}R_{\alpha\beta\rho\sigma}, \cdots D_{(\alpha_1}...D_{\alpha_m)}R_{\mu\nu\alpha\beta}, \cdots ,\\
& F_{\mu\nu}, \, D_{\alpha_1}F_{\mu\nu}, \cdots ,D_{(\alpha_1}...D_{\alpha_n)}F_{\mu\nu}, \cdots \Big\} \, .
\end{split}
\end{equation}
It is important to note that, following our arguments given before, $E_R^{\mu\nu\alpha\beta}$ and ${\cal T}^{\mu\alpha_1\alpha_2...\alpha_k}$ will receive contributions from the gauge field terms but will always be gauge invariant.

We will follow exactly the same steps worked out in \S 3.2.2 of \cite{Bhattacharyya:2021jhr} to analyze the terms written above. In repeating those steps, however, we have to be careful whenever we need to use ``Result: 1", which has been modified in eq.\eqref{modresult1f}. We write down below the new contributions to the generic structure of a tensor $\mathbb{T}_{k+1}$ with boost weight $k+1 >0$, 
\begin{equation} \label{modresult1N}
    \mathbb{T}_{k+1} \sim \, \partial^k_v ( F_{vi} \, D^i_{(0)} ) + E_{(0)} \, \partial^k_v ( F_{vi} \, G^i_{(0)} ) + \mathcal{O}(\epsilon^2) \, .
\end{equation}
Since, the other terms have already been taken care of in \cite{Bhattacharyya:2021jhr}, we will just need to track down these extra contributions in eq.\eqref{modresult1N} through the steps outlined in \S 3.2.2 of \cite{Bhattacharyya:2021jhr}. We skip the details of that calculation and present them in Appendix-\ref{app:detailTh}. Once we do that, we indeed obtain the final result as expected and $\Theta^r$ has the following structure 
\begin{equation} \label{Thexpr1}
    \Theta^r = (1+v \, \partial_v) \mathcal{A}_{(1)} + v \, \partial^2_v \mathcal{B}_{(0)} + \mathcal{O}(\epsilon^2) \, ,
\end{equation}
where the new modifications due to eq.\eqref{modresult1N} are contained within the expression above as follows
\begin{equation}\label{toprove2result}
    \begin{split}
        \mathcal{A}_{(1)} &= \partial_v \left[ \sum^{k-1}_{m=0} (-1)^m X_{(-k+m)} \, \partial^{k-(m+1)}_v ( F_{vi} Y^i_{(0)} ) \right]  + (-1)^k (k+1) X_{(0)} F_{vi} Y^i_{(0)} \, , \\
        \mathcal{B}_{(0)} &= \sum^{k-1}_{m=0} (-1)^m (k-m) X_{(-k+m)} \, \partial^{k-(m+1)}_v (F_{vi} Y^i_{(0)}) \, .
    \end{split}
\end{equation}
We note that eq.\eqref{Thexpr1} matches exactly with eq.(3.56) of \cite{Bhattacharyya:2021jhr} with the additional modifications given in eq.\eqref{toprove2result}.

\subsubsection*{Contribution of the gauge field terms in $Q^{\mu\nu}$ to the entropy current:}
\label{QEntropycurrent}

We will now consider the contributions from the gauge field terms in $Q^{\mu\nu}$ to the entropy current. It is obvious from eq.\eqref{finalformtheta}, eq.\eqref{Thgenform1} and eq.\eqref{Thgenform2}, that the term with a different structure that involves contributions from the gauge fields is the following \footnote{There are obviously other terms in eq.\eqref{finalformtheta} that involves contributions from gauge fields, but structurally they are not different from what has already been analyzed in \cite{Bhattacharyya:2021jhr}. So they will be taken care of following the same steps.}
\begin{equation}
\Theta^\mu \sim \sum^{l-2}_{i=0} \widetilde{U}^{\mu\alpha_1 \dots \alpha_i \rho \sigma}_i \, \delta D_{(\alpha_1} \dots D_{\alpha_i)} F_{\rho\sigma} \, .
\end{equation}
We would like to obtain the contribution that this term will make to $Q^{\mu\nu}$. For that one needs to rearrange it in the following way 
\begin{equation}
    \begin{split}
    \Theta^\mu & \sim    \widetilde{U}^{\mu\alpha_1 \dots \alpha_i \rho \sigma}_i \mathcal{L}_{\xi} D_{(\alpha_1} \dots D_{\alpha_i)} F_{\rho\sigma} \\
        &= \widetilde{U}^{\mu\alpha_1 \dots \alpha_i \rho \sigma}_i \xi^{\beta} D_{\beta} D_{(\alpha_1} \dots D_{\alpha_i)} F_{\rho\sigma} + D_{\alpha_1} \left( \widetilde{U}^{\mu\alpha_1 \dots \alpha_i \rho \sigma}_i \xi^{\beta} D_{(\beta} \dots D_{\alpha_i)} F_{\rho\sigma} \right) \\
        & + D_{\alpha_2} \left( \widetilde{U}^{\mu\alpha_1 \dots \alpha_i \rho \sigma}_i \xi^{\beta} D_{(\alpha_1} D_{\beta} \dots D_{\alpha_i)} F_{\rho\sigma} \right) + \dots + D_{\alpha_i} \left( \widetilde{U}^{\mu\alpha_1 \dots \alpha_i \rho \sigma}_i \xi^{\beta} D_{(\alpha_1} \dots D_{\beta)} F_{\rho\sigma} \right) \\
        & + D_{\rho} \left( \widetilde{U}^{\mu\alpha_1 \dots \alpha_i \rho \sigma}_i \xi^{\beta} D_{(\alpha_1} \dots D_{\alpha_i)} F_{\beta\sigma} \right) + D_{\sigma} \left( \widetilde{U}^{\mu\alpha_1 \dots \alpha_i \rho \sigma}_i \xi^{\beta} D_{(\alpha_1} \dots D_{\alpha_i)} F_{\rho\beta} \right) \\
        & - \dots \text{Terms from Integration by parts} \, .
    \end{split}
\end{equation}
Thus, following the proposition 4.1 of \cite{Iyer:1994ys}. it is clear that the contribution of this term to $Q^{\mu\nu}$ is of the form given in eq.\eqref{exprQmunu1}, such that 
\begin{equation}\label{eq:wgauge}
    W^{\mu\nu\alpha} = \widetilde{U}^{\mu\alpha_1 \dots \nu \dots \alpha_i \rho \sigma}_i D_{(\alpha_1} \dots D^{\alpha} \dots D_{\alpha_i)} F_{\rho\sigma} \, ,
\end{equation}
and therefore we can now just apply the similar steps following \S 3.2.3 in \cite{Bhattacharyya:2021jhr}, to obtain the desired conclusion (see eq.(3.60) in \cite{Bhattacharyya:2021jhr})
\begin{equation}  \label{Qrmuexpr1}
Q^{r \mu} = \widetilde{Q}^{r \mu} + v \, W_v^{r \mu} \, ,
\end{equation}
where both $\widetilde{Q}^{r \mu}$ and $W_v^{r \mu}$ involve no explicit $v$-dependence. From eq.\eqref{eq:wgauge}, we can see that it will definitely contribute to $W_v^{r \mu}$, with $\mu = v, i$.

\subsubsection*{The final expression for the components of entropy current:}

Since we have already shown that for the new contributions due to gauge field coupling to gravity (including non-minimal couplings), the general structure of both $\Theta^r$ and $Q^{r\nu}$ are no different to what was obtained in \cite{Bhattacharyya:2021jhr}. We can straightforwardly adapt the steps in \S 3.2.4 of \cite{Bhattacharyya:2021jhr} to combine the individual contributions from $\Theta^r$ and $Q^{r\nu}$ in eq.\eqref{maineq} to obtain the expressions for the components of entropy current as well as the algorithm to calculate them. 

Once we know the Lagrangian of the theory, we must calculate $\Theta^r$ expressed in the form given by eq.\eqref{Thexpr1}. This will define for us $\mathcal{A}_{(1)}$ and $\mathcal{B}_{(0)}$. Next, we should calculate the $Q^{r\mu}$ in a form given by eq.\eqref{Qrmuexpr1}, which will define for us $\widetilde{Q}^{r \mu}$  and $W_v^{r \mu}$. From $W_v^{r i} = \partial_v J^i_{(1)}$, we can also obtain $J^i_{(1)}$. Finally, as was shown in \S 3.2.4 of \cite{Bhattacharyya:2021jhr}, we conclude that the following holds true for a theory of gauge fields coupled to gravity,
\begin{equation} \label{eq:AmuTvv}
(E_{vv}+\widetilde{T}_{vv})|_{r=0}=\partial_v\left[\frac{1}{\sqrt{h}}\partial_v\left(\sqrt{h}\, ( {\cal J}^v + \widetilde{\cal J}^v )\right)+\nabla_i ( {\cal J}^i + \widetilde{\cal J}^i)\right]+{\cal O}(\epsilon^2) \, ,
\end{equation}
with the following expressions for $( {\cal J}^v + \widetilde{\cal J}^v )$ and $( {\cal J}^i + \widetilde{\cal J}^i )$
\begin{equation} \label{finJvJiAmu}
\begin{split}
( {\cal J}^v + \widetilde{\cal J}^v ) =-\frac{1}{2}\left(\widetilde{Q}^{rv}+{\cal B}_{(0)}\right)\, ,\quad \text{and} \quad ( {\cal J}^i + \widetilde{\cal J}^i)=-\frac{1}{2}\left(\widetilde{Q}^{ri}-J^i_{(1)}\right) \, .
\end{split}
\end{equation}
Here, all the ingredients $\mathcal{B}_{(0)}, \, \widetilde{Q}^{rv}, \, \widetilde{Q}^{ri}$, and $ {\cal J}^i$ are defined above. This completes the proof that for gauge fields non-minimally coupled to gravity, the ${\cal O}(\epsilon)$ contribution $(\widetilde{T}_{vv})$ does indeed have the desired structure of an entropy current. 

It is also important to highlight that the $( {\cal J}^v + \widetilde{\cal J}^v )$ and $( {\cal J}^i + \widetilde{\cal J}^i )$ obtained above are $U(1)$ gauge-invariant expressions. This can be traced back to the fact that, as discussed in \S\ref{ssec:ThQcontr} and \S\ref{app:ThQNoGinv}, the only possible source of gauge non-invariant contribution to $\Theta^r$ and hence to $Q^{r\nu}$, is of ${\cal O}(\epsilon^2)$. Therefore, the quantities $\mathcal{B}_{(0)}, \, \widetilde{Q}^{rv}, \, \widetilde{Q}^{ri}$, and $ {\cal J}^i$, appearing in eq.\eqref{finJvJiAmu}, will only involve gauge invariant quantities, i.e., $F_{\mu\nu}$, for calculations up to ${\cal O}(\epsilon)$ and with the Lagrangian given in eq.\eqref{eq:gaugelagrangian}.  

\subsection{Verification of the abstract proof: An example}

This sub-section aims to provide a consistency check for the abstract algorithm for constructing the entropy current given in the previous sub-section. We want to check that $\mathcal{J}^v$ and $\mathcal{J}^i$ for a particular model theory with gauge fields non-minimally coupled to gravity matches what one would obtain from their abstract definitions. 
We consider a Lagrangian of the following form
\begin{equation}
    \mathcal{L}_{(A_\mu g_{\mu\nu})} = R_{\mu\nu\rho\sigma} F^{\mu\nu} F^{\rho\sigma} \, .
\end{equation}
The variation of $\mathcal{L}_{(A_\mu g_{\mu\nu})}$ can be computed following standard procedure \footnote{The convention used for symmetrization of indices are as follows: $A^{(\alpha}B^{\beta)}= (1/2)(A^{\alpha}B^{\beta}+A^{\beta}B^{\alpha})$.}
\begin{equation}\label{eq:varl3}
    \begin{split}
        \delta (\sqrt{-g} \, \mathcal{L}_{(A_\mu g_{\mu\nu})}) &= \sqrt{-g} \Bigg[\dfrac{1}{2} g^{\alpha\beta} R_{\mu\nu\rho\sigma} F^{\mu\nu} F^{\rho\sigma} - 6 R^{(\alpha}_{~~\nu\rho\sigma} F^{\beta)\nu} F^{\rho\sigma} + 2 D_{\nu} D_{\rho} (F^{\alpha\nu}F^{\rho\beta}) \Bigg] \delta g_{\alpha\beta} \\
        & ~-4 \sqrt{-g} D_{\rho} \left( R^{\rho\sigma}_{\hspace{0.4cm}\mu\nu} F^{\mu\nu} \right) \delta A_{\sigma} \\
        &+\sqrt{-g} \, D_{\rho} \Bigg[ 2 F^{\mu\nu} F^{\rho\sigma} D_{\nu} \delta g_{\sigma\mu} - 2 D_{\nu}(F^{\mu\rho}F^{\nu\sigma})\delta g_{\sigma\mu} + 4 R^{\rho\sigma}_{\hspace{0.4cm}\mu\nu} F^{\mu\nu} \delta A_{\sigma} \Bigg] \, . 
    \end{split}
\end{equation}
The EoM for the metric and the gauge field are given by
\begin{equation}
\begin{split}
\widetilde{T}^{\alpha\beta} = & \, \dfrac{1}{2} g^{\alpha\beta} R_{\mu\nu\rho\sigma} F^{\mu\nu} F^{\rho\sigma} - 6 R^{(\alpha}_{~~\nu\rho\sigma} F^{\beta)\nu} F^{\rho\sigma} + 2 D_{\nu} D_{\rho} (F^{\alpha\nu}F^{\rho\beta}) \, , \\
    G^{\rho} = & -4 D_{\sigma} \left( R^{\rho\sigma}_{\hspace{0.4cm}\mu\nu} F^{\mu\nu} \right) \, .
    \end{split}
\end{equation}
Therefore, a brute force calculation in our chosen metric gauge eq.\eqref{eq:metric} would result in 
\begin{equation}
  \widetilde{T}_{vv} = \partial_v \left[ \dfrac{1}{\sqrt{h}} \partial_v (\sqrt{h} \, 2 F^{rv} F^{rv}) + \nabla_i (4 F^{rv} F^{ri} ) \right] \, .
\end{equation}
such that we obtain
\begin{equation} \label{exAmuJvJi}
\widetilde{\cal J}^v  = 2 F^{rv} F^{rv}\, ,\quad \text{and} \quad \widetilde{\cal J}^i = 4 F^{rv} F^{ri} \, .
\end{equation}

On the other hand, from eq.\eqref{eq:varl3} we also get the expression for $\Theta^{\rho}$ as
\begin{equation}
    \Theta^{\rho} = 2 F^{\mu\nu} F^{\rho\sigma} D_{\nu} \delta g_{\sigma\mu} - 2 D_{\nu}(F^{\mu\rho}F^{\nu\sigma})\delta g_{\sigma\mu} + 4 R^{\rho\sigma}_{\hspace{0.4cm}\mu\nu} F^{\mu\nu} \delta A_{\sigma} \, .
\end{equation}
Substituting from eq.\eqref{eq:Fvariation}, one can also obtain the Noether charge as the following 
\begin{equation}
    \begin{split}
        Q^{\rho\nu} &= 2 F^{\rho\nu} F^{\mu\sigma}D_{\sigma}\xi_{\mu} + 2 D_{\sigma}(F^{\rho\mu}F^{\sigma\nu}-F^{\rho\sigma}F^{\mu\nu})\xi_{\mu}  \, .
    \end{split}
\end{equation}
From which we obtain the following expressions, 
\begin{equation}
\begin{split}
{\cal B}_{(0)} &= 0 \, ,\quad \widetilde{Q}^{rv} = 4 F^{rv} F^{rv} \, ,\quad \widetilde{Q}^{ri} = 4 F^{ri} F^{rv} \, ,\quad J^i_{(1)} = -4 F^{ri} F^{rv} \, .
\end{split}
\end{equation}
With these we can check that 
\begin{equation}
    \begin{split}
        -\Theta^r + D_{\nu}Q^{r\nu}= -v\partial_v\left[\frac{1}{\sqrt{h}}\partial_v\left(\sqrt{h}~4~ F_{vr} F_{vr}\right)+\nabla_i(8~F_{vr} F_v{}^i)\right] \, .
    \end{split}
\end{equation}
which confirms the expressions of $\mathcal{J}^v$, $\mathcal{J}^i$ obtained in eq.\eqref{exAmuJvJi} obtained by a brute force calculation. Also, these expressions turn out to be gauge invariant, as expected. 

Let us also mention that our expression for $\mathcal{J}^v$ obtained above exactly matches with the same obtained in \cite{Wang:2021zyt}, for the same theory that we are focussing on in this sub-section. However, we cannot compare the spatial component $\mathcal{J}^i$ of the entropy current as in \cite{Wang:2021zyt} a version of the linearized second law was examined for entropy integrated over a compact spatial slice of the horizon, and hence it looses the $\nabla_i \mathcal{J}^i$ term.

\section{Physical process version of the first law with non-minimal interactions}
\label{sec:ppflproof}

As in \cite{Bhattacharyya:2021jhr}, once we have proved the linearized second law via the consequence of the structure of entropy current in the EOM in eq.\eqref{eq:AmuTvv}, we can straightforwardly use this setup to argue the ``physical process" version of the first law. In this section, we will prove the physical process version of the first law \cite{Jacobson:1995uq, Gao:2001ut, Amsel:2007mh, Bhattacharjee:2014eea, Chakraborty:2017kob, Chatterjee:2011wj, PhysRevD.86.021501} for theories with arbitrary non-minimal gauge couplings generalizing the results of \cite{Gao:2001ut}. We consider theories described by the general Lagrangians of the form (eq.\eqref{eq:gaugelagrangian})
\begin{equation}\label{eq:laggao}
\mathcal{L}_{(A_\mu g_{\mu\nu})}=\mathcal{L}_{(A_\mu g_{\mu\nu})} (g_{\alpha\beta},R_{\alpha\beta\rho\sigma},D_{\alpha_1}R_{\alpha\beta\rho\sigma},\cdots,F_{\mu\nu},D_{\alpha_1} F_{\mu\nu},D_{(\alpha_1}D_{\alpha_2)}F_{\mu\nu},\cdots) \, ,
\end{equation}
and focus on stationary, charged black hole solutions of the theory. We consider $(g_{\mu\nu},A_\mu)$ to be a solution of the equations of motion ${\cal E}_{\mu\nu}=0$ and $G_\mu=0$, (see eq.\eqref{eq:Lvariation} for the definitions of ${\cal E}_{\mu\nu}$ and $G_\mu$) and let $(\delta g_{\mu\nu}, \delta A_\mu)$ be a solution of linearized equations of  motion with sources $\delta T_{\mu\nu}$ and $\delta J_\mu$
\begin{equation}\label{eq:sourceeom}
	\begin{split}
		\delta \mathcal{E}^{\mu\nu} = \dfrac{1}{2}\delta T^{\mu\nu}, \quad \delta G^{\mu} = \delta J^{\mu} \, .
	\end{split}
\end{equation}

We aim to establish the physical process version of the first law given by
\begin{equation}\label{eq:firstlaw}
    \dfrac{\kappa}{2\pi} \delta S = \delta M - \Omega_H \delta I - \Phi_{bh}\delta q \, ,
\end{equation}
where the first order variations in the mass $\delta M$, angular momentum $\delta I$ and the charge $\delta q$ are generated by perturbations in the stress tensor $\delta T_{\mu\nu}$ and the source current density $\delta J^{\mu}$. Following section II of \cite{Gao:2001ut}, the formulae for the first order variation of mass $M$ and angular momentum $I$ can be written as follows \footnote{The determinant factor and the measure factor are included in the integral.}
\begin{equation}\label{eq:dmass}
	\begin{split}
		\delta M& = \int_{\Sigma} n_{\mu} \left( \, \delta T^{\mu\nu} t_{\nu} + \delta J^{\mu} (A^{\nu}t_{\nu} + \Lambda) \right) + \int_{\partial \Sigma} (\delta Q^{\mu\nu}[t] - t^{\mu}\Theta^{\nu})\epsilon_{\mu\nu} \, ,\\
		\delta I& = -\int_{\Sigma} n_{\mu} \left( \, \delta T^{\mu\nu} \varphi_{\nu} + \delta J^{\mu} (A^{\nu}\varphi_{\nu} + \Lambda) \right) - \int_{\partial \Sigma} (\delta Q^{\mu\nu}[\varphi] - \varphi^{\mu}\Theta^{\nu})\epsilon_{\mu\nu} \, ,
	\end{split}
\end{equation}
where, $\Sigma$ is a hypersurface that extends from some ``interior boundary" $\partial \Sigma$ to asymptotic infinity, $n_\mu$ is the normal to the hypersurface $\Sigma$, $\epsilon_{\mu\nu}$ is the binormal of $\partial \Sigma$, $\xi^\mu=t^\mu$ is asymptotic time translation and $\xi^\mu=\varphi^\mu$ is asymptotic rotation. Note that $\Theta^{\mu}$ and $Q^{\mu\nu}$ are those given in eq.\eqref{eq:zetareln1}. In eq.\eqref{eq:dmass} there is an additional term involving $\Lambda$ in comparison to \cite{Gao:2001ut} because we consider variations of the form eq.\eqref{eq:Fvariation}. In eq.\eqref{eq:firstlaw}, $S$ is the entropy of the black hole, $\kappa$ is the surface gravity of the black hole, $\Omega_H$ is the angular velocity of the horizon and $\Phi_{bh}$ (see eq.\eqref{eq:deltaq} for definition of $\Phi_{bh}$) is the electrostatic potential.

Though the analysis of section 5 of \cite{Bhattacharyya:2021jhr} (which discussed the physical process first law) seemingly carries over, minor subtleties remain when viewed from the perspective of \cite{Gao:2001ut}. As we are considering the charged case, there are two different sources in the form of $\delta T_{\mu\nu}$ (which contributes to $\delta S$, $\delta M$, and $\delta I$) and $\delta J^{\mu}$ (which contributes to $\delta q$). Since \cite{Gao:2001ut} considered a simple minimally coupled Einstein-Maxwell theory, there was a clean separation between the sources which contributed to $\delta S$ and $\delta q$. However, we are primarily interested in non-minimally coupled theories of the form given by eq.\eqref{eq:laggao} and such terms will contribute both to the stress tensor and the source current. It is thus not clear how to differentiate the contributions to $\delta S$ and $\delta q$ apriori, and the analysis of this section sheds some light on this subtlety. 

Let $\xi^{\mu} = t^{\mu} + \Omega_H \varphi^{\mu}$ be the horizon generating Killing field. We consider a linearized perturbation $(\delta g_{\mu\nu},\delta A^{\mu})$ on some time slice $\Sigma_0$ \footnote{$\Sigma_0$ is a hypersurface that extends from the black hole horizon to asymptotic infinity.} with sources $\delta T_{\mu\nu}$, $\delta J^{\mu}$ such that $\delta T_{\mu\nu}$ and $\delta J^{\mu}$ vanish near spatial infinity and in a small neighbourhood of black hole horizon on $\Sigma_0$, i.e., the perturbation vanishes on $\partial \Sigma_0$. Thus from eq.\eqref{eq:dmass}, we have
\begin{equation}
    \delta M - \Omega_H \delta I = \int_{\Sigma_0} n_{\mu} \left( \, \delta T^{\mu\nu} \, \xi_{\nu} + \delta J^{\mu} (A^{\nu}\xi_{\nu} + \Lambda) \right) \, .
\end{equation}
In terms of the conserved current $\alpha^\mu=2\,\delta {\cal E}^{\mu\nu} \,  \xi_{\nu} + \delta G^{\mu} (A^{\nu}\xi_{\nu} + \Lambda)$ after using the linearized EOM given in eq.\eqref{eq:sourceeom} (see Appendix \ref{app:consalpha} for details), the above equation becomes
\begin{equation}\label{eq:alphamu}
    \delta M - \Omega_H \delta I = \int_{\Sigma_0} n_{\mu} \alpha^{\mu} \, .
\end{equation}
Assuming that all the matter eventually falls into the black hole and using the fact that $\alpha^{\mu}$ is conserved,
\begin{equation}\label{eq:intres3}
    \delta M - \Omega_H \delta I = \int_{\mathcal{H}} k_{\mu} \alpha^{\mu} \, ,
\end{equation}
where $k^{\mu}$ is the affinely parameterized null generators of the horizon \footnote{In our setup, $k^{\mu} = (\partial_v)^{\mu}$.}. Consider the following integral
\begin{equation}\label{eq:deltaq}
    \int_{\mathcal{H}} k_{\mu} \, \delta J^{\mu} (A^{\nu} \, \xi_{\nu} + \Lambda) = - \int_{\mathcal{H}} \Phi_{bh} \, k_{\mu} \, \delta J^{\mu} = - \Phi_{bh} \int_{\mathcal{H}} k_{\mu} \, \delta J^{\mu} = \Phi_{bh} \, \delta q \, .
\end{equation}
where $\Phi_{bh} = -(A^{\nu}\xi_{\nu} + \Lambda)|_{\mathcal{H}} $ is the electrostatic potential of the horizon and $\delta q$ is the net change in the charge of the black hole due to the current source. We could write the final step because we have shown in Appendix \ref{app:genzerothlaw} that this quantity remains constant throughout the horizon in our setup. At this point, there is a departure from the analysis of \cite{Gao:2001ut}. To prove that $\Phi_{bh}$ is constant, \cite{Gao:2001ut} had to use the fact that they were considering a minimally coupled Einstein-Maxwell theory where the stress tensor satisfied the Null energy condition (NEC). NEC was a crucial input in the proof. The main theme of our present paper is to consider non-minimally coupled theories of gravity that generically violate NEC. Thus, it is unclear how to prove the constancy of $\Phi_{bh}$ for the theories we are interested in when working within the setup of \cite{Gao:2001ut}. Our proof in Appendix \ref{app:genzerothlaw} automatically takes care of this subtle issue. Note that the non-minimal terms contribute to $\delta J^{\mu}$, and thus to $\delta q$ through the source current, but this will not affect the constancy of $\Phi_{bh}$ \footnote{For instance, the constancy of $\Phi_{bh}$ will not depend on whether we consider the non-minimal terms to be on the RHS or the LHS of eq.\eqref{eq:sourceeom}.}. Thus, using eq.\eqref{eq:deltaq} in eq.\eqref{eq:intres3}, we have
\begin{equation}\label{eq:firstlaw1}
    \delta M - \Omega_H \delta I - \Phi_{bh} \delta q = \int_{\mathcal{H}} \, k_{\mu} \, \delta T^{\mu\nu} \, \xi_{\nu} \, .
\end{equation}

To prove the physical process version of the first law, we have to relate the RHS of the above equation to the change in entropy.
%
%We know that the definition of entropy for an arbitrary theory described by the Lagrangian of the form eq.\eqref{eq:laggao} is given by the Wald entropy. Moreover, in dynamical situations, the Wald Entropy is supplemented by JKM ambiguities \cite{Wall:2015raa, Bhattacharya:2019qal, Bhattacharyya:2021jhr}. Thus, the entropy has the following structure \footnote{Henceforth, we will be working in our gauge described in \cite{Bhattacharya:2019qal, Bhattacharyya:2021jhr}.}
%\begin{equation}\label{eq:entdef}
%    S = 4 \pi \int_{\mathcal{H}} d^{d-2}x \, \sqrt{h} \, (1 + s_w) \, ,
%\end{equation}
%
\begin{equation}\label{eq:kT}
\int_{\mathcal{H}} \, k_{\mu} \, \delta T^{\mu\nu} \xi_{\nu}=2\,\kappa\,\int_{-\infty}^\infty dv\int_{{\cal H}_v} d^{d-2}x \sqrt{h} \,v\, \delta {\cal E}_{vv} \, ,
\end{equation}
where ${\cal H}_v$ is the constant v-slices of the horizon in our gauge eq.\eqref{eq:metric}. For the last equality, we have used the equation of motion $\delta T^{\mu\nu}=2\,\delta {\cal E}^{\mu\nu}$, along with $k^v=1$ and in general $\xi^v|_{\mathcal{H}}=\kappa\, v$.

Now, let us recollect the main result of this paper, given in eq.\eqref{eq:nonmin3}, which is to write ${\cal E}_{vv}$ in the following form \footnote{Note that in eq.\eqref{eq:nonmin3} we had an explicit contribution in the form of $T_{vv}$. While writing the RHS of eq.\eqref{eq:EvvPPFL} we have absorbed that within the ${\cal O}(\epsilon^2)$ pieces. This is justified since $T_{vv}$, by construction, signifies contribution from matter couplings that satisfies the NEC, and hence $T_{vv} \sim {\cal O}(\epsilon^2)$.}
\begin{equation} \label{eq:EvvPPFL}
{\cal E}_{vv}|_{r=0}=\partial_v\left[\frac{1}{\sqrt{h}}\partial_v\left(\sqrt{h}\, ( {\cal J}^v + \widetilde{\cal J}^v )\right)+\nabla_i ( {\cal J}^i + \widetilde{\cal J}^i)\right]+{\cal O}(\epsilon^2) \, .
\end{equation}
This result makes the contribution of the non-minimal terms explicit via $\widetilde{\cal J}^v$ 
\footnote{At this point, it is imperative to note that eq.\eqref{eq:EvvPPFL} has been written following the conventions of \cite{Bhattacharyya:2021jhr}, which is slightly different compared to \cite{Gao:2001ut}. Following the stationary background and fluctuation split given in eq.\eqref{eq:metdecomp}, it is obvious that what appears on the RHS of eq.\eqref{eq:EvvPPFL} is actually the linearized ${\cal O}(\epsilon)$ piece of ${\cal E}_{vv}$. On the contrary, in eq.\eqref{eq:sourceeom}, we have explicitly denoted the linearized piece as $\delta \mathcal{E}^{\mu\nu}$ (following the conventions of \cite{Gao:2001ut}), which should be understood as equivalent to eq.\eqref{eq:EvvPPFL}.}.
Then, exactly following section 5 of \cite{Bhattacharyya:2021jhr}, we can write eq.\eqref{eq:kT} as
\begin{equation}\label{eq:kT1}
\begin{split}
\int_{\mathcal{H}} \, k_{\mu} \, \delta T^{\mu\nu} \xi_{\nu}&=2\,\kappa\, \int_{-\infty}^\infty dv\int _{{\cal H}_v} d^{d-2}x\left[ \partial_v\left(\sqrt{h}\, ( {\cal J}^v + \widetilde{\cal J}^v )\right)+\nabla_i\left(\sqrt{h}\, ( {\cal J}^i + \widetilde{\cal J}^i)\right)\right]\\
&=2\,\kappa\, \int_{-\infty}^\infty dv\int _{{\cal H}_v} d^{d-2}x\left[ \partial_v\left(\sqrt{h}\, ( {\cal J}^v + \widetilde{\cal J}^v )\right)\right] \, .
\end{split}
\end{equation}
The second term in the above integral, being a total derivative, vanishes for compact horizons. Now, using the following expression of entropy \footnote{This definition of entropy was justified in \cite{Bhattacharyya:2021jhr}.}
\begin{equation}\label{eq:entropydef}
S=4\pi\int_{{\cal H}_v} d^{d-2}x  \sqrt{h}\, \left( {\cal J}^v + \widetilde{\cal J}^v \right) \, ,
\end{equation}
in eq.\eqref{eq:kT1}, we get
\begin{equation}\label{eq:2}
\int_{\mathcal{H}} \, k_{\mu} \, \delta T^{\mu\nu} \xi_{\nu}=\frac{\kappa}{2\pi}\delta S \, .
\end{equation}
Finally, comparing eq.\eqref{eq:firstlaw1} and eq.\eqref{eq:2}, we get the physical process version of the first law
\begin{equation}\label{eq:firstlawfinal}
    \dfrac{\kappa}{2\pi} \delta S = \delta M - \Omega_H \delta I - \Phi_{bh}\delta q \, .
\end{equation}
In eq.\eqref{eq:entropydef}, $\widetilde{\cal J}^v$ takes into account the contribution of non-minimal terms to the entropy, and we have seen above how the source-current corresponding to the non-minimal term contributes to $\delta q$. One should note that $\kappa$ being a constant for stationary black hole solutions of such theories was recently addressed in \cite{Bhattacharyya:2022nqa}, and $\Phi_{bh}$ being constant has been argued in Appendix \ref{app:genzerothlaw}. This completes the proof of the physical process first law for non-minimally coupled theories of the form eq.\eqref{eq:laggao}. We have also included an alternate perspective on the physical process first law along the lines of \cite{Chakraborty:2017kob} with an explicit example in Appendix \ref{app:ppflproofalt}.

\section{Summary and outlook}\label{sec:summary}

Recently, in \cite{Bhattacharyya:2021jhr}, a local version of the linearized second law has been argued for diffeomorphism invariant theories of gravity. In such theories, it was facilitated by constructing an entropy current with non-negative divergences associated with small dynamical perturbations around stationary black holes. The crucial element was to constrain the off-shell structure of the $vv$-component of the equations of motion using boost symmetry of the near horizon space-time of the black hole. This boost symmetry is exact for stationary black holes, but it gets slightly broken by the small dynamical fluctuations caused due to some external sources. The whole analysis was performed in a linearized amplitude expansion (denoted by $\epsilon$) around stationary black hole solutions. 
In achieving that, an important assumption was made in \cite{Bhattacharyya:2021jhr} about how matter fields can possibly couple to gravity. The related matter energy-momentum tensor was taken to obey the null energy conditions in that analysis. Consequently, up to linearized order in the amplitude expansion, the $vv$-component of the energy-momentum tensor drops out of the equations of motion ($T_{vv} \sim \mathcal{O}(\epsilon^2)$). However, even classically, the null energy condition can be violated when the non-minimal coupling of matter fields to gravity is considered. The energy-momentum tensor, obtained from such non-minimal couplings, can contribute at linear order in amplitude expansion ($\sim \mathcal{O}(\epsilon)$) to the EoM. Therefore, the analysis in \cite{Bhattacharyya:2021jhr} does not apply to such theories with possible non-minimal coupling between matter fields and gravity. 

In this paper, we have extended the formalism of \cite{Bhattacharyya:2021jhr} mentioned above by including the arbitrary non-minimal coupling of scalar and $U(1)$ gauge fields with gravity. For gauge fields coupled to gravity, we considered only gauge-invariant Lagrangians. Working with a particular choice of metric gauge, we have shown that the off-shell structure of the EoM (or to the energy-momentum tensor) from this arbitrary coupling of the scalar or gauge fields with gravity which has $\mathcal{O}(\epsilon)$ contributions can always be cast in the form given in eq.\eqref{eq:nonmin3} and eq.\eqref{entcur_nonmin}. It should be noted that this is a statement about the structure of EoM in a theory of gravity coupled to matter fields and thus holds even without any reference to the linearized second law. 

However, we have still assumed that the $\mathcal{O}(\epsilon^2)$ contributions, denoted by $T_{vv}$, satisfies the NEC (similar to what was assumed in \cite{Bhattacharyya:2021jhr}) and hence do not participate in our analysis, valid till linear order in $\epsilon$-expansion. This splitting of energy-momentum tensor for matter fields coupling to gravity is seemingly getting decided based on whether they contribute to $\mathcal{O}(\epsilon)$ or $\mathcal{O}(\epsilon^2)$ in the EOM. Accordingly, they also either satisfy or violate the NEC. 

Our analysis in this paper also abstractly proves that an entropy current with non-negative divergence could be constructed, even when matter field couplings give non-vanishing $\mathcal{O}(\epsilon)$ contributions to the EOM as shown in eq.\eqref{cond3modf}. This means that these $\mathcal{O}(\epsilon)$ contributions from the matter couplings merely add to the existing entropy current obtained from the pure gravity modes, already studied in \cite{Bhattacharyya:2021jhr}. Interestingly, in precisely the same manner as in \cite{Bhattacharyya:2021jhr}, a spatial current on the horizon is present for out-of-equilibrium dynamical situations. It should be noted that, although we started with an aim to address issues regarding non-minimal matter couplings to gravity, in our abstract arguments, we did not need to use that. Therefore, our result cares only about all possible $\mathcal{O}(\epsilon)$ contributions to EoM from scalar or gauge field coupling to gravity. 

Another interesting aspect of the arbitrary coupling of gauge fields to gravity is worth emphasizing. The $\mathcal{O}(\epsilon)$ contributions from the non-minimally coupled gauge fields to the components of entropy current, denoted by $\widetilde{\cal J}^v$ and $\widetilde{\cal J}^i$ (see eq.\eqref{finJvJiAmu}), are invariant under the $U(1)$ gauge transformations. In our opinion, this is a non-trivial statement. It is very non-trivial to manipulate the $vv$-component of EoM ($\mathcal{E}_{vv}$) by organizing and distributing exactly two $\partial_v$s as they appear on the RHS of eq.\eqref{eq:AmuTvv}. As a consequence, for non-minimally coupled gauge fields, there is no a priori reason to believe that $\widetilde{\cal J}^v$ and $\widetilde{\cal J}^i$ will only involve $F_{\mu\nu}$ and no explicit factors of $A_{\mu}$. However, we have been able to argue that for a gauge-invariant Lagrangian eq.\eqref{eq:gaugelagrangian}, the components of entropy current are indeed gauge-invariant.

Expressions of entropy density are already available in the literature for specific theories involving non-minimal coupling of matter with higher derivative theories of gravity. The answer obtained from our abstract arguments presented in this paper comes out to be consistent with them. However, the spatial components of the entropy current that we have obtained in this paper were not worked out before. Thus our result widens the applicability of the result of \cite{Bhattacharyya:2021jhr} to more general cases. The analysis of \cite{Bhattacharyya:2021jhr} had not been careful in incorporating the matter field sector, which might contribute non-trivially at $\mathcal{O}(\epsilon)$ and our result rectifies that issue. 

Following the ideas of section 5 of \cite{Bhattacharyya:2021jhr}, we could construct a proof of the physical process version of the first law for arbitrary non-minimally coupled gauge theories. Our analysis in section \S\ref{sec:ppflproof} generalized the results of \cite{Gao:2001ut} clearing some of the subtleties that one might encounter when attempting to prove the first law following the outline of \cite{Gao:2001ut}. We have also given an alternative perspective on the same, along the lines of \cite{Chakraborty:2017kob} in Appendix \ref{app:ppflproofalt}.

%xxxx As discussed in \cite{Bhattacharyya:2021jhr}, to prove the physical process version of the first law, one needs a similar framework used for the linearized version of the second law. Therefore, our results in this paper can be easily used to construct a proof for the physical process version of the first law for theories, including the arbitrary non-minimal coupling of matter fields with gravity. 

There are a few directions of interest to pursue in the future. In considering the form of the Lagrangian given in eq.\eqref{eq:gaugelagrangian}, we have chosen to ignore terms that may involve the gauge fields $A_\mu$ explicitly, like the Cherns-Simmons type terms. So far, we have considered diffeomorphism invariant theories, but Cherns-Simmons theories are diffeomorphism invariant up to total derivatives only. Also, in this regard, one can consider Proca Lagrangians involving massive gauge fields, where it is not gauge-invariant anymore. It would be interesting to see if this formalism can be extended further to such theories. 

The proof of the existence of entropy current studied in this paper, including arbitrary matter coupling, following \cite{Bhattacharyya:2021jhr} relies on choosing the gauge given by eq.\eqref{eq:metric}. In other words, this depends on the particular slicing of the null horizon, i.e., constant $v$ slices. Understanding how the entropy current transforms under arbitrary re-definition of the constant $v$ slices is important. Recently, \cite{Bhattacharyya:2022njk} and \cite{Hollands:2022fck} have particularly looked at this. With our results in this paper, it would be interesting to study whether matter fields can be incorporated into their story. In \cite{Hollands:2022fck} this has been looked at for scalar fields coupled to gravity. It would be a nice exercise to extend their analysis to understand the invariance of the entropy current under the choice of metric gauge in the presence of additional fields having internal symmetries, e.g., $U(1)$ gauge fields.

The canonical definition of energy-momentum tensor is basically obtained by differentiating the matter Lagrangian with respect to the metric field. This definition was assumed to satisfy the null energy condition $(T_{vv} \geq 0)$ in \cite{Bhattacharyya:2021jhr}, which immediately implies that $T_{vv}$ is of the order ${\cal O}(\epsilon^2)$. We have, however, seen in this paper that there exist terms that are $\mathcal{O}(\epsilon)$, and hence, no longer satisfy the null energy condition. This is well known in examples of non-minimally coupled Lagrangians \cite{Barcelo:2000zf, Flanagan_1996, Wall:2018ydq, Chatterjee:2015uya}. It will be interesting to see if we can come up with a prescription for a covariant definition of energy-momentum tensor that satisfies the null energy condition. This definition would apply to arbitrary matter coupling with gravity but desirably would not pick up the $\mathcal{O}(\epsilon)$ contributions in $T_{vv}$.

\acknowledgments

We would like to thank Sayantani Bhattacharyya  for collaboration at the initial stage and for comments on the draft. We are also thankful to   Jyotirmoy Bhattacharya, Sayantani Bhattacharyya, Diptarka Das, Anirban Dinda, Alok Laddha, R. Loganayagam, Milan Patra, Shuvayu Roy, Amitabh Virmani for several discussions and important inputs. We would also like to thank \'{A}ron Kov\'{a}cs and Harvey Reall for comments on the draft. P.B. would like to acknowledge the support provided by the grant SB/SJF/2019-20/08. P.D. would like to duly acknowledge the Council of Scientific and Industrial Research (CSIR), New Delhi for financial assistance through the Senior Research Fellowship (SRF) scheme. We finally would like to thank the people of India for their steady support for research in basic sciences.

\appendix

\section{A short review of the gauge choice and boost weight analysis} \label{app:boost}

This appendix serves as a quick recapitulation of the gauge choice and boost weight analysis of \cite{Bhattacharyya:2021jhr}. Thus, we will be brief here and we refer the reader to \cite{Bhattacharyya:2021jhr} for more details.

The most general dynamical black hole metric can always be expressed by the following metric \footnote{This fact has been explicitly argued for in the appendix A of \cite{Bhattacharyya:2016xfs}.}
\begin{equation}\label{eq:metricAP}
ds^2=2~ dv~ dr-r^2 X(r,v,x^i) ~ dv^2+2~r~\omega_i(r,v,x^i)~dv~ dx^i+h_{ij}(r,v,x^i)dx^i dx^j \, .
\end{equation}
As noted in \cite{Bhattacharya:2019qal,Bhattacharyya:2021jhr}, the above metric choice does not completely fix the gauge freedom. We are allowed to do two types of coordinate redefinition:
\begin{eqnarray}
v &\rightarrow &  \tilde{v} = f_1 (x^i)\,  v + f_2(x^i), \quad \text{with an appropriate re-definition of} \, r \, ,\label{residual1AP} \\
x^i &\rightarrow & \tilde{x}^i = g^i(x^j) \, . \label{residual2AP}
\end{eqnarray}
The second coordinate freedom eq.\eqref{residual2AP} is just a relabelling of the spatial tangents and it represents the residual diffeomorphism invariance of the spatial metric $h_{ij}$. This freedom can be implemented by lifting the partial derivatives $\partial_i$ to covariant derivatives $\nabla_i$ compatible with $h_{ij}$. The first coordinate freedom eq.\eqref{residual1AP} is non-trivial and the transformation under this coordinate change constrains the structure of the relevant components of covariant tensors. We will only be interested in a sub-class of the full residual symmetry of the gauge given by,
\begin{equation} \label{boosttransfAP}
	r\rightarrow\tilde r = \lambda ~r,~~v\rightarrow\tilde v = {v\over \lambda} \, .
\end{equation}
This will be called the boost transformation \footnote{The nomenclature follows that of \cite{Wall:2015raa,Bhattacharya:2019qal,Bhattacharyya:2021jhr}.}. The infinitesimal version of this transformation is generated by the vector field
\begin{equation} \label{BT_generatorAP}
	\xi = \xi^\mu\partial_\mu = v\partial_v - r\partial_r \, .
\end{equation}
Now the metric in our gauge eq.\eqref{eq:metricAP} that remains stationary under this vector field eq.\eqref{BT_generatorAP} is given by
\begin{equation}\label{eq:metequlAP} 
	ds^2 = 2 \, dv \, dr - r^2 \, X(rv,x^i)\, dv^2 + 2 \, r \,\omega_i(rv,x^i)\, dv \, dx^i + h_{ij}(rv,x^i) \, dx^i \, dx^j \, .
\end{equation}
The vector field $\xi^{\mu}$ is a Killing vector which generates the Killing horizon of this stationary space-time. This form of the stationary metric constrains the dynamics away from it.

To describe the dynamics, we will employ a perturbative expansion in the amplitude of the dynamics and we will be working up to linearized order in the dynamics. Mathematically, this means that we can decompose the metric $g_{\mu\nu}$ given in eq.\eqref{eq:metricAP} as follows \footnote{Here $\epsilon$ is very small.}
\begin{equation}\label{eq:metdecompAP}
    g_{\mu\nu} = g^{eq}_{\mu\nu} + \epsilon \, \delta g_{\mu\nu} \, ,
\end{equation}
where $g^{eq}_{\mu\nu}$ is the equilibrium metric given by eq.\eqref{eq:metequlAP}. Notice that here the metric components arising in $g^{eq}_{\mu\nu}$ namely, $X,\omega_i,h_{ij}$ have a specific functional dependence on the coordinates (i.e) $X(rv,x^i), \omega_i(rv,x^i), h_{ij}(rv,x^i)$. The components in $\delta g_{\mu\nu}$ have arbitrary dependence. This is summarized below:
\begin{equation}
    \begin{split}
        X &= X(rv,x^i) + \epsilon \, \delta X(r,v,x^i) \, ,\\
        \omega_i &= \omega_i(rv,x^i) + \epsilon \, \delta \omega_i(r,v,x^i) \, , \\
        h_{ij} &= h_{ij}(rv,x^i) + \epsilon \, \delta h_{ij}(r,v,x^i) \, .
    \end{split}
\end{equation}
Such a Background plus fluctuation split enforces structures on covariant tensors. Suppose one considers a generic covariant tensor of the form
\begin{equation} \label{eqlmTensAP}
{\cal C} \sim (\partial_r)^{q_r}(\partial_v)^{q_v}{\cal D} \, ,
\end{equation}
constructed out of the metric components, where $\mathcal{D}$ \footnote{$\mathcal{D}$ is typically constructed out of  the metric components $X,\omega_i,h_{ij}$ and $\nabla_i$.} doesn't have explicit $\partial_v$ and $\partial_r$ derivatives, and evaluate it on the full metric eq.\eqref{eq:metdecompAP}, $\mathcal{D}$ will have the following structure
\begin{equation}
    \mathcal{D} = \mathcal{D}(rv,x^i) + \epsilon \, \mathcal{D}(r,v,x^i) \, .
\end{equation}
Thus, we can use the following criterion to distinguish between equilibrium configuration and non-equilibrium configuration
\begin{equation} \label{defEQvsNEQAP}
\begin{split}
\text{Equilibrium configuration} \rightarrow ~& (\partial_r)^{q_r}(\partial_v)^{q_v}{\cal D}(rv,x) \vert_{r=0}  =0 ~\text{(for $q_v>q_r$)} \, , \\
\text{Non-equilibrium configuration} \rightarrow ~& (\partial_r)^{q_r}(\partial_v)^{q_v}{\cal D}(r,v,x) \vert_{r=0} \neq 0 ~\text{(for $q_v>q_r$)} \, .
\end{split}
\end{equation}
The non-equilibrium configuration essentially evaluates to $\mathcal{O}(\epsilon)$. Thus, whenever extra $\partial_v$ derivatives are distributed between product of two or more different factors of $\mathcal{D}$, the resulting term is non-linear in dynamics, i.e., $\mathcal{O}(\epsilon^2)$. This type of analysis carries over straightforwardly to fields propagating on this space-time and this is discussed in detail in section \S\ref{sec:stationarity}.

To study these dynamics of small fluctuations away from stationary solutions, it is very useful to describe how covariant quantities transform under the boost transformation given in eq.\eqref{boosttransfAP}. We define a quantity $w$ called boost weight as the power of $\lambda$ to which a covariant tensor transforms under the boost transformation eq.\eqref{boosttransfAP}: 
\begin{equation} \label{defBW0AP}
{\cal A} \rightarrow \widetilde{\cal A} = \lambda^w {\cal A} \, , ~~\text{under}~~ \{r\rightarrow\tilde r = \lambda ~r,~v\rightarrow\tilde v = \lambda^{-1} v\} ~~ \Rightarrow ~~ \text{boost weight of ${\cal A}$ is $w$} \, .
\end{equation}
Now under this transformation, the operators $\partial_v$ and $\partial_r$ transform non-trivially as follows:
\begin{equation}
\partial_r \rightarrow \partial_{\tilde{r}} = \lambda^{-1} \, \partial_r \, , ~~ \partial_v \rightarrow \partial_{\tilde{v}} =\lambda  \, \partial_v \, .
\end{equation}
Thus, if a covariant tensor has positive boost weight, the structure has more number of $\partial_v$ derivatives than $\partial_r$ derivatives. This implies that, when the tensor is evaluated on the horizon using the full metric eq.\eqref{eq:metdecompAP}, the equilibrium contribution vanishes by the argument of eq.\eqref{defEQvsNEQAP}. Hence, it is a non-equilibrium configuration, i.e., $\mathcal{O}(\epsilon)$. This helps in characterizing whether terms are $\mathcal{O}(1)$ (equilibrium) or $\mathcal{O}(\epsilon)$ (non-equilibrium). 

\section{A derivation of the generalized zeroth law} \label{app:genzerothlaw}
In this appendix, our aim is to explicitly show how to derive eq.\eqref{eq:genzerothlaw} from eq.\eqref{eq:gaugestat}. Working in our gauge for the metric eq.\eqref{eq:metric}, the derivation goes as follows,
\begin{equation}
    \begin{split}
        \mathcal{L}_{\xi} A^{eq}_{\mu} + D_{\mu} \Lambda &= \xi^{\alpha} D_{\alpha} A^{eq}_{\mu} + A^{eq}_{\alpha} D_{\mu} \xi^{\alpha} + D_{\mu} \Lambda \\
        &= \xi^{\alpha} F^{eq}_{\alpha\mu} + D_{\mu} ( A^{eq}_{\alpha}\xi^{\alpha} + \Lambda ) \, .
    \end{split}
\end{equation}
Further, contracting with $\xi^{\mu}$ on both sides and setting the RHS to zero by demanding eq.\eqref{eq:gaugestat}, we get
\begin{equation}
    \begin{split}\label{gaugestep0}
        \xi^{\mu} D_{\mu} ( A^{eq}_{\alpha}\xi^{\alpha} + \Lambda ) = 0 \, ,
    \end{split} 
\end{equation}
such that one can solve it using $\xi$ from eq.\eqref{BT_generator} and obtain
\begin{equation}
    \begin{split}\label{gaugestep1}
    A^{eq}_{\alpha}\xi^{\alpha} + \Lambda = f(rv,x^i) \, .
    \end{split} 
\end{equation}
We now assume that the gauge potential is smooth at $r=0$. 
%freedom $\Lambda$ at hand to remove the poles of $A^{eq}_{\alpha}\xi^{\alpha}$ at $r=0$. 
Thus, 
\begin{equation}
    f(rv,x^i)|_{r=0} = f^{(0)}(x^i) \, ,
\end{equation}
and $f^{(0)}(x^i)$ is non-singular \footnote{An example involving the Reissner-Nordstorm black hole, where the gauge transformation is used to remove the singular part of the gauge field at the horizon, is given in Section I of \cite{Gao:2003ys}}. Now, we use the $x^i$ component of the stationarity condition given by eq.\eqref{eq:gaugestat} to set this to a constant as follows
\begin{equation}\label{gaugestep2}
    \begin{split}
        \mathcal{L}_{\xi} A^{eq}_i + D_i \Lambda^{eq} = & \xi^{\alpha} F^{eq}_{\alpha i} + D_i (A^{eq}_{\alpha}\xi^{\alpha} + \Lambda) = 0 \, 
         \implies D_i (A^{eq}_{\alpha}\xi^{\alpha} + \Lambda) |_{r=0} = 0 \, , \\
        & \implies f^{(0)}(x^i) = c \, .
    \end{split}  
\end{equation}
In the second step, we have evaluated the expression of the RHS on $r=0$: $\xi^{\alpha} F^{eq}_{\alpha i} |_{r=0} = v F^{eq}_{vi} |_{r=0} = 0$. This follows because $F_{vi}$ is a component of a covariant tensor that has positive boost weight. Thus, an equilibrium configuration evaluates to zero. In the third step, $c$ is a constant and it can be set to zero by using the gauge freedom in $\Lambda$. Putting the results in eq.\eqref{gaugestep1} and eq.\eqref{gaugestep2} together with setting $c$ to zero,
\begin{equation}
    (A^{eq}_{\alpha}\xi^{\alpha} + \Lambda) |_{r=0} = 0 \, .
\end{equation}
This is a simplified version of the argument presented in Proposition 3.1 of \cite{Prabhu:2015vua}. This is also explained in section III of \cite{Gao:2001ut}. This completes the proof of eq.\eqref{eq:genzerothlaw}.

\section{Modified generic structure of any covariant tensor with positive boost weight}
\label{app:ModRes1}
%\label{sec:modresult1}

In this appendix, we will show that ``Result:1" of appendix-E of \cite{Bhattacharyya:2021jhr} will be generically modified once we consider gauge fields coupled to gravity. Our aim will be to give a detailed analysis justifying eq.\eqref{modresult1f} once we have additional building blocks in the form of gauge field components $A_v,A_r,A_i$. In the scalar case, the explicit $v$ dependence of $E_{vv}$ was through boost invariant quantities like the metric functions $(X,\omega_i,h_{ij})$ and the scalar field $\phi$. The structure of $E_{vv}$ was constrained by how $\partial_v$ and $\partial_r$ derivatives act on them. Without the gauge fields in \cite{Bhattacharyya:2021jhr} and in the scalar case of \S\ref{ssec:scalJvJi}, $\partial_v$ and $\partial_r$ were the only objects that had non-trivial boost weights. Thus, from Result:1 of appendix E of \cite{Bhattacharyya:2021jhr}, we have that, any covariant tensor of boost weight $a+1>0$ that doesn't explicitly contain $\xi^{\mu}$ can always be re-expressed as follows
\begin{equation}\label{result1ap}
    \begin{split}
        t&^{(k)}_{(a+1)}|_{r=0} = \Tilde{T}_{(-k)}\partial^{(k+a+1)}_v T_{(0)}|_{r=0} + \mathcal{O}(\epsilon^2) \\
        &= \partial^{a+1}_v \left[ \sum^{k-1}_{m=0} (-1)^m ~ (^{m+a}C_m) ~ \Tilde{T}_{(-k+m)} \partial^{k-m}_v T_{(0)} \right] 
        + (-1)^k ~ ^{k+a}C_a ~ \Tilde{T}_{(0)}\partial^{a+1}_v T_{(0)} + \mathcal{O}(\epsilon^2) \, ,
    \end{split}
\end{equation}
where $\tilde{T}_{(-k+m)} = \partial^m_v \tilde{T}_{(-k)}$ and we are not being very careful with the indices and the numbers in the $()$ of subscripts denote the boost weights. But once we introduce a gauge field $A_{\mu}$, we have $A_v$ and $A_r$ as additional components that transform non-trivially under boost transformation. Thus, Result:1 must be modified by accounting for these extra fields. However, $A_v$ and $A_r$ cannot arbitrarily appear in the covariant tensors constructed out of a gauge-invariant Lagrangian of the form considered in eq.\eqref{eq:gaugelagrangian}. The only place where $A_{\mu}$ explicitly appears is in the term $U^{r\mu}_0 \delta A_{\mu} $ of $\Theta^r$ and we have seen in \S\ref{sssec:amutheta} that this term doesn't contribute to eq.\eqref{maineq}. Thus, for the modification of Result:1, we can completely ignore this term. Since the Lagrangian is gauge invariant, in all other terms of $\Theta^{\mu}$ and $Q^{\mu\nu}$, $A_v$ and $A_r$ always arise in the form of $F_{vr}$, $F_{vi}$, $F_{ri}$ (where $F_{\mu\nu}$ is the field strength tensor). Of these $F_{vr}$ is boost invariant but $F_{vi}$ and $F_{ri}$ have non-trivial boost weights.

From the details of appendix-\ref{app:boost}, we know that $F_{vi}|_{r=0} \sim \mathcal{O}(\epsilon)$ as it is a covariant tensor of boost weight $+1$. From the structure of $\Theta^{\mu}$ and $Q^{\mu\nu}$ worked out in section \S\ref{ssec:gaugeEvv}, it is clear that, Result:1 will be generically modified by incorporating new non zero boost weight elements in $F_{vi}$ and $F_{ri}$ as \footnote{The '$i$' index can appear in $\Tilde{T}_{(-k)}$ as well but we consider this form without loss of generality. Also, we are not taking into account the $U^{\mu\nu}_0 \delta A_{\nu}$ term. }
\begin{equation}
    \begin{split}
        t^{(k)}_{(a+1)}|_{r=0} = &\Tilde{T}_{(-k)} \, \partial^{k+a+1}_v \, T_{(0)}|_{r=0} + \Tilde{T}_{1(-k)} \, (F_{vi})^{m_1} \partial^{k-m_1+a+1}_v \, T^i_{1(0)}|_{r=0} \\
        &+ \Tilde{T}_{2(-k)} \, \partial^{k-m_2+a+1}_v \, ((F_{vi})^{m_2} \, T^i_{2(0)} )|_{r=0} \\
        &+ \Tilde{T}_{3(-k)} \, (F_{ri})^{m_3} \, \partial^{k+m_3+a+1}_v \, T^i_{3(0)} |_{r=0} \\
        &+ \Tilde{T}_{4(-k)} \, \partial^{k+m_4+a+1}_v \, ( (F_{ri})^{m_4} \, T^i_{4(0)} ) |_{r=0} + \mathcal{O}(\epsilon^2) \, ,
    \end{split}
\end{equation}
where \footnote{We have separated out all the occurrences of $F_{vi}$ and $F_{ri}$, i.e., $\tilde{T}_{n(-k)}$ and $T_{n(0)}$ can only contain gauge field terms in the form $F_{vr}$ and $F_{ij}$.} we assume $k+a+1>m_1,m_2$. Now, since $F_{vi}|_{r=0} \sim \mathcal{O}(\epsilon)$,
\begin{equation}
    \Tilde{T}_{1(-k)} \, (F_{vi})^{m_1} \, \partial^{k - m_1 + a+1}_v \, T^i_{1(0)} |_{r=0} \sim \mathcal{O}(\epsilon^2) \, ,
\end{equation}
and \footnote{The only term of type $T_2$ that survives is $m_2 = 1$.}
\begin{equation}
    \begin{split}
        \Tilde{T}_{2(-k)} \, \partial^{k-m_2+a+1}_v ( (F_{vi})^{m_2} \, T^i_{2(0)} ) |_{r=0} &= \Tilde{T}_{2(-k)} \, \partial^{k-1+a+1}_v ( F_{vi} \, T^i_{2(0)} ) |_{r=0} + \mathcal{O}(\epsilon^2) \\
        &= \Tilde{T}_{2(-k)} \, \partial^{k+a}_v (F_{vi} \, T^i_{2(0)} ) |_{r=0} \, .
    \end{split}
\end{equation}
Now, $\Tilde{T}_{3(-k)} \, (F_{ri})^{m_3} \, \partial^{k+m_3+a+1}_v \, T^i_{3(0)} |_{r=0}$ is already of the form accounted for by Result:1. So, there is no new analysis that is needed for this term. Similarly, by using the $m_4$ $\partial_v$ derivatives, we can easily see that
\begin{equation}
    \Tilde{T}_{4(-k)} \, \partial^{k+m_4+a+1}_v \, ( (F_{ri})^{m_4} \, T^i_{4(0)} ) |_{r=0} = \Tilde{T}_{4(-k)} \, \partial^{k+a+1}_v ( \mathcal{C}_{(0)} ) + \mathcal{O}(\epsilon^2) \, .
\end{equation}
Thus, this term also doesn't require new analysis. As a result, any covariant tensor with boost weight $a+1$ takes the following form 
\begin{equation}
    t^{(k)}_{(a+1)}|_{r=0} = \Tilde{T}_{5(-k)} \, \partial^{k+a+1}_v \, T_{5(0)} |_{r=0} + \Tilde{T}_{2(-k)}\,\partial^{k+a}_v (F_{vi} \, T^i_{2(0)}) + \mathcal{O}(\epsilon^2) \, .
\end{equation}
Now, the  term with $F_{vi}$ can be re-organized in the following manner (if $a>0$) using eq.\eqref{result1ap} where we substitute $a-1$ for $a$ and $F_{vi}T^i_{2(0)}$ for $T_{(0)}$:
\begin{equation}\label{extratermap}
    \begin{split}
        \Tilde{T}_{2(-k)} \, \partial^{k+a}_v \, ( F_{vi} \, T^i_{2(0)} )|_{r=0} = & \partial^a_v \left[ \sum^{k-1}_{m=0} (-1)^m ~ ^{m+a-1} C_m \, \Tilde{T}_{2(-k+m)} \partial^{k-m}_v ( F_{vi} \, T^i_{2(0)} ) \right] \\
        & ~ + (-1)^k ~\, ^{k+a-1}C_{a-1} \, \Tilde{T}_{2(0)} \, \partial^a_v ( F_{vi} \, T^i_{2(0)} ) + \mathcal{O}(\epsilon^2) \, .
    \end{split}
\end{equation}
Thus, Result:1 is modified as
\begin{equation} \label{eq:res1detAP}
    \begin{split}
        t^{(k)}_{(a+1)} &= \partial^{a+1}_v \left[ \sum^{k-1}_{m=0} (-1)^m ~ (^{m+a}C_m) ~ \Tilde{T}_{5(-k+m)} \partial^{k-m}_v T_{5(0)} \right] \\ 
        & ~~ + (-1)^k ~ ^{k+a}C_a ~ \Tilde{T}_{5(0)}\partial^{a+1}_v T_{5(0)} \\
        & ~~ +\partial^a_v \left[ \sum^{k-1}_{m=0} (-1)^m ~ ^{m+a-1} C_m \, \Tilde{T}_{2(-k+m)} \partial^{k-m}_v ( F_{vi} \, T^i_{2(0)} ) \right] \\
        & ~~ + (-1)^k ~\, ^{k+a-1}C_{a-1} \, \Tilde{T}_{2(0)} \, \partial^a_v ( F_{vi} \, T^i_{2(0)} ) + \mathcal{O}(\epsilon^2) \, .
    \end{split}
\end{equation}
What it essentially states is that, ignoring the  co-efficients, we have
\begin{equation}\label{modresult1fap}
    {\mathbb T}_{(k+1)}= \partial_v^{k+1} C_{(0)} + A_{(0)}~\partial_v^{(k+1)}B_{(0)} + \partial^k_v ( F_{vi} \, D^i_{(0)} ) + E_{(0)} \, \partial^k_v ( F_{vi} \, G^i_{(0)} ) + \mathcal{O}(\epsilon^2) \, .
\end{equation}
This is a generalization of eq.(3.33) of \cite{Bhattacharyya:2021jhr} with gauge fields. This completes the proof of eq.\eqref{modresult1f}.

\section{Details regarding the structure of $\Theta^r$ for gauge theories} \label{app:detailTH}

\subsection{Gauge invariant terms in $\Theta^r$}
\label{Ap:structureu}

In this appendix, we show that $\Theta^{\mu}$ inherits some additional structure from the gauge-invariant Lagrangian given by eq.\eqref{eq:gaugelagrangian}. In this regard, we will explicitly prove eq.\eqref{eq:structureu} and this can be argued for by working out $\Theta^{\mu}$ for the Lagrangian given by eq.\eqref{eq:gaugelagrangian}. For example, if the Lagrangian is of the form 
\begin{equation}
    L = L(F_{\mu\nu},D_{\alpha_1}F_{\mu\nu},D_{(\alpha_1}D_{\alpha_2)}F_{\mu\nu},...) \, ,
\end{equation}
then we consider the following term in the standard variation,
\begin{equation}\label{eq:thetaterm1}
	\begin{split}
		\dfrac{\partial L}{\partial (D_{\alpha_1} F_{\mu\nu})} \delta D_{\alpha_1} F_{\mu\nu} &= \dfrac{\partial L}{\partial (D_{\alpha_1} F_{\mu\nu})} \left[ D_{\alpha_1} \delta F_{\mu\nu} - \delta \Gamma^{\beta}_{\alpha_1\mu} F_{\beta\nu} - \delta \Gamma^{\beta}_{\alpha_1\nu} F_{\mu\beta} \right] \\
		&= \dfrac{\partial L}{\partial (D_{\alpha_1} F_{\mu\nu})} \left[ D_{\alpha_1} \delta F_{\mu\nu} + \text{Terms proportional to $D_{\alpha_1} \delta g_{\mu\gamma}$} \right] \, .
	\end{split}
\end{equation}
The Terms proportional to $D_{\alpha_1} \delta g_{\mu\gamma}$ clearly contribute to Terms proportional to $\delta g_{\mu\gamma}$ in $\Theta^{\rho}$. These terms are gauge invariant and they have no role to play in the proof of eq.\eqref{eq:structureu}. For example, consider the following term in eq.\eqref{eq:thetaterm1}:
\begin{equation}
	\begin{split}
		-\dfrac{\partial L}{\partial (D_{\alpha_1} F_{\mu\nu})} \delta \Gamma^{\beta}_{\alpha_1 \mu} F_{\beta \nu} &=  -\dfrac{\partial L}{\partial (D_{\alpha_1} F_{\mu\nu})} F_{\beta\nu} \left[ \dfrac{1}{2}g^{\beta\gamma} D_{\alpha_1} \delta g_{\mu\gamma} + \dots \right] \\
		&= -D_{\alpha_1} \left[ \dfrac{1}{2} \dfrac{\partial L}{\partial (D_{\alpha_1} F_{\mu\nu})} F^{\gamma}_{~\nu} \delta g_{\mu\gamma} \right]  + \dfrac{1}{2} D_{\alpha_1} \left[\dfrac{\partial L}{\partial (D_{\alpha_1} F_{\mu\nu})} F^{\gamma}_{~\nu} \right] \delta g_{\mu\gamma} + \dots \, . 
	\end{split}
\end{equation}
(i.e.) such terms contribute a term of the form $S^{\nu\alpha\beta}\delta g_{\alpha\beta}$ in $\Theta^{\nu}$ where $S^{\nu\alpha\beta}$ is gauge invariant. Thus, we only need to analyze
\begin{equation}
    \begin{split}
        \dfrac{\partial L}{\partial (D_{\alpha_1} F_{\mu\nu})} D_{\alpha_1} \delta F_{\mu\nu} &= 2 \dfrac{\partial \, L}{\partial (D_{\alpha_1} F_{\mu\nu} )} D_{\alpha_1} \delta D_{\mu} A_{\nu} \\
        &= 2 D_{\alpha_1} \left[\dfrac{\partial \, L}{\partial (D_{\alpha_1} F_{\mu\nu})} \delta D_{\mu}A_{\nu} \right] - 2 D_{\alpha_1} \left[ \dfrac{\partial \, L}{\partial (D_{\alpha_1} F_{\mu\nu})} \right] \delta D_{[\mu} A_{\nu]} \\
        & = 2 D_{\alpha_1} \left[\dfrac{\partial \, L}{\partial (D_{\alpha_1} F_{\mu\nu})} \delta D_{\mu}A_{\nu} \right] - 2 D_{\alpha_1} \left[ \dfrac{\partial \, L}{\partial (D_{\alpha_1} F_{\mu\nu})} \right] D_{[\mu} \delta A_{\nu]}\\
        &= 2 D_{\alpha_1} \left[\dfrac{\partial \, L}{\partial (D_{\alpha_1} F_{\mu\nu})} \delta D_{\mu}A_{\nu} \right] - 2 D_{\mu} \left[ D_{\alpha_1} \left[ \dfrac{\partial \, L}{\partial (D_{\alpha_1} F_{\mu\nu})} \right] \delta A_{\nu} \right] \\
        &~~ + 2 D_{\mu} D_{\alpha_1} \left[ \dfrac{\partial \, L}{\partial (D_{\alpha_1} F_{\mu\nu})} \right] \delta A_{\nu} \, .
    \end{split}
\end{equation}
In the third step, we have crucially used the fact that the indices $\mu$ and $\nu$ are antisymmetric to interchange $\delta$ and $D_{\mu}$. We can do this because
\begin{equation}
	\begin{split}
		\delta D_{\mu} A_{\nu} &= D_{\mu} \delta A_{\nu} - \delta \Gamma^{\rho}_{\mu\nu}\delta A_{\rho} \\
		\implies \delta D_{[\mu} A_{\nu]} &= D_{[\mu} \delta A_{\nu]} \, .
	\end{split}
\end{equation}
Thus, the contribution of the first term in eq.\eqref{eq:thetaterm1} to $\Theta^{\rho}$ is
\begin{equation}\label{eq:theta1}
    \begin{split}
        \Theta^{\rho} &= 2 \dfrac{\partial \, L}{\partial (D_{\rho} F_{\mu\nu})} \delta D_{\mu} A_{\nu} - D_{\alpha_1} \left[ \dfrac{\partial \, L}{\partial (D_{\alpha_1} F_{\rho\nu})} \right] \delta A_{\nu} ~+  S^{\rho\alpha\beta}_1 \delta g_{\alpha\beta} \\
        &= \dfrac{\partial \, L}{\partial (D_{\rho} F_{\mu\nu})} \delta F_{\mu\nu} - D_{\alpha_1} \left[ \dfrac{\partial \, L}{\partial (D_{\alpha_1} F_{\rho\nu})} \right] \delta A_{\nu} +  S^{\rho\alpha\beta}_1 \delta g_{\alpha\beta} \, .
    \end{split}
\end{equation}
Notice the anti-symmetry predicted by eq.\eqref{eq:structureu} is present in the first term of the above equation. The lessons learned from the structures arising from the simplest term can be straightforwardly carried over to more complicated examples. In what follows, we won't carefully work out the exact form of the terms that contribute to $S^{\nu\alpha\beta}\delta g_{\alpha\beta}$ in $\Theta^{\nu}$. We will indicate those terms which will eventually contribute to $S^{\mu\alpha\beta}$ by ``$\dots$". For example, another possible term in the standard variation gives
\begin{equation}
    \begin{split}
       & \dfrac{\partial \, L}{\partial ( D_{(\alpha_1} D_{\alpha_2)} F_{\mu\nu} )} \delta D_{(\alpha_1} D_{\alpha_2)} F_{\mu\nu} = 2 \dfrac{\partial \, L}{\partial ( D_{(\alpha_1} D_{\alpha_2)} F_{\mu\nu} )} D_{(\alpha_1} \delta D_{\alpha_2)} D_{\mu} A_{\nu} + \dots \\
        &= 2 D_{\alpha_1} \left[ \dfrac{\partial \, L}{\partial ( D_{(\alpha_1} D_{\alpha_2)} F_{\mu\nu} )} \delta D_{\alpha_2} D_{\mu} A_{\nu} \right] - 2 D_{\alpha_1} \left[ \dfrac{\partial \, L}{\partial ( D_{(\alpha_1} D_{\alpha_2)} F_{\mu\nu} )} \right] \delta D_{\alpha_2} D_{\mu} A_{\nu} \\
        &= 2 D_{\alpha_1} \left[ \dfrac{\partial \, L}{\partial ( D_{(\alpha_1} D_{\alpha_2)} F_{\mu\nu} )} \delta D_{\alpha_2} D_{\mu} A_{\nu} \right] - 2 D_{\alpha_2} \left[ D_{\alpha_1} \left[ \dfrac{\partial \, L}{\partial ( D_{(\alpha_1} D_{\alpha_2)} F_{\mu\nu} )} \right] \delta D_{\mu} A_{\nu}  \right] \\
        & ~~ + 2 D_{\alpha_2} D_{\alpha_1} \left[ \dfrac{\partial \, L}{\partial ( D_{(\alpha_1} D_{\alpha_2)} F_{\mu\nu} )} \right] \delta D_{\mu} A_{\nu} + \dots \, .
    \end{split}
\end{equation}
Thus from a similar analysis of the term of eq.\eqref{eq:thetaterm1}, the contribution of this term to $\Theta^{\rho}$ can be worked out to be
\begin{equation}\label{eq:theta2}
    \begin{split}
        \Theta^{\rho} &= 2 \dfrac{\partial \, L}{\partial ( D_{(\rho} D_{\alpha_2)} F_{\mu\nu} )} \delta D_{\alpha_2} D_{\mu} A_{\nu} - 2 D_{\alpha_1} \left[ \dfrac{\partial \, L}{\partial ( D_{(\alpha_1} D_{\rho)} F_{\mu\nu} )} \right] \delta D_{\mu} A_{\nu} \\ 
        & ~~ + 2 D_{\alpha_2} D_{\alpha_1} \left[ \dfrac{\partial \, L}{\partial ( D_{(\alpha_1} D_{\alpha_2)} F_{\rho\nu} )} \right] \delta A_{\nu} + ~ S^{\rho\alpha\beta}_2 \delta g_{\alpha\beta} \\
        &= \dfrac{\partial \, L}{\partial ( D_{(\rho} D_{\alpha_2)} F_{\mu\nu} )} \delta D_{\alpha_2} F_{\mu\nu} -  D_{\alpha_1} \left[ \dfrac{\partial \, L}{\partial ( D_{(\alpha_1} D_{\rho)} F_{\mu\nu} )} \right] \delta F_{\mu\nu} \\ & ~~ + 2 D_{\alpha_2} D_{\alpha_1} \left[ \dfrac{\partial \, L}{\partial ( D_{(\alpha_1} D_{\alpha_2)} F_{\rho\nu} )} \right] \delta A_{\nu}  +S^{\rho\alpha\beta}_2 \delta g_{\alpha\beta} \, .
    \end{split}
\end{equation}
Here $S^{\rho\alpha\beta}_2$ is again gauge invariant. The anti-symmetry of eq.\eqref{eq:structureu} is again seen in the first 2 terms of the above equation. These observations can be generalized straightforwardly to a term containing any number of derivatives acting on the gauge field given by
\begin{equation}
    \dfrac{\partial \, L}{\partial ( D_{(\alpha_1} D_{\alpha_2} \dots D_{\alpha_n)} F_{\mu\nu} )} \delta D_{(\alpha_1} D_{\alpha_2} \dots D_{\alpha_n)} F_{\mu\nu} \, .
\end{equation}
As we repeatedly interchange $\delta$ and $D_{\alpha_i}$ to extract $\Theta^{\mu}$, not only do we get terms of the form $S^{\mu\alpha\beta}\delta g_{\alpha\beta}$ in $\Theta^{\mu}$, but also we get terms of the form given in eq.\eqref{eq:structureu} by integration by parts manipulation. In the end, there will be one term of the form $\delta D_{\mu} A_{\nu}$. But since we vary the Lagrangian with respect to gauge invariant quantities like $D_{(\alpha_1} \dots D_{\alpha_i)} F_{\mu\nu}$, there will always be an antisymmetry in the indices of the gauge field and the covariant derivative acting on it. This is a direct consequence of the gauge invariant structure of the Lagrangian given by eq.\eqref{eq:gaugelagrangian}.

\subsection{Non gauge invariant terms in $\Theta^r$} \label{app:ThQNoGinv}

In this appendix, we will detail the structure of non gauge invariant terms in $\Theta^r$. First we will prove eq.\eqref{U0antisym}. From the arguments of appendix \ref{Ap:structureu}, we note an additional structure in the $U^{\mu\nu}_0\delta A_{\nu}$ term of $\Theta^{\mu}$. From eq.\eqref{eq:theta1} and eq.\eqref{eq:theta2}, we see that
\begin{equation}
    U^{\mu\nu}_0 = - U^{\nu\mu}_0 \, .
\end{equation}
This is because the term with $\delta A_{\nu}$ is contracted with a term which contains $F_{\rho\nu}$. Thus, the only non-gauge invariant term in $\Theta^{\mu}$ still has additional structure due to gauge invariance of the Lagrangian. 

We will now detail the calculations of \S\ref{sssec:amutheta} and show that this non-gauge invariant term drops out from our main equation eq.\eqref{maineq}. In $\Theta^r(\delta_{\xi}A_{\mu})|_{r=0}$ (here $\delta_{\xi} A_{\mu}$ is given by eq.\eqref{eq:Fvariation}, i.e., $\delta_{\xi} A_{\mu} = \mathcal{L}_{\xi} A_{\mu} + D_{\mu} \Lambda$), we have
\begin{equation}\label{eq:thetarinitial}
    U^{r\mu}_0 \delta_{\xi} A_{\mu} |_{r=0} = U^{rv}_0 \delta_{\xi} A_v |_{r=0} + U^{ri}_0 \delta_{\xi} A_i |_{r=0} \, .
\end{equation}
The RHS can be straightforwardly evaluated using eq.(2.20) of \cite{Bhattacharyya:2021jhr} with the additional gauge transformation contribution $D_{\mu}\Lambda$ given according to eq.\eqref{eq:Fvariation}. As a result, we have the following
\begin{equation}
    \begin{split}
        U^{rv}_0\delta_{\xi} A_v |_{r=0} &= U^{rv}_0 (1 + v \partial_v) A_v + U^{rv}_0 \partial_v \Lambda = U^{rv}_0 \partial_v (v A_v + \Lambda) \\
        &= \partial_v \left[ U^{rv}_0 (v A_v + \Lambda) \right] + \mathcal{O}(\epsilon^2) \, ,
    \end{split}
\end{equation}
where we have used the eq.\eqref{eq:STgauge1}, i.e., $(v A_v + \Lambda)|_{r=0} \sim \mathcal{O}(\epsilon)$. Similarly, for the other component we have,
\begin{equation}
    \begin{split}
        U^{ri}_0\delta_{\xi}A_i |_{r=0} &= U^{ri}_0 \left( v F_{vi} + \partial_i [ v A_v + \Lambda ] \right) = \mathcal{O}(\epsilon^2) \, ,
    \end{split}
\end{equation}
where in the first step, we have used eq.\eqref{eq:Fvariation} and in the second step, we have used $F_{vi}|_{r=0} \sim \mathcal{O}(\epsilon)$ and $(v A_v + \Lambda)|_{r=0} \sim \mathcal{O}(\epsilon)$. Thus, the $\Theta^r$ contribution of $U^{r\nu}_0 \delta A_{\nu}$ is
\begin{equation}\label{eq:nongaugetheta}
    \partial_v \left[ U^{rv}_0 (v A_v + \Lambda) \right] + \mathcal{O}(\epsilon^2) \, .
\end{equation}

In $Q^{\mu\nu}$ analogously, we only need to consider $Q^{\mu\nu} = U^{\mu\nu}_0 ( A^{\alpha}\xi_{\alpha} + \Lambda )$,
\begin{equation}
    Q^{rv}|_{r=0} = U^{rv}_0 ( v \, A_v + \Lambda ) \, , \hspace{2cm} Q^{ri} |_{r=0} = U^{ri}_0 ( v \, A_v + \Lambda ) \sim \mathcal{O}(\epsilon^2) \, .
\end{equation}
\begin{equation}
    \begin{split}
        D_{\mu} Q^{r\mu} |_{r=0} &= \dfrac{1}{\sqrt{h}} \partial_v ( \sqrt{h} \, Q^{rv} ) + \nabla_i Q^{ri} = \dfrac{1}{\sqrt{h}} \partial_v \left[ \sqrt{h} \, ( U^{rv}_0 [ v \, A_v + \Lambda ] ) \right] \\
        &= \partial_v ( U^{rv}_0 [ v \, A_v + \Lambda ]) + \mathcal{O}(\epsilon^2) \, .
    \end{split}
\end{equation}
From eq.\eqref{eq:nongaugetheta}, we know that the $\Theta^r$ contribution of $U^{r\nu}_0 \delta A_{\nu}$ is
\begin{equation}
    \partial_v \left[ U^{rv}_0 (v A_v + \Lambda) \right] + \mathcal{O}(\epsilon^2) \, .
\end{equation}
Thus,
\begin{equation}
    -\Theta^r + D_{\mu} Q^{r\mu} = \mathcal{O}(\epsilon^2) \, ,
\end{equation}
for $U^{\mu\nu}_0 \delta A_{\nu}$. The only non-gauge invariant term drops out within our approximation from eq.\eqref{maineq}.

\subsection{Final form of $\Theta^r$}
\label{app:detailTh}

In this appendix, we show that the additional terms that arise due to the modification of Result:1 with gauge fields in eq.\eqref{modresult1N} will not affect the arguments following eqn.(3.33) up to eqn.(3.41) of \cite{Bhattacharyya:2021jhr} (arguments of Section 3.2.2 of \cite{Bhattacharyya:2021jhr}), i.e., we must show that
\begin{equation}\label{toprove1}
    \begin{split}
        \sum_k \Tilde{\mathbb{T}}_{(-k)}(k+1 + v \partial_v) \mathbb{T}_{(k+1)} = (1+v\partial_v) \mathcal{M}_{(1)} + v \partial^2_v \mathcal{N}_{(0)} + \mathcal{O}(\epsilon^2) \, ,
    \end{split}
\end{equation}
where $\mathbb{T}_{(k+1)}$ is given by the extra terms present in eq.\eqref{modresult1f}:
\begin{equation}
    \mathbb{T}_{k+1} = \partial^k_v ( F_{vi} \, D^i_{(0)} ) + E_{(0)} \, \partial^k_v ( F_{vi} \, G^i_{(0)} ) + \mathcal{O}(\epsilon^2) \, .
\end{equation}
Now we collect a couple of results that we need to prove eq.\eqref{toprove1}: \footnote{Note that these results are not contradicting eq.\eqref{result1ap} since here we are pulling out an extra $\partial_v$ derivative when compared to eq.\eqref{result1ap}.}
\begin{equation}\label{result2}
    \begin{split}
        \Tilde{T}_{(-n)} \partial^n_v ( F_{vi} T^i_{(0)} ) &= \partial_v \left[ \sum^{n-1}_{m=0} (-1)^m  \Tilde{T}_{(-n+m)} \partial^{n-(m+1)}_v (F_{vi} T^i_{(0)}) \right] \\
        &~~ + (-1)^n \Tilde{T}_{(0)} F_{vi} T^i_{(0)} + \mathcal{O}(\epsilon^2) \, ,
    \end{split}
\end{equation}
\begin{equation}\label{result3}
    \begin{split}
        \Tilde{T}_{(-n)} \partial^{n+1}_v ( F_{vi} T^i_{(0)} ) &= \partial^2_v \left[ \sum^{n-1}_{m=0} (-1)^m (m+1) \Tilde{T}_{(-n+m)} \partial^{n-(m+1)}_v (F_{vi} T^i_{(0)}) \right] \\ 
        & ~~ + (-1)^n (n+1) \partial_v ( \Tilde{T}_{(0)} F_{vi} T^i_{(0)} ) + \mathcal{O}(\epsilon^2) \, .
    \end{split}
\end{equation}
(where $\Tilde{T}_{(-n+m)} = \partial^m_v \Tilde{T}_{(-n)}$) These can be straightforwardly proved by method of induction (see the last part of this appendix \ref{app:detailTh} for the proofs of eq.\eqref{result2} and eq.\eqref{result3})

Now,
\begin{equation}
    \begin{split}
        \Tilde{\mathbb{T}}_{(-k)} ( k+1 + v \partial_v ) \mathbb{T}_{(k+1)} &= \Tilde{\mathbb{T}}_{(-k)} ( k+1 + v \partial_v ) \left[ \partial^k_v ( F_{vi} \, D^i_{(0)} ) + E_{(0)} \, \partial^k_v ( F_{vi} \, G^i_{(0)} ) \right] \\
        &= (k+1) \Tilde{\mathbb{T}}_{(-k)} \partial^k_v ( F_{vi} \, D^i_{(0)} ) + v \Tilde{\mathbb{T}}_{(-k)} \partial^{k+1}_v ( F_{vi} \, D^i_{(0)} ) \\
        & \, + (k+1) \Tilde{\mathbb{T}}_{(-k)} E_{(0)} \partial^k_v ( F_{vi} \, G^i_{(0)} ) + v \Tilde{\mathbb{T}}_{(-k)} E_{(0)} \partial^{k+1}_v ( F_{vi} \, G^i_{(0)} ) + \mathcal{O}(\epsilon^2) \\
        & \sim (k+1) X_{(-k)} \partial^k_v ( F_{vi} \, Y^i_{(0)} ) + v X_{(-k)} \partial^{k+1}_v ( F_{vi} \, Y^i_{(0)} ) \, .
    \end{split}
\end{equation}
Using the results, eq.\eqref{result2} and eq.\eqref{result3},
\begin{equation}
    \begin{split}
        X_{(-k)} \partial^k_v ( F_{vi} Y^i_{(0)} ) &= \partial_v \left[ \sum^{k-1}_{m=0} (-1)^m X_{(-k+m)} \partial^{k-(m+1)}_v ( F_{vi} \, Y^i_{(0)}) \right] \\
        & ~~ + (-1)^k X_{(0)} F_{vi} Y^i_{(0)} + \mathcal{O}(\epsilon^2) \, ,
    \end{split}
\end{equation}
\begin{equation}
    \begin{split}
        X_{(-k)} \partial^{k+1}_v ( F_{vi} Y^i_{(0)} ) &= \partial^2_v \left[ \sum^{k-1}_{m=0} (-1)^m (m+1) X_{(-k+m)} \partial^{k-(m+1)}_v ( F_{vi} \, Y^i_{(0)}) \right] \\
        & ~~ + (-1)^k (k+1) \, \partial_v ( X_{(0)} F_{vi} Y^i_{(0)} ) + \mathcal{O}(\epsilon^2) \, ,
    \end{split}
\end{equation}
Using these, one can straightforwardly show that the terms in $\Theta^r$ are of the desired form:
\begin{equation}
    \begin{split}
        (k+1) & X_{(-k)} \partial^k_v ( F_{vi} Y^i_{(0)} ) + v X_{(-k)} \partial^{k+1}_v ( F_{vi} Y^i_{(0)} ) \\ 
        &= (k+1) \partial_v \left[ \sum^{k-1}_{m=0} (-1)^m X_{(-k+m)} \partial^{k-(m+1)}_v ( F_{vi} Y^i_{(0)} ) \right] \\
        & ~ + (-1)^k (k+1) X_{(0)} F_{vi} Y^i_{(0)} \\
        & ~ + v \partial^2_v \left[ \sum^{k-1}_{m=0} (-1)^m (m+1) X_{(-k+m)} \partial^{k-(m+1)}_v (F_{vi} Y^i_{(0)}) \right] \\
        & ~ + (-1)^k (k+1) \partial_v \left[ X_{(0)} F_{vi} Y^i_{(0)} \right] \, , \\
        & = (1 + v \partial_v) \left[ (-1)^k (k+1) X_{(0)} F_{vi} Y^i_{(0)} \right] \\
        & ~ + (k+1) \partial_v \left[ \sum^{k-1}_{m=0} (-1)^m X_{(-k+m)} \partial^{k-(m+1)}_v ( F_{vi} Y^i_{(0)} ) \right] \\
        & ~ + v \partial^2_v \left[ \sum^{k-1}_{m=0} (-1)^m (k+1+m-k) X_{(-k+m)} \partial^{k-(m+1)}_v (F_{vi} Y^i_{(0)}) \right] \, , \\
        &= (1 + v \partial_v) \left[ \partial_v \left[ \sum^{k-1}_{m=0} (-1)^m X_{(-k+m)} \partial^{k-(m+1)}_v ( F_{vi} Y^i_{(0)} ) \right] \right. \\
        & ~ \left. + (-1)^k (k+1) X_{(0)} F_{vi} Y^i_{(0)} \right] \\
        &- v \partial^2_v \left[ \sum^{k-1}_{m=0} (-1)^m (k-m) X_{(-k+m)} \partial^{k-(m+1)}_v (F_{vi} Y^i_{(0)}) \right] \, , \\
        & \sim (1 + v \partial_v) \mathcal{M}_{(1)} + v \partial^2_v \mathcal{N}_{(0)} \, .
    \end{split}
\end{equation}
where
\begin{equation}
    \begin{split}
        \mathcal{M}_{(1)} &= \partial_v \left[ \sum^{k-1}_{m=0} (-1)^m X_{(-k+m)} \partial^{k-(m+1)}_v ( F_{vi} Y^i_{(0)} ) \right] \\
        & ~ + (-1)^k (k+1) X_{(0)} F_{vi} Y^i_{(0)} \, , \\
        \mathcal{N}_{(0)} &= \sum^{k-1}_{m=0} (-1)^m (k-m) X_{(-k+m)} \partial^{k-(m+1)}_v (F_{vi} Y^i_{(0)}) \, .
    \end{split}
\end{equation}
This completes the proof of eq.\eqref{toprove1}. 

We have already seen how $U^{r\nu}\delta A_{\nu}$ term drops out from the eq.\eqref{maineq} in \S\ref{sssec:amutheta}. Hence, even if we have non-zero $A_v$ and $A_r$, $\Theta^r$ given by eq.\eqref{finalformtheta} takes the required form needed for our proof, i.e., $\Theta^r$ takes the form
\begin{equation}
    \Theta^r = (1+v\partial_v) \mathcal{A}_{(1)} + v \partial^2_v \mathcal{B}_{(0)} + \mathcal{O}(\epsilon^2) \, .
\end{equation}
Most importantly note that $\mathcal{N}_{(0)}$ given by eq.\eqref{toprove2result}, which contributes $\mathcal{J}^v$ (according to eq.\eqref{finJvJiAmu}) is gauge-invariant.

It is interesting to note that the anti-symmetry in $U^{\mu\nu}_0$ was crucial for ensuring the gauge invariance of entropy current. Suppose, we didn't have anti-symmetry, then we would have a term of the form $U^{rr}_0$. Then, we have
\begin{equation}
    \begin{split}
        U^{rr}_0 \mathcal{L}_{\xi} A_r &= U^{rr}_0 (-1 + v \partial_v ) A_r \\
        &= (1 + v \partial_v) \mathcal{C}_{(1)} + v \partial^2_v \mathcal{D}_{(0)} + \mathcal{O}(\epsilon^2) \, ,
    \end{split}
\end{equation}
where we used eqn(3.51) of \cite{Bhattacharyya:2021jhr} in the second step. Note that $\mathcal{D}_{(0)}$ will explicitly contain $A_r$ type terms and it will contribute to $\mathcal{J}^v$. This will spoil the gauge invariance of $\mathcal{J}^v$.

\subsection*{Proof of intermediary result eq.(\ref{result2})}
\label{intermediaryresults2}

We will prove eq.\eqref{result2} using the method of induction. First we check for $n=1$:
\begin{equation}
    \Tilde{T}_{(-1)} \partial_v ( F_{vi} T^i_{(0)} ) = \partial_v \left[ \Tilde{T}_{(-1)} F_{vi} T^i_{(0)} \right] - \Tilde{T}_{(0)} F_{vi} T^i_{(0)} \, .
\end{equation}
This is trivially of the form predicted in eq.\eqref{result2}. Now we assume that the result is valid for $n$:
\begin{equation}\label{result2n}
    \begin{split}
        \Tilde{T}_{(-n)} \partial^n_v \left[ F_{vi} T^i_{(0)} \right] = \partial_v \left[ \Tilde{T}_{(-n)} \partial^{n-1}_v ( F_{vi} T^i_{(0)} ) - (\partial_v \Tilde{T}_{(-n)}) \partial^{n-2}_v ( F_{vi} T^i_{(0)} ) \right. \\
        + \left. (\partial^2_v \Tilde{T}_{(-n)}) \partial^{n-3}_v  ( F_{vi} T^i_{(0)} ) - \dots \right] + (-1)^n \Tilde{T}_{(0)} F_{vi} T^i_{(0)} + \mathcal{O}(\epsilon^2) \, .
    \end{split}
\end{equation}
Consider the expression for $n+1$:
\begin{equation}
    \begin{split}
        \Tilde{T}_{(-n-1)} \partial^{n+1}_v ( F_{vi} T^i_{(0)} ) &= \partial_v \left[ \Tilde{T}_{(-n-1)} \partial^n_v ( F_{vi} T^i_{0}) \right] - \Tilde{T}_{(-n)} \partial^n_v ( F_{vi} T^i_{(0)} ) \, , \\
        & = \partial_v \left[ \Tilde{T}_{(-n-1)} \partial^n_v ( F_{vi} T^i_{0}) \right] - \partial_v \left[ \Tilde{T}_{(-n)} \partial^{n-1}_v ( F_{vi} T^i_{(0)} ) \right. \\
        & ~ \left. - \Tilde{T}_{(-n+1)} \partial^{n-2}_v ( F_{vi} T^i_{(0)} ) + \Tilde{T}_{(-n+2)} \partial^{n-3}_v ( F_{vi} T^i_{(0)} ) - \dots \right] \\
        &~ - (-1)^n \Tilde{T}_{(0)} F_{vi} T^i_{(0)} + \mathcal{O}(\epsilon^2) \, , \\
        &= \partial_v \left[ \Tilde{T}_{(-n-1)} \partial^n_v ( F_{vi} T^i_{(0)} ) - \Tilde{T}_{(-n)} \partial^{n-1}_v (F_{vi} T^i_{(0)} ) \right. \\
        & ~ + \left. \Tilde{T}_{(-n+1)} \partial^{n-2}_v ( F_{vi} T^i_{(0)} ) - \Tilde{T}_{(-n+2)} \partial^{n-3}_v ( F_{vi} T^i_{(0)} ) + \dots \right] \\
        & ~ + (-1)^{n+1} \Tilde{T}_{(0)} F_{vi} T^i_{(0)} + \mathcal{O}(\epsilon^2) \, ,
    \end{split}
\end{equation}
where in the second step, we have used the result for $n$ in eq.\eqref{result2n}. This is of the required form in eq.\eqref{result2} we tried to prove. 

\subsection*{Proof of intermediary result eq.(\ref{result3})}
\label{intermediaryresults3}

We will prove eq.\eqref{result3} by using method of induction. For $n=1$, we have
\begin{equation}
    \begin{split}
        \Tilde{T}_{(-1)} \partial^2_v ( F_{vi} T^i_{(0)} ) &= \partial_v \left[ \Tilde{T}_{(-1)} \partial_v ( F_{vi} T^i_{(0)} ) \right] - \partial_v \Tilde{T}_{(-1)} \partial_v ( F_{vi} T^i_{(0)} ) \\
        &= \partial^2_v ( \Tilde{T}_{(-1)} F_{vi} T^i_{(0)} ) - 2 \partial_v ( \Tilde{T}_{(0)} F_{vi} T^i_{(0)} ) + \mathcal{O}(\epsilon^2) \, .
    \end{split}
\end{equation}
This is trivially of the form given by eq.\eqref{result3}. Now we assume that the result is valid for $n$:
\begin{equation}
    \begin{split}
        \Tilde{T}_{(-n)} \partial^{n+1}_v ( F_{vi} T^i_{(0)} ) &= \partial^2_v \left[ \Tilde{T}_{(-n)} \partial^{n-1}_v ( F_{vi} T^i_{(0)} ) - 2 \, \Tilde{T}_{(-n+1)} \partial^{n-2}_v ( F_{vi} T^i_{(0)} ) \right. \\
        & + \left. 3 \, \Tilde{T}_{(-n+2)} \partial^{n-3}_v ( F_{vi} T^i_{(0)} ) + \dots \right] + (-1)^n (n+1) \partial_v ( \Tilde{T}_{(0)} F_{vi} T^i_{(0)} ) + \mathcal{O}(\epsilon^2) \, .
    \end{split}
\end{equation}
For $n+1$, we have
\begin{equation}
    \begin{split}
        \Tilde{T}_{(-n-1)} \partial^{n+2}_v ( F_{vi} T^i_{(0)} ) &= \partial_v \left[ \Tilde{T}_{(-n-1)} \partial^{n+1}_v ( F_{vi} T^i_{(0)} ) \right] - \Tilde{T}_{(-n)} \partial^{n+1}_v ( F_{vi} T_{(0)} ) \, , \\
        & = \partial_v \left[ \Tilde{T}_{(-n-1)} \partial^{n+1}_v ( F_{vi} T^i_{(0)} ) \right] - \partial^2_v \left[ \Tilde{T}_{(-n)} \partial_v (F_{vi} T^i_{(0)} )  \right. \\
        & \left. - 2 \, \Tilde{T}_{(-n+1)} \partial^{n-2}_v ( F_{vi} T^i_{(0)} ) + 3 \, \Tilde{T}_{(-n+2)} \partial^{n-3}_v ( F_{vi} T^i_{(0)} ) - \dots \right] \\
        & - (-1)^n (n+1) \partial_v \left[ \Tilde{T}_{(0)} F_{vi} T^i_{(0)} \right] + \mathcal{O}(\epsilon^2) \, ,
    \end{split}
\end{equation}
where in the second equality, we have used the previous equation. Now, we will use the result of eq.\eqref{result2} for $n+1$ in the above equation since we already proved it for all $n$ in appendix \ref{intermediaryresults2}. RHS of the above equation thus becomes
\begin{equation}
    \begin{split}
        &= \partial^2_v \left[ \Tilde{T}_{(-n-1)} \partial^{n-1}_v ( F_{vi} T^i_{(0)} ) - \Tilde{T}_{(-n)} \partial^{n-1}_v ( F_{vi} T^i_{(0)} ) + \Tilde{T}_{(-n+1)} \partial^{n-2}_v ( F_{vi} T^i_{(0)} ) - \dots \right] \\
        & ~~ + (-1)^{n+1} \partial_v \left[ \Tilde{T}_{(0)} F_{vi} T^i_{(0)} \right] \\
        & + \partial^2_v \left[ -\Tilde{T}_{(-n)} \partial^{n-1}_v ( F_{vi} T^i_{(0)} ) + 2 \Tilde{T}_{(-n+1)} \partial^{n-2}_v ( F_{vi} T^i_{(0)} ) - 3 \Tilde{T}_{(-n+2)} \partial^{n-3}_v ( F_{vi} T^i_{(0)} ) - \dots \right] \\
        & ~~ - (-1)^{n} (n+1) \partial_v \left[ \Tilde{T}_{(0)} F_{vi} T^i_{(0)} \right] + \mathcal{O}(\epsilon^2) \, , \\
        & = \partial^2_v \left[ \Tilde{T}_{(-n-1)} \partial^{n}_v ( F_{vi} T^i_{(0)} ) - 2 \Tilde{T}_{(-n)} \partial^{n-1}_v ( F_{vi} T^i_{(0)} ) + 3 \Tilde{T}_{(-n+1)}\partial^{n-2}_v ( F_{vi} T^i_{(0)} )  \right. \\
        & \left. - 4 \Tilde{T}_{(-n+2)} \partial^{n-3}_v ( F_{vi} T^i_{(0)} ) + \dots \right] + (-1)^{n+1} (n+2) \partial_v \left[ \Tilde{T}_{(0)} F_{vi} T^i_{(0)} \right] + \mathcal{O}(\epsilon^2) \, , \\
        & = \partial^2_v \left[ \sum^{n}_{m=0} (-1)^m (m+1) \Tilde{T}_{(-n-1+m)} \partial^{n+1-(m+1)}_v ( F_{vi} T^i_{(0)} ) \right] \\
        & ~~ + (-1)^{n+1} (n+2) \, \partial_v \left[ \Tilde{T}_{(0)} F_{vi} T^i_{(0)} \right] + \mathcal{O}(\epsilon^2) \, ,
    \end{split}
\end{equation}
This is of the required form in eq.\eqref{result3} we tried to prove.

\section{Details of the proof of the physical process first law}

\subsection*{Conservation of $\alpha^{\mu}$:}
\label{app:consalpha}

Here, we will quickly show why $\alpha^{\mu}$ given in eq.\eqref{eq:alphamu} is conserved. For this, we will need the following expression derived in eq.\eqref{eq:zetareln1}:
\begin{equation}\label{eq:alphaeq1}
	\Theta^{\mu} - \zeta^{\mu} \mathcal{L}_{(A_\mu g_{\mu\nu})} = - 2 \mathcal{E}^{\mu\nu}\zeta_{\nu} - \, G^{\mu} (A^{\nu}\zeta_{\nu} + \Lambda ) + D_{\nu} Q^{\mu\nu} \, .
\end{equation}
The LHS is the usual Noether current $N^{\mu} = \Theta^{\mu} - \zeta^{\mu} \mathcal{L}_{(A_\mu g_{\mu\nu})}$ in the language of \cite{doi:10.1063/1.528801}. The variation of the LHS of eq.\eqref{eq:alphaeq1} results in the following identity derived in \cite{Iyer:1994ys}:
\begin{equation}\label{eq:deltanmu}
	\delta N^{\mu} = \omega^{\mu}(\psi,\delta \psi, \mathcal{L}_{\zeta} \psi) + \dfrac{1}{2}D_{\nu}( \zeta^{\mu} \Theta^{\nu} - \zeta^{\nu} \Theta^{\mu} ) \, .
\end{equation}
Here $\psi$ collectively denotes the fields $A_{\mu}$ and $g_{\mu\nu}$. $\omega^{\mu}$ is defined by the anti-symmetrized variation of $\Theta^{\mu}$ \cite{doi:10.1063/1.528801} \footnote{Here we substitute $\delta_2 = \mathcal{L}_{\xi}$ to obtain eq.\eqref{eq:deltanmu}.}:
\begin{equation}
	\omega^{\mu}(\psi,\delta_1 \psi, \delta_2 \psi) = \delta_1 \Theta^{\mu}(\psi,\delta_2 \psi) - \delta_2 \Theta^{\mu}(\psi,\delta_1 \psi) \, .
\end{equation}
If we now restrict to the case $\zeta^{\mu} = \xi^{\mu}$ of eq.\eqref{BT_generator}, which is a Killing symmetry of the background spacetime and the fields $\psi$, then we have $\mathcal{L}_{\xi} \psi = 0$, and this implies
\begin{equation}\label{eq:symres1}
	\omega^{\mu}(\psi,\delta \psi, \mathcal{L}_{\xi}\psi) = 0 \, .
\end{equation}
Substituting eq.\eqref{eq:symres1} in eq.\eqref{eq:deltanmu}, we have
\begin{equation}\label{eq:ndiv}
	\delta N^{\mu} = \dfrac{1}{2}D_{\nu}( \xi^{\mu} \Theta^{\nu} - \xi^{\nu} \Theta^{\mu} ) \implies D_{\mu} \delta N^{\mu} = 0 \, .
\end{equation}
The variation of the RHS of eq.\eqref{eq:deltanmu} results in
\begin{equation}\label{eq:deltanrhs}
	\delta N^{\mu} = - \alpha^{\mu} + D_{\nu} \delta Q^{\mu\nu} \, ,
\end{equation}
where $\alpha^{\mu} = 2 \, \delta \mathcal{E}^{\mu\nu}\zeta_{\nu} + \, \delta G^{\mu} (A^{\nu}\zeta_{\nu} + \Lambda )$. Using eq.\eqref{eq:ndiv} in eq.\eqref{eq:deltanrhs}, we straightforwardly have
\begin{equation}
	D_{\mu} \alpha^{\mu} = 0 \, .
\end{equation}

\subsection*{Alternate perspective on the physical process first law:}
\label{app:ppflproofalt}

In this section, we wish to emphaisze the point of view taken in \cite{Chakraborty:2017kob} to prove the physical process version of the first law. This serves as an alternative perspective of the same. We start with the following generic structure of the entropy
\begin{equation}\label{eq:entdef}
	S = 4 \pi \int_{\mathcal{H}} d^{d-2}x \, \sqrt{h} \, (1 + s_w) \, ,
\end{equation}
where $s_w = s^{\text{HD}}_w + s_{cor}$. $s^{\text{HD}}_w$ denotes the Wald entropy contribution from the higher derivative terms and $s_{cor}$ denotes the out-of-equilibrium extension of the entropy in dynamical situations \cite{Wall:2015raa,Bhattacharya:2019qal,Bhattacharyya:2021jhr}. For concreteness, we consider a theory of the form
\begin{equation}\label{eq:applagppfl}
	\mathcal{L}_{(A_{\mu}g_{\mu\nu})} = \sqrt{-g} \left( R - \dfrac{1}{4} F_{\mu\nu}F^{\mu\nu} + L_{\text{non-min}} \right) \, .
\end{equation}
Thus, it is clear why we chose to explicitly mention the area term (contributed by the Einstein-Maxwell term) in eq.\eqref{eq:entdef}; the $L_{\text{non-min}}$ is of the form given in eq.\eqref{eq:laggao} and it will contribute to $s_w$. The equations of motion are thus
\begin{equation}\label{eq:eom}
	\mathcal{E}_{\mu\nu} = - R_{\mu\nu} + \dfrac{1}{2} g_{\mu\nu} R + T^{\text{EM}}_{\mu\nu} + \widetilde{T}_{\mu\nu} = 0 \, ,
\end{equation}
where $T^{\text{EM}}_{\mu\nu}$ denotes the contribution from the Maxwell Lagrangian and $\widetilde{T}_{\mu\nu}$ denotes the contribution from the higher derivative non-minimal couplings. We have to relate the change in entropy of eq.\eqref{eq:entdef} to the source term given in eq.\eqref{eq:firstlaw1}. The variation of the entropy given by eq.\eqref{eq:entdef} due to some infalling matter is given by \footnote{Here we are using our setup described in the gauge eq.\eqref{eq:metric}.}
\begin{equation}\label{eq:intres4}
	\delta S = 4 \pi \int_{\mathcal{H}} d^{d-2}x \, \dfrac{d}{dv}[\sqrt{h} \, (1 + s_w)] \, dv = 4 \pi \int_{\mathcal{H}} \sqrt{h} \, d^{d-2}x \int dv \, \vartheta_k \, ,
\end{equation}
where $\vartheta_k$ is the generalized expansion given by
\begin{equation}
	\vartheta_k = \theta_k + s_w \, \theta_k + \dfrac{d \, s_w}{d v} \,.
\end{equation}
Here $\theta_k = \dfrac{1}{\sqrt{h}}\partial_v \sqrt{h}$ is the expansion of null generators $k^{\mu} = (\partial_v)^{\mu}$ of the horizon within General Relativity. The generalized expansion $\vartheta_k$ reduces to $\theta_k$ when $s_w = 0$ as for GR. Using integration by parts in eq.\eqref{eq:intres4} to find the change in entropy between two different stationary slices $v= v_1$ and $v = v_2$ of the horizon, we have
\begin{equation}\label{eq:intres5}
	\delta S = 4 \pi \left( \int_{\mathcal{H}} \sqrt{h} \, d^{d-2}x \, v \, \vartheta_k \right)^{v_2}_{v_1} - 4 \pi \int_{\mathcal{H}} dA \, dv \, v \dfrac{d \, \vartheta_k}{dv} \, .
\end{equation}
Now,
\begin{equation}\label{eq:intres6}
	\begin{split}
		\dfrac{d \, \vartheta_k}{dv}|_{\mathcal{H}} &= (1+ s_w) \dfrac{d \, \theta_k}{dv} + \theta_k \dfrac{d \, s_w}{dv} + \dfrac{d^2 \, s_w}{dv^2} \\
		&= (1 + s_w) \left( - \dfrac{(d-3)}{d-2} \theta^2_k - \sigma^2_k - R_{\mu\nu} k^{\mu} k^{\nu} \right) + \theta_k \dfrac{d \, s_w}{dv} + \dfrac{d^2 \, s_w}{dv^2} \\
		& \approx - R_{\mu\nu} k^{\mu} k^{\nu} + \dfrac{d^2 \, s_w}{dv^2} - s_w \, R_{\mu\nu} k^{\mu} k^{\nu} + \mathcal{O}(\epsilon^2) \\
		&= \dfrac{d^2 \, s_w}{dv^2} - s_w \, R_{\mu\nu} k^{\mu} k^{\nu} - \widetilde{T}_{\mu\nu} k^{\mu}k^{\nu} - T^{\text{EM}}_{\mu\nu} k^{\mu}k^{\nu} - \dfrac{1}{2}\delta T_{\mu\nu} k^{\mu} k^{\nu} \\
		& \approx \dfrac{d^2 \, s_w}{dv^2} - s_w \, R_{\mu\nu} k^{\mu} k^{\nu} - \widetilde{T}_{\mu\nu} k^{\mu}k^{\nu} - \dfrac{1}{2}\delta T_{\mu\nu} k^{\mu} k^{\nu} + \mathcal{O}(\epsilon^2) \, .
	\end{split}
\end{equation}
Here, in the second step we have used the $d$ dimensional Raychaudhuri equation for the congruence of null generators $k^{\mu}$. In the third step, we have used the amplitude approximation of our gauge in eq.\eqref{eq:metdecomp} to cast away all the terms that are quadratic in the amplitude. In the fourth step, we have basically used the EOM eq.\eqref{eq:eom} with a source term $\dfrac{1}{2}\delta T_{\mu\nu}$. In the last step, we have used the fact that $T^{\text{EM}}_{\mu\nu} k^{\mu} k^{\nu} \approx \mathcal{O}(\epsilon^2) $ in our gauge \footnote{This essentially amounts to the fact that the Maxwell stress tensor satisfies the null energy condition.}.

Thus, using the final result of eq.\eqref{eq:intres6} in eq.\eqref{eq:intres5},
\begin{equation}\label{eq:intres7}
	\begin{split}
		\delta S &= 4 \pi \left( \int_{\mathcal{H}} \sqrt{h} \, d^{d-2}x \, v \, \vartheta_k \right)^{v_2}_{v_1} + 2 \pi \int_{\mathcal{H}} dA \, dv \, v \, \delta T_{\mu\nu} k^{\mu} k^{\nu} \\
		& ~~ - 4 \pi \int_{\mathcal{H}} dA \, dv \, v \, \left( \dfrac{d^2 \, s_w}{dv^2} - s_w \, R_{\mu\nu} k^{\mu} k^{\nu} - \widetilde{T}_{\mu\nu} k^{\mu}k^{\nu} \right) \, .
	\end{split}
\end{equation}
Since we have assumed that the black hole was stationary before the matter fell in and it finally settles down to a stationary state after the perturbation has ended, we can set the boundary term (first term in the above equation) to zero. Now we use the main result of our paper eq.\eqref{eq:AmuTvv} which proves that even when we consider a non-minimally coupled gauge theory, $\widetilde{T}_{vv}$ still has the entropy current structure of the following form when evaluated on the horizon:
\begin{equation}\label{eq:Tstruc}
	\widetilde{T}_{\mu\nu}k^{\mu}k^{\nu}|_{\mathcal{H}} =\widetilde{T}_{vv}|_{r=0}=\partial_v\left[\frac{1}{\sqrt{h}}\partial_v\left(\sqrt{h} \, \widetilde{\cal J}^v\right)+\nabla_i \widetilde{\cal J}^i\right]+{\cal O}(\epsilon^2) \, .
\end{equation}
We can thus use the results of section 5 of \cite{Bhattacharyya:2021jhr} to basically state that
\begin{equation}\label{eq:ppflres}
	\int_{\mathcal{H}} dA \, dv \, v \, \left( \dfrac{d^2 \, s_w}{dv^2} - s_w \, R_{\mu\nu} k^{\mu} k^{\nu} - \widetilde{T}_{\mu\nu} k^{\mu}k^{\nu} \right) = \mathcal{O}(\epsilon^2) \, .
\end{equation}
This has been argued for in eq.(5.7) to eq.(5.10) of \cite{Bhattacharyya:2021jhr} and the main requirement was the fact that $\widetilde{T}_{vv}$ had the entropy current structure given by eq.\eqref{eq:Tstruc}. Finally, eq.\eqref{eq:intres7} becomes
\begin{equation}
	\delta S =  2 \pi \int_{\mathcal{H}} dA \, dv \, v \, \delta T_{\mu\nu} k^{\mu} k^{\nu} + \mathcal{O}(\epsilon^2) \, .
\end{equation}
The Killing field $\xi^{\mu}$ and the null generators of the horizon $k^{\mu}$ are related as $\xi^{\mu} = \kappa \, v \, k^{\mu}$, where $\kappa$ is the surface gravity. Thus, finally we get
\begin{equation}\label{eq:firstlaw2}
	\delta S = \dfrac{2 \pi}{\kappa} \int_{\mathcal{H}} \delta T_{\mu\nu} k^{\mu} \xi^{\nu} \, .
\end{equation}
Combining eq.\eqref{eq:firstlaw1} and eq.\eqref{eq:firstlaw2}, we get the desired first law relation that we originally set out to prove
\begin{equation}
	\delta M - \Omega_H \delta I - \Phi_{bh} \delta q = \dfrac{\kappa}{2 \pi} \delta S \, .
\end{equation}
The changes in the mass $M$, angular momentum $I$, charge $q$ and the entropy $S$ are induced by some physical matter falling into the black hole. This completes the proof of the physical process first law for arbitrary non-minimally coupled gauge theories. We emphasize that the non-trivial part of the proof was basically the structure of the entropy current eq.\eqref{eq:Tstruc} leading to the result of eq.\eqref{eq:ppflres}. The result of eq.\eqref{eq:ppflres} was first established for theories of gravity in \cite{Bhattacharyya:2021jhr} and we have extended it to non-minimally coupled gauge theories here. 

\subsection*{Verification of eq.(\ref{eq:ppflres}) through an example:}
\label{app:ppfleg}

In the proof given above, eq.\eqref{eq:ppflres} was the crucial step. We argued that this relation is true even for arbitrary non-minimal coupling of gauge fields with gravity. To illustrate this through an explicit example, let us consider a gauge invariant Lagrangian of the form given in eq.\eqref{eq:applagppfl}
\begin{equation}\label{eq:gaugeexlag}
	\mathcal{L} = \sqrt{-g} \left( R - \dfrac{1}{4} F_{\mu\nu} F^{\mu\nu} + \alpha \, R_{\mu\nu\rho\sigma} F^{\mu\nu} F^{\rho\sigma} \right)\, .
\end{equation}
The standard variation gives \footnote{The symmetrization of indices $\alpha$ and $\beta$ has a factor of $\dfrac{1}{2}$.}
\begin{equation}
	\begin{split}
		\delta \mathcal{L} &= \sqrt{-g} \left[ -R^{\alpha\beta} + \dfrac{1}{2}g^{\alpha\beta} R + \dfrac{1}{2} \left( F^{\alpha\nu} F^{\beta}_{~\nu} - \dfrac{1}{4} g^{\alpha\beta} F_{\mu\nu} F^{\mu\nu} \right) \right] \delta g_{\alpha\beta} \\
		&+ \alpha \, \sqrt{-g} \Bigg[\dfrac{1}{2} g^{\alpha\beta} R_{\mu\nu\rho\sigma} F^{\mu\nu} F^{\rho\sigma} - 3 R^{(\alpha}_{~~\nu\rho\sigma} F^{\beta)\nu} F^{\rho\sigma} + 2 D_{\nu} D_{\rho} (F^{\alpha\nu}F^{\rho\beta}) \Bigg] \delta g_{\alpha\beta} \\
		&+ ~\sqrt{-g}\left[ D_{\mu} F^{\mu\sigma} -4 \alpha D_{\rho} \left( R^{\rho\sigma}_{\hspace{0.4cm}\mu\nu} F^{\mu\nu} \right) \right] \delta A_{\sigma} \\
		& + \sqrt{-g} D_{\rho} \left[ g^{\rho\alpha} D^{\sigma} \delta g_{\alpha\sigma} - g^{\alpha\sigma} D^{\mu} \delta g_{\alpha\sigma} - F^{\rho\nu} \delta A_{\nu} \right]  \\
		&+ \alpha \, \sqrt{-g} D_{\rho} \Bigg[ 2 F^{\mu\nu} F^{\rho\sigma} D_{\nu} \delta g_{\sigma\mu} - 2 D_{\nu}(F^{\mu\rho}F^{\nu\sigma})\delta g_{\sigma\mu} + 4 R^{\rho\sigma}_{\hspace{0.4cm}\mu\nu} F^{\mu\nu} \delta A_{\sigma} \Bigg] \, .
	\end{split}
\end{equation}
The total derivative term, $\theta^{\rho}$ is given by
\begin{equation}\label{eq:thetaexgauge}
	\begin{split}
		\theta^{\rho} &= g^{\rho\alpha} D^{\sigma} \delta g_{\alpha\sigma} - g^{\alpha\sigma} D^{\mu} \delta g_{\alpha\sigma} - F^{\rho\nu} \delta A_{\nu} \\
		&~~ + \alpha \left[ 2 F^{\mu\nu} F^{\rho\sigma} D_{\nu} \delta g_{\sigma\mu} - 2 D_{\nu}(F^{\mu\rho}F^{\nu\sigma})\delta g_{\sigma\mu} + 4 R^{\rho\sigma}_{\hspace{0.4cm}\mu\nu} F^{\mu\nu} \delta A_{\sigma} \right] \,.
	\end{split}
\end{equation}
The Noether charge $Q^{\rho\nu}$ is given by
\begin{equation}\label{eq:qexgauge}
	\begin{split}
		Q^{\rho\nu} &= D^{\nu} \xi^{\rho} - D^{\rho} \xi^{\nu} - F^{\rho\nu} ( A_{\alpha}\xi^{\alpha} + \Lambda ) \\
		& ~~ +\alpha \, \left[ 2 F^{\rho\nu} F^{\mu\sigma}D_{\sigma}\xi_{\mu} + 2 D_{\sigma}(F^{\rho\mu}F^{\sigma\nu}-F^{\rho\sigma}F^{\mu\nu})\xi_{\mu} \right. \\
		& ~~ + \left. 2 D_{\mu} (F^{\rho\nu}F^{\mu\sigma})\xi_{\sigma} + 4 R^{\rho\nu}_{\hspace{0.4cm}\alpha\beta}F^{\alpha\beta} (A_{\gamma}\xi^{\gamma} + \Lambda) \right] \, .
	\end{split}
\end{equation}
From these expressions, one can readily obtain the following
\begin{equation}\label{eq:currenteg}
	\mathcal{E}_{vv}|_{r=0} = \partial_v \left[ \dfrac{1}{\sqrt{h}} \partial_v (\sqrt{h} \, \left[ 1 - 2 \alpha \, F^{rv} F^{rv} \right]) - \nabla_i (4 \alpha \, F^{rv} F^{ri} ) \right] \,.
\end{equation}
Thus, we have the components of the Entropy current
\begin{equation}
	\mathcal{J}^v = 1 - 2 \alpha \, F^{rv} F^{rv}, \hspace{3cm} \mathcal{J}^i = -4 \alpha \, F^{ri} F^{rv} \,.
\end{equation}
If we filter out the non-minimal component of the EOM,
\begin{equation}\label{eq:currentegnonmin}
	\widetilde{T}_{vv}|_{r=0} = \partial_v \left[ \dfrac{1}{\sqrt{h}} \partial_v (\sqrt{h} \, \left[ - 2 \alpha \, F^{rv} F^{rv} \right]) - \nabla_i (4 \alpha \, F^{rv} F^{ri} ) \right] + \mathcal{O}(\epsilon^2) \,.
\end{equation}
From $\mathcal{J}^v$ of eq.\eqref{eq:currenteg}, we can read off the expression for entropy from eq.\eqref{eq:entdef} (or from eq.(5.6) of \cite{Bhattacharyya:2021jhr}) as
\begin{equation}
	S = 4 \pi \int_{\mathcal{H}} \sqrt{h} \, d^{d-2}x \, \left( 1 - 2 \alpha \, F^{rv} F^{rv} \right) \, ,
\end{equation}
and $s_w = - 2 \alpha \, F^{rv} F^{rv}$. Thus, we have 
\begin{equation}
	\begin{split}
		\dfrac{d^2 \, s_w}{dv^2} - s_w \, R_{\mu\nu} k^{\mu} k^{\nu} - \widetilde{T}_{\mu\nu} k^{\mu}k^{\nu}  & = \partial_v \left[ \dfrac{1}{\sqrt{h}} \partial_v ( \sqrt{h} \, s_w ) \right] - \widetilde{T}_{vv} \\
		& = \partial_v \left[ \nabla_i (4 \alpha \, F^{rv} F^{ri}) \right] + \mathcal{O}(\epsilon^2) \, ,
	\end{split}
\end{equation}
where we have used eq.\eqref{eq:currentegnonmin} in the second step. Finally, we have eq.\eqref{eq:ppflres}:
\begin{equation}
	\begin{split}
		\int_{\mathcal{H}} dA \, dv \, v \, \left( \dfrac{d^2 \, s_w}{dv^2} \right. & \left. - s_w \, R_{\mu\nu} k^{\mu} k^{\nu} - \widetilde{T}_{\mu\nu} k^{\mu}k^{\nu} \right) \\
		& = \int_{\mathcal{H}} dA \, dv \, v \, \partial_v \left[ \nabla_i (4 \alpha \, F^{rv} F^{ri}) \right] + \mathcal{O}(\epsilon^2) = \mathcal{O}(\epsilon^2) \, ,
	\end{split}
\end{equation}
where we have used the assumption of compact horizons in the final step to set the total derivative term to zero. This illustrates the proof of the physical process first law from the perspective of \cite{Chakraborty:2017kob} for a particular example.

\bibliographystyle{JHEP}
\bibliography{nonminimal}

\providecommand{\href}[2]{#2}\begingroup\raggedright\begin{thebibliography}{10}

\bibitem{Hawking:1971tu}
S.~W. Hawking, \emph{{Gravitational radiation from colliding black holes}},
  \href{http://dx.doi.org/10.1103/PhysRevLett.26.1344}{\emph{Phys. Rev. Lett.}
  {\bfseries 26} (1971) 1344--1346}.

\bibitem{Bardeen:1973gs}
J.~M. Bardeen, B.~Carter and S.~W. Hawking, \emph{{The Four laws of black hole
  mechanics}}, \href{http://dx.doi.org/10.1007/BF01645742}{\emph{Commun. Math.
  Phys.} {\bfseries 31} (1973) 161--170}.

\bibitem{Bekenstein:1973ur}
J.~D. Bekenstein, \emph{{Black holes and entropy}},
  \href{http://dx.doi.org/10.1103/PhysRevD.7.2333}{\emph{Phys. Rev. D}
  {\bfseries 7} (1973) 2333--2346}.

\bibitem{Hawking:1974sw}
S.~W. Hawking, \emph{{Particle Creation by Black Holes}},
  \href{http://dx.doi.org/10.1007/BF02345020}{\emph{Commun. Math. Phys.}
  {\bfseries 43} (1975) 199--220}.

\bibitem{Wald:1993nt}
R.~M. Wald, \emph{{Black hole entropy is the Noether charge}},
  \href{http://dx.doi.org/10.1103/PhysRevD.48.R3427}{\emph{Phys. Rev. D}
  {\bfseries 48} (1993) R3427--R3431},
  [\href{https://arxiv.org/abs/gr-qc/9307038}{{\ttfamily gr-qc/9307038}}].

\bibitem{Iyer:1994ys}
V.~Iyer and R.~M. Wald, \emph{{Some properties of Noether charge and a proposal
  for dynamical black hole entropy}},
  \href{http://dx.doi.org/10.1103/PhysRevD.50.846}{\emph{Phys. Rev.} {\bfseries
  D50} (1994) 846--864}, [\href{https://arxiv.org/abs/gr-qc/9403028}{{\ttfamily
  gr-qc/9403028}}].

\bibitem{Jacobson:1995uq}
T.~Jacobson, G.~Kang and R.~C. Myers, \emph{{Increase of black hole entropy in
  higher curvature gravity}},
  \href{http://dx.doi.org/10.1103/PhysRevD.52.3518}{\emph{Phys. Rev.}
  {\bfseries D52} (1995) 3518--3528},
  [\href{https://arxiv.org/abs/gr-qc/9503020}{{\ttfamily gr-qc/9503020}}].

\bibitem{Gao:2001ut}
S.~Gao and R.~M. Wald, \emph{{The `Physical process' version of the first law
  and the generalized second law for charged and rotating black holes}},
  \href{http://dx.doi.org/10.1103/PhysRevD.64.084020}{\emph{Phys. Rev.}
  {\bfseries D64} (2001) 084020},
  [\href{https://arxiv.org/abs/gr-qc/0106071}{{\ttfamily gr-qc/0106071}}].

\bibitem{Amsel:2007mh}
A.~J. Amsel, D.~Marolf and A.~Virmani, \emph{{The Physical Process First Law
  for Bifurcate Killing Horizons}},
  \href{http://dx.doi.org/10.1103/PhysRevD.77.024011}{\emph{Phys. Rev.}
  {\bfseries D77} (2008) 024011},
  [\href{https://arxiv.org/abs/0708.2738}{{\ttfamily 0708.2738}}].

\bibitem{Bhattacharjee:2014eea}
S.~Bhattacharjee and S.~Sarkar, \emph{{Physical process first law and caustic
  avoidance for Rindler horizons}},
  \href{http://dx.doi.org/10.1103/PhysRevD.91.024024}{\emph{Phys. Rev.}
  {\bfseries D91} (2015) 024024},
  [\href{https://arxiv.org/abs/1412.1287}{{\ttfamily 1412.1287}}].

\bibitem{Chakraborty:2017kob}
A.~Mishra, S.~Chakraborty, A.~Ghosh and S.~Sarkar, \emph{{On the physical
  process first law for dynamical black holes}},
  \href{http://dx.doi.org/10.1007/JHEP09(2018)034}{\emph{JHEP} {\bfseries 09}
  (2018) 034}, [\href{https://arxiv.org/abs/1709.08925}{{\ttfamily
  1709.08925}}].

\bibitem{Chatterjee:2011wj}
A.~Chatterjee and S.~Sarkar, \emph{{Physical process first law and increase of
  horizon entropy for black holes in Einstein-Gauss-Bonnet gravity}},
  \href{http://dx.doi.org/10.1103/PhysRevLett.108.091301}{\emph{Phys. Rev.
  Lett.} {\bfseries 108} (2012) 091301},
  [\href{https://arxiv.org/abs/1111.3021}{{\ttfamily 1111.3021}}].

\bibitem{Kolekar:2012tq}
S.~Kolekar, T.~Padmanabhan and S.~Sarkar, \emph{{Entropy Increase during
  Physical Processes for Black Holes in Lanczos-Lovelock Gravity}},
  \href{http://dx.doi.org/10.1103/PhysRevD.86.021501}{\emph{Phys. Rev.}
  {\bfseries D86} (2012) 021501},
  [\href{https://arxiv.org/abs/1201.2947}{{\ttfamily 1201.2947}}].

\bibitem{Bhattacharyya:2021jhr}
S.~Bhattacharyya, P.~Dhivakar, A.~Dinda, N.~Kundu, M.~Patra and S.~Roy,
  \emph{{An entropy current and the second law in higher derivative theories of
  gravity}}, \href{http://dx.doi.org/10.1007/JHEP09(2021)169}{\emph{JHEP}
  {\bfseries 09} (2021) 169},
  [\href{https://arxiv.org/abs/2105.06455}{{\ttfamily 2105.06455}}].

\bibitem{Zerothwald}
I.~R{\'{a}}cz and R.~M. Wald, \emph{Global extensions of spacetimes describing
  asymptotic final states of black holes},
  \href{http://dx.doi.org/10.1088/0264-9381/13/3/017}{\emph{Classical and
  Quantum Gravity} {\bfseries 13} (mar, 1996) 539--552}.

\bibitem{Bhattacharyya:2022nqa}
S.~Bhattacharyya, P.~Biswas, A.~Dinda and N.~Kundu, \emph{{The zeroth law of
  black hole thermodynamics in arbitrary higher derivative theories of
  gravity}},  \href{https://arxiv.org/abs/2205.01648}{{\ttfamily 2205.01648}}.

\bibitem{Ghosh:2020dkk}
R.~Ghosh and S.~Sarkar, \emph{{Black Hole Zeroth Law in Higher Curvature
  Gravity}}, \href{http://dx.doi.org/10.1103/PhysRevD.102.101503}{\emph{Phys.
  Rev. D} {\bfseries 102} (2020) 101503},
  [\href{https://arxiv.org/abs/2009.01543}{{\ttfamily 2009.01543}}].

\bibitem{Reall:2021voz}
H.~S. Reall, \emph{{Causality in gravitational theories with second order
  equations of motion}},
  \href{http://dx.doi.org/10.1103/PhysRevD.103.084027}{\emph{Phys. Rev. D}
  {\bfseries 103} (2021) 084027},
  [\href{https://arxiv.org/abs/2101.11623}{{\ttfamily 2101.11623}}].

\bibitem{Xie:2021bur}
Y.~Xie, J.~Zhang, H.~O. Silva, C.~de~Rham, H.~Witek and N.~Yunes, \emph{{Square
  Peg in a Circular Hole: Choosing the Right Ansatz for Isolated Black Holes in
  Generic Gravitational Theories}},
  \href{http://dx.doi.org/10.1103/PhysRevLett.126.241104}{\emph{Phys. Rev.
  Lett.} {\bfseries 126} (2021) 241104},
  [\href{https://arxiv.org/abs/2103.03925}{{\ttfamily 2103.03925}}].

\bibitem{Sang:2021rla}
A.~Sang and J.~Jiang, \emph{{Black hole zeroth law in the Horndeski gravity}},
  \href{http://dx.doi.org/10.1103/PhysRevD.104.084092}{\emph{Phys. Rev. D}
  {\bfseries 104} (2021) 084092},
  [\href{https://arxiv.org/abs/2110.00903}{{\ttfamily 2110.00903}}].

\bibitem{Jacobson:1993xs}
T.~Jacobson and R.~C. Myers, \emph{{Black hole entropy and higher curvature
  interactions}},
  \href{http://dx.doi.org/10.1103/PhysRevLett.70.3684}{\emph{Phys. Rev. Lett.}
  {\bfseries 70} (1993) 3684--3687},
  [\href{https://arxiv.org/abs/hep-th/9305016}{{\ttfamily hep-th/9305016}}].

\bibitem{Jacobson:1993vj}
T.~Jacobson, G.~Kang and R.~C. Myers, \emph{{On black hole entropy}},
  \href{http://dx.doi.org/10.1103/PhysRevD.49.6587}{\emph{Phys. Rev.}
  {\bfseries D49} (1994) 6587--6598},
  [\href{https://arxiv.org/abs/gr-qc/9312023}{{\ttfamily gr-qc/9312023}}].

\bibitem{Wall:2011hj}
A.~C. Wall, \emph{{A proof of the generalized second law for rapidly changing
  fields and arbitrary horizon slices}},
  \href{http://dx.doi.org/10.1103/PhysRevD.87.069904,
  10.1103/PhysRevD.85.104049}{\emph{Phys. Rev.} {\bfseries D85} (2012) 104049},
  [\href{https://arxiv.org/abs/1105.3445}{{\ttfamily 1105.3445}}].

\bibitem{Sarkar:2013swa}
S.~Sarkar and A.~C. Wall, \emph{{Generalized second law at linear order for
  actions that are functions of Lovelock densities}},
  \href{http://dx.doi.org/10.1103/PhysRevD.88.044017}{\emph{Phys. Rev.}
  {\bfseries D88} (2013) 044017},
  [\href{https://arxiv.org/abs/1306.1623}{{\ttfamily 1306.1623}}].

\bibitem{Bhattacharjee:2015yaa}
S.~Bhattacharjee, S.~Sarkar and A.~C. Wall, \emph{{Holographic entropy
  increases in quadratic curvature gravity}},
  \href{http://dx.doi.org/10.1103/PhysRevD.92.064006}{\emph{Phys. Rev.}
  {\bfseries D92} (2015) 064006},
  [\href{https://arxiv.org/abs/1504.04706}{{\ttfamily 1504.04706}}].

\bibitem{Wall:2015raa}
A.~C. Wall, \emph{{A Second Law for Higher Curvature Gravity}},
  \href{http://dx.doi.org/10.1142/S0218271815440149}{\emph{Int. J. Mod. Phys.}
  {\bfseries D24} (2015) 1544014},
  [\href{https://arxiv.org/abs/1504.08040}{{\ttfamily 1504.08040}}].

\bibitem{Bhattacharjee:2015qaa}
S.~Bhattacharjee, A.~Bhattacharyya, S.~Sarkar and A.~Sinha, \emph{{Entropy
  functionals and c-theorems from the second law}},
  \href{http://dx.doi.org/10.1103/PhysRevD.93.104045}{\emph{Phys. Rev.}
  {\bfseries D93} (2016) 104045},
  [\href{https://arxiv.org/abs/1508.01658}{{\ttfamily 1508.01658}}].

\bibitem{Bhattacharyya:2016xfs}
S.~Bhattacharyya, F.~M. Haehl, N.~Kundu, R.~Loganayagam and M.~Rangamani,
  \emph{{Towards a second law for Lovelock theories}},
  \href{http://dx.doi.org/10.1007/JHEP03(2017)065}{\emph{JHEP} {\bfseries 03}
  (2017) 065}, [\href{https://arxiv.org/abs/1612.04024}{{\ttfamily
  1612.04024}}].

\bibitem{Wall:2018ydq}
A.~C. Wall, \emph{{A Survey of Black Hole Thermodynamics}},
  \href{https://arxiv.org/abs/1804.10610}{{\ttfamily 1804.10610}}.

\bibitem{Sarkar:2019xfd}
S.~Sarkar, \emph{{Black Hole Thermodynamics: General Relativity and Beyond}},
  \href{http://dx.doi.org/10.1007/s10714-019-2545-y}{\emph{Gen. Rel. Grav.}
  {\bfseries 51} (2019) 63},
  [\href{https://arxiv.org/abs/1905.04466}{{\ttfamily 1905.04466}}].

\bibitem{Bhattacharya:2019qal}
J.~Bhattacharya, S.~Bhattacharyya, A.~Dinda and N.~Kundu, \emph{{An entropy
  current for dynamical black holes in four-derivative theories of gravity}},
  \href{http://dx.doi.org/10.1007/JHEP06(2020)017}{\emph{JHEP} {\bfseries 06}
  (2020) 017}, [\href{https://arxiv.org/abs/1912.11030}{{\ttfamily
  1912.11030}}].

\bibitem{Hollands:2022fck}
S.~Hollands, A.~D. Kov\'acs and H.~S. Reall, \emph{{The second law of black
  hole mechanics in effective field theory}},
  \href{https://arxiv.org/abs/2205.15341}{{\ttfamily 2205.15341}}.

\bibitem{carroll_2019}
S.~M. Carroll, \emph{Spacetime and Geometry: An Introduction to General
  Relativity}.
\newblock Cambridge University Press, 2019,
  \href{http://dx.doi.org/10.1017/9781108770385}{10.1017/9781108770385}.

\bibitem{blau}
M.~Blau, \emph{{Lecture notes on General Relativity}}.
\newblock 2022,
  \href{http://dx.doi.org/http://www.blau.itp.unibe.ch/newlecturesGR.pdf}{http://www.blau.itp.unibe.ch/newlecturesGR.pdf}.

\bibitem{Barcelo:2000zf}
C.~Barcelo and M.~Visser, \emph{{Scalar fields, energy conditions, and
  traversable wormholes}},
  \href{http://dx.doi.org/10.1088/0264-9381/17/18/318}{\emph{Class. Quant.
  Grav.} {\bfseries 17} (2000) 3843--3864},
  [\href{https://arxiv.org/abs/gr-qc/0003025}{{\ttfamily gr-qc/0003025}}].

\bibitem{Flanagan_1996}
E.~E. Flanagan and R.~M. Wald, \emph{Does back reaction enforce the averaged
  null energy condition in semiclassical gravity?},
  \href{http://dx.doi.org/10.1103/physrevd.54.6233}{\emph{Physical Review D}
  {\bfseries 54} (nov, 1996) 6233--6283}.

\bibitem{Chatterjee:2015uya}
S.~Chatterjee, M.~Parikh and J.~P. van~der Schaar, \emph{{On Coupling
  NEC-Violating Matter to Gravity}},
  \href{http://dx.doi.org/10.1016/j.physletb.2015.03.020}{\emph{Phys. Lett. B}
  {\bfseries 744} (2015) 34--37},
  [\href{https://arxiv.org/abs/1503.07950}{{\ttfamily 1503.07950}}].

\bibitem{Wang:2020svl}
X.-Y. Wang and J.~Jiang, \emph{{Investigating the Linearized Second Law in
  Horndeski Gravity}},
  \href{http://dx.doi.org/10.1103/PhysRevD.102.084020}{\emph{Phys. Rev. D}
  {\bfseries 102} (2020) 084020},
  [\href{https://arxiv.org/abs/2008.09774}{{\ttfamily 2008.09774}}].

\bibitem{Wang:2021zyt}
X.-Y. Wang and J.~Jiang, \emph{{Generalized proof of the linearized second law
  in general quadric corrected Einstein-Maxwell gravity}},
  \href{http://dx.doi.org/10.1103/PhysRevD.104.064007}{\emph{Phys. Rev. D}
  {\bfseries 104} (2021) 064007},
  [\href{https://arxiv.org/abs/2108.04402}{{\ttfamily 2108.04402}}].

\bibitem{Gao:2003ys}
S.~Gao, \emph{{The First law of black hole mechanics in Einstein-Maxwell and
  Einstein-Yang-Mills theories}},
  \href{http://dx.doi.org/10.1103/PhysRevD.68.044016}{\emph{Phys. Rev. D}
  {\bfseries 68} (2003) 044016},
  [\href{https://arxiv.org/abs/gr-qc/0304094}{{\ttfamily gr-qc/0304094}}].

\bibitem{Prabhu:2015vua}
K.~Prabhu, \emph{{The First Law of Black Hole Mechanics for Fields with
  Internal Gauge Freedom}},
  \href{http://dx.doi.org/10.1088/1361-6382/aa536b}{\emph{Class. Quant. Grav.}
  {\bfseries 34} (2017) 035011},
  [\href{https://arxiv.org/abs/1511.00388}{{\ttfamily 1511.00388}}].

\bibitem{Elgood:2020svt}
Z.~Elgood, P.~Meessen and T.~Ort\'\i{}n, \emph{{The first law of black hole
  mechanics in the Einstein-Maxwell theory revisited}},
  \href{http://dx.doi.org/10.1007/JHEP09(2020)026}{\emph{JHEP} {\bfseries 09}
  (2020) 026}, [\href{https://arxiv.org/abs/2006.02792}{{\ttfamily
  2006.02792}}].

\bibitem{PhysRevD.86.021501}
S.~Kolekar, T.~Padmanabhan and S.~Sarkar, \emph{Entropy increase during
  physical processes for black holes in lanczos-lovelock gravity},
  \href{http://dx.doi.org/10.1103/PhysRevD.86.021501}{\emph{Phys. Rev. D}
  {\bfseries 86} (Jul, 2012) 021501}.

\bibitem{Bhattacharyya:2022njk}
S.~Bhattacharyya, P.~Jethwani, M.~Patra and S.~Roy, \emph{{Reparametrization
  Symmetry of Local Entropy Production on a Dynamical Horizon}},
  \href{https://arxiv.org/abs/2204.08447}{{\ttfamily 2204.08447}}.

\bibitem{doi:10.1063/1.528801}
J.~Lee and R.~M. Wald, \emph{Local symmetries and constraints},
  \href{http://dx.doi.org/10.1063/1.528801}{\emph{Journal of Mathematical
  Physics} {\bfseries 31} (1990) 725--743},
  [\href{https://arxiv.org/abs/https://doi.org/10.1063/1.528801}{{\ttfamily
  https://doi.org/10.1063/1.528801}}].

\end{thebibliography}\endgroup

\end{document}